\documentclass[pdflatex,sn-mathphys-num]{sn-jnl}

\usepackage{graphicx}%
\usepackage{multirow}%
\usepackage{amsmath,amssymb,amsfonts}%
\usepackage{amsthm}%
\usepackage{mathrsfs}%
\usepackage[title]{appendix}%
\usepackage{xcolor}%
\usepackage{textcomp}%
\usepackage{manyfoot}%
\usepackage{booktabs}%
\usepackage{algorithm}%
\usepackage{algorithmicx}%
\usepackage{algpseudocode}%
\usepackage{listings}%

\usepackage{subfigure} 
\usepackage{enumitem}
\usepackage{url}
\usepackage{epsfig}
\usepackage{caption}
\usepackage{float}
\usepackage{array}
\usepackage{soul}
\usepackage{dcolumn}
\usepackage{tabularx}
\usepackage{here}
\usepackage{hyperref}
\usepackage{pslatex}          
\usepackage{makecell}
\usepackage{microtype}
\usepackage[normalem]{ulem}
\usepackage[utf8]{inputenc}
\usepackage[T1]{fontenc}
\usepackage{makecell}
\usepackage{color}
\usepackage{natbib}
\usepackage{setspace}
\usepackage{lineno}
\modulolinenumbers[5]

\theoremstyle{thmstyleone}%
\theoremstyle{thmstyletwo}%
\theoremstyle{thmstylethree}%
\begin{document}
\title[]{Unveiling the magnetic behavior of C$_{\rm3}$N$_{\rm4}$ 2D material by defect creation, defect passivation, and transition metal adsorption}

\author*[1]{\fnm{Taoufik} \sur{Sakhraoui}}\email{taoufik.sakhraoui@osu.cz}
\author[1]{\fnm{František} \sur{Karlický}}\email{frantisek.karlicky@osu.cz}
\affil[1]{\orgdiv{Department of Physics}, \orgname{Faculty of Science, University of Ostrava}, \orgaddress{\street{30. dubna
22}, \city{Ostrava}, \postcode{70103}, \country{Czech Republic}}}
\abstract{ Using the density functional tight binding method (DFTB) and the GFN1-xTB (Geometries, Frequencies, and Noncovalent interactions Tight Binding) Hamiltonian, we have investigated the structural, electronic and magnetic properties of vacancy defects, hydrogen and oxygen passivated defects, and Fe adsorption in two-dimensional (2D) graphitic carbon nitride (gt-C$_{\rm3}$N$_{\rm4}$) 2D material. The ring shape is the most preferred vacancy evolution path, with significant stability of the semicircle fourfold C-N-C-N vacancy. We found that bare gt-C$_{\rm3}$N$_{\rm4}$ which is non-magnetic becomes magnetic by 2-, and 5-defects creation, hydrogen/oxygen passivation of the defects, and upon Fe adsorption. Interestingly, Fe atoms interact with the gt-C$_{\rm3}$N$_{\rm4}$ sheet and result in a ground ferromagnetic (FM) state. In addition, we investigate the effects of passivating the vacancies by hydrogen in gt-C$_{\rm3}$N$_{\rm4}$ on its structural, electrical, and magnetic properties. We found that substituting the 1, 2, and 3 vacancies with hydrogen and passivating the 6-defect with oxygen turns on magnetism in the system. Due to structural distortion, the passivated defects do not have a well-ordered magnetic orientation. However, passivating the remaining defected structures maintains the nonmagnetic state. It is also shown that passivation leads to a semiconductor with a band gap value higher than that of the bare material. We also calculate the electronic and magnetic properties of transition metal (TM) atoms, including V, Cr, Mn, Fe, Co, Ni-adsorbed gt-C$_{\rm3}$N$_{\rm4}$ monolayer. All TM atoms show slight lattice distortion, and the adsorbed system almost maintains the original structure type. Moreover, a FM alignment was observed with a total magnetic moments of 2.89 $\mu_{\rm B}$, 2 $\mu_{\rm B}$, and 1 $\mu_{\rm B}$ for V, Fe, and Co atoms, respectively. The Cr, Mn, and Ni atoms induce no magnetism to the non magnetic gt-C$_{\rm3}$N$_{\rm4}$ system. 
}


\maketitle

\section{\label{sec1}Introduction}
After the great success of graphene \cite{Novoselov2004, Geim2007} since 2004, 2D materials have attracted remarkable interest. Among 2D structures, graphitic carbon-nitride (C$_{\rm x}$N$_{\rm y}$) have been extensively explored in recent years \cite{Mortzavi2020, Algara-Siller2014, Sakhraoui_2023, LuisFrancisco2020}. C$_{\rm x}$N$_{\rm y}$ consists of a porous structure made of carbon and nitrogen covalently bonded \cite{Mortzavi2020}. In particular, s-triazine and tri-triazine graphitic carbon nitride (g-C$_{\rm3}$N$_{\rm4}$) 2D materials have been successfully synthesized \cite{Algara-Siller2014, LuisFrancisco2020} and show fascinating characteristics \cite{Zheng2011, Thomas2008, Zhu2014}. 
g-C$_{{\rm3}}$N$_{{\rm4}}$, analogous to graphene material, is arranged in a hexagonal grid and has nitrogen atoms bound through strong covalent sp$^{{\rm2}}$ bonds. At room temperature, two geometrical structural have been widely proposed for the graphitic phase (g-C$_{{\rm3}}$N$_{{\rm4}}$) of carbon nitride, namely, gh-C$_{{\rm3}}$N$_{{\rm4}}$ (based on tri-s-triazine or s-heptazine units (C$_{{\rm6}}$N$_{{\rm7}}$)) and gt-C$_{{\rm3}}$N$_{{\rm4}}$ (based on s-triazine units (C$_{{\rm3}}$N$_{{\rm3}}$)) \cite{Botari2017, Miller2017}. The s-triazine in gt-C$_{{\rm3}}$N$_{{\rm4}}$ units and the tri-s-triazine subunits in gh-C$_{{\rm3}}$N$_{{\rm4}}$ are linked by a trigonal nitrogen atom.

On the other hand, the formation of defects in 2D material structures was shown to be unavoidable, such as graphene \cite{Banhart2011, Vicarelli2015, Lopez-Polin2015}, BN \cite{Wong2015, Liu2012, Bourrellier2016}, MXene \cite{Sang2016, Karlsson2015, Sakhraoui_2022, Halim2016, Lipatov2016}, and MoS$_{\rm 2}$ \cite{Zhou2013, Martin2015}. The influence of different types of defects on the electrical, electrochemical, electronic, optoelectronic, and mechanical properties of 2D materials is extensively studied \cite{Zhou2013, Banhart2011, Sang2016, Vicarelli2015, Liu2012}. The properties of 2D materials \cite{Lopez-Polin2015, Zhou2013} could be tailored by controlled defect formation. The formation of defects in graphene was found to lead to the weakening of its mechanical strength \cite{Lopez-Polin2015, Wei2012} and its electronic performance \cite{Yazyev2010}. Compared to graphene and 2D materials, defect formation was observed in graphitic carbon nitride \cite{Yuhan2023, Dong2020, Xue2019}. 

Understanding the effect of defects in C$_{\rm3}$N$_{\rm4}$ material should be of great importance. The presence of defects in C$_{\rm3}$N$_{\rm4}$ may be the origin of the regulation of electronic and optical properties \cite{Liu2024}. Authors reported for the first time the successful fabrication of C-defected gCN (Cv-gCN) {\it via} thermal treatment of pristine gCN in CO$_{\rm2}$ atmosphere using electron paramagnetic resonance (EPR) and X-ray photoelectron spectroscopy (XPS) \cite{YuhanLi2018}. Y. Li et al. reviewed the studies carried out on the optimization of the carrier mobility dynamics of g-C$_{{\rm3}}$N$_{{\rm4}}$ compound {\it via} the internal and external modification strategies \cite{YuhanLi2020}. Defects in C$_{\rm3}$N$_{\rm4}$ can be generated mainly by C-\cite{Peng2021, Wang2025, Kumar2021}, and N-vacancies \cite{Tay2015, Daopeng2024, Peng2021, Wang2025, Kumar2021} and also by metal and/or non-metal doping \cite{Wang2023, Nur2021, Zhu2017}. The experimental technique for the production of the defected C$_{\rm3}$N$_{\rm4}$ was conducted  via a calcination-solvothermal-calcination method \cite{Quanhao2020, Daopeng2024}. L. Yong et al. showed, using a combined experimental and theoretical study, that N defects can be introduced into the g-C$_{\rm3}$N$_{\rm4}$ compound in all nitrogen sites \cite{Liu2024}. To our knowledge, most studies reported the use of defected C$_{\rm3}$N$_{\rm4}$ for  catalytic/photocatalytic applications \cite{Wang2025, Peng2021, Daopeng2024, Liu2024}. H. Shaoqi et al. reviewed the advancement of defect‑engineered g‑C$_{\rm3}$N$_{\rm4}$ for solar catalytic applications \cite{Hou2024}. S. Jiao et al. reviewed the formation process of the defects, along with the emerging preparation methods to generate imperfections in the electrocatalysts. They also presented a basic-theory-and-practice-combined introduction of the advanced characterization methods for the identification and analysis of defects \cite{Shilong2020}. D. Zhong et al. reported an experimental investigation of nitrogen defects in C$_{\rm3}$N$_{\rm4}$ for  highly efficient H$_{\rm2}$ evolution \cite{Daopeng2024}. The effect of vacancy defects on the magnetic properties of g-C$_{\rm3}$N$_{\rm4}$ has not been extensively studied \cite{Peng2021}. On the other hand, most studies report defects in gh-C$_{\rm3}$N$_{\rm4}$ \cite{Liu2024, Sun2022, Peng2021, Kumar2021}. Hence, a detailed understanding of the effect of different types of defects on the electronic structure and magnetic properties of gt-C$_{\rm3}$N$_{\rm4}$ material may provide new insight into the design of efficient C/N defect-C$_{\rm3}$N$_{\rm4}$-based magnetic applications. 

Moreover, the substitution of vacancies with different species has been shown to be an effective method to modify electronic and optical properties \cite{Aiqing2020}. The passivation of nitrogen and carbon vacancy defects is essential for improving material performance. It may allow for controllable electronic and magnetic properties of the material. Different experimental methods for defect passivation have been proposed, including thermal annealing \cite{Nan2014, Tongay2013}, and chemical treatments \cite{Kim2018, Atallah2017, ZhangSiyuan2018}. Recently, a dangling bond passivation process was shown for graphite material \cite{Xiang2024} and ZnO nanoparticles \cite{Keliang_2023}. The authors in ref. \cite{Keliang_2023} show that the passivation of oxygen vacancy defects in ZnO nanoparticles allows the improvement of the optical reflectance and the electrical conductivity stability of the nanoparticles. To our knowledge, the hydrogen passivation of vacancies in gt-C$_{\rm3}$N$_{\rm4}$ 2D material and how it impacts the electronic and magnetic properties has not yet been investigated. 

In the present work, we performed DFTB calculations based on the xTB Hamiltonian to study the influence of vacancy defects, hydrogen and oxygen passivation of the defects, and Fe adsorption on the structural stability, electronic behavior, and magnetic properties of gt-C$_{\rm3}$N$_{\rm4}$ 2D material.

\section{Theoretical details}
The DFTB+ software package \cite{dftb_2007, dftb_2020}, with the GFN-xTB methodology \cite{grimme-xTB1_2017}, was used to perform DFTB calculations of the structural, electronic and magnetic properties of gt-C$_{\rm3}$N$_{\rm4}$. GFN-xTB (shortly, xTB) is a new extended tight-binding method, recently developed by Grimme et al. \cite{grimme-xTB1_2017}, that covers most elements of the periodic table up to Z = 86. It allows for efficient quantum mechanical simulations of large systems and long-time scales as an efficient alternative to density functional theory (DFT). Please, see refs \cite{grimme-xTB1_2017, grimme-xTB2_2021} for a detailed description of the method and refs \cite{Taoufik2024, Vincent2021, Maryam2022} for its application on some systems. Validation of the reliability of the DFTB calculations was already performed in carbon-based structures \cite{Xinru_2024, Zhang_2023, Sakhraoui_2023, Sakhraoui_2021, Kang_2017}. In ref. \cite{Kang_2017}, authors show that DFTB calculations were appropriate to predict the band gap of $\alpha$-graphyne nanotubes ($\alpha$-GNTs), with a level of precision comparable to that of the conventional Perdew-Burke-Ernzerhof (PBE) functional of the DFT method. The Brillouin zone of the C$_{\rm3}$N$_{\rm4}$ unit cell was sampled using a $\Gamma$-point centered 12$\times$12$\times$1 Monkhorst-Pack grid \cite{Monkhorst_Pack_1976}. The supercell structures of each defected, defect-passivated, and Fe-adsorbed system were fully optimized with a Brillouin zone containing $\Gamma$-point only. All atoms were free to move under periodic boundary conditions. As the convergence threshold, the electronic steps and the force are set at 10$^{-5}$ eV and 10$^{-3}$ eV/\AA{}, respectively. A vacuum layer of 20 \AA{}{} was inserted along the z direction to prevent spurious interaction of periodic images.
\begin{table}[h]
\centering
\begin{tabular*}{\textwidth}{@{\extracolsep\fill}lccccc}
\toprule%
     \multicolumn{3}{@{}c@{}}{gh-C$_{\rm3}$N$_{\rm4}$} & \multicolumn{3}{@{}c@{}}{gt-C$_{\rm3}$N$_{\rm4}$} \\
\hline
    & (DFTB-xTB) & DFT/HSE && (DFTB-xTB) & DFT/HSE \\
a=b & 7.06 & 7.15\cite{Zhu2017}, 7.10\cite{Aspera_2010} && 4.74 & 4.74\cite{Bafekry2019}, 4.79\cite{Wang2014} \\ [0.3em]
    & & HSE: 7.10\cite{Yali2019}7.13\cite{Liu2015} && & 4.76\cite{Yali2019}, 4.79\cite{Pratap2022} \\ [0.3em]
    & & 7.13 (exp.)\cite{Wang2009} && & \\ [0.3em]
C1-N1 & 1.46 & 1.47\cite{Zhu2017}, 1.48\cite{Aspera_2010} && 1.45 & 1.44\cite{Bafekry2019}, 1.47\cite{Wang2014} \\ [0.3em]
    & & HSE: 1.46\cite{Yali2019} && & 1.46\cite{Yali2019}, 1.47\cite{Pratap2022} \\ [0.3em]
C1-N2 & 1.32 & 1.33\cite{Aspera_2010, Zhu2017} && 1.31 & 1.32\cite{Bafekry2019}, 1.33\cite{Wang2014} \\ [0.3em]
   & & HSE: 1.33\cite{Yali2019} && & 1.32\cite{Yali2019}, 1.33\cite{Pratap2022} \\ [0.3em]
C2-N2 & 1.31 & 1.34\cite{Zhu2017}, 1.33\cite{Aspera_2010} && - & - \\ [0.3em]
 &  & HSE: 1.32\cite{Yali2019} && - & - \\ [0.3em]
C2-N3 & 1.38 & 1.39\cite{Aspera_2010, Zhu2017} && - & - \\ [0.3em]
 &  & HSE: 1.39\cite{Yali2019} && - & - \\ [0.3em]
${\rm E_g}$ & 1.85 & 1.18\cite{Zhu2017}, 1.00\cite{Aspera_2010}, 2.17\cite{Fiorentin2021} && 2.28 & 1.45\cite{Bafekry2019}, 2.70\cite{Wang2014}, 2.48\cite{Fiorentin2021} \\ [0.3em]
 &  & HSE: 2.80\cite{Mortzavi2020}, 2.70\cite{Yali2019, Liu2016}, 2.73\cite{Liu2015} &&  & 3.17\cite{Mortzavi2020}, 2.98\cite{Yali2019} \\ [0.3em]
 & & 2.70 (exp.)\cite{Han2017, Wang2009} && & \\
    
\botrule
\end{tabular*} 
\caption{Calculated lattice parameters a, bond lengths (C1-N1, C1-N2, C2-N2, and C2-N3) (in units of \AA{}) and the electronic band gap (${\rm E_g}$, in units of eV) values of gh-C$_{\rm3}$N$_{\rm4}$ and gt-C$_{\rm3}$N$_{\rm4}$ calculated by xTB-DFTB method and compared to DFT/HSE data.}
\label{tab:geom}
\end{table}
\section{Results and discussion}
\subsection{Structure of gt-C$_{{\rm3}}$N$_{{\rm4}}$}
\begin{figure}[H]
{\centering
\captionsetup{justification=centering} 
\subfigure[ ] {\epsfig{figure=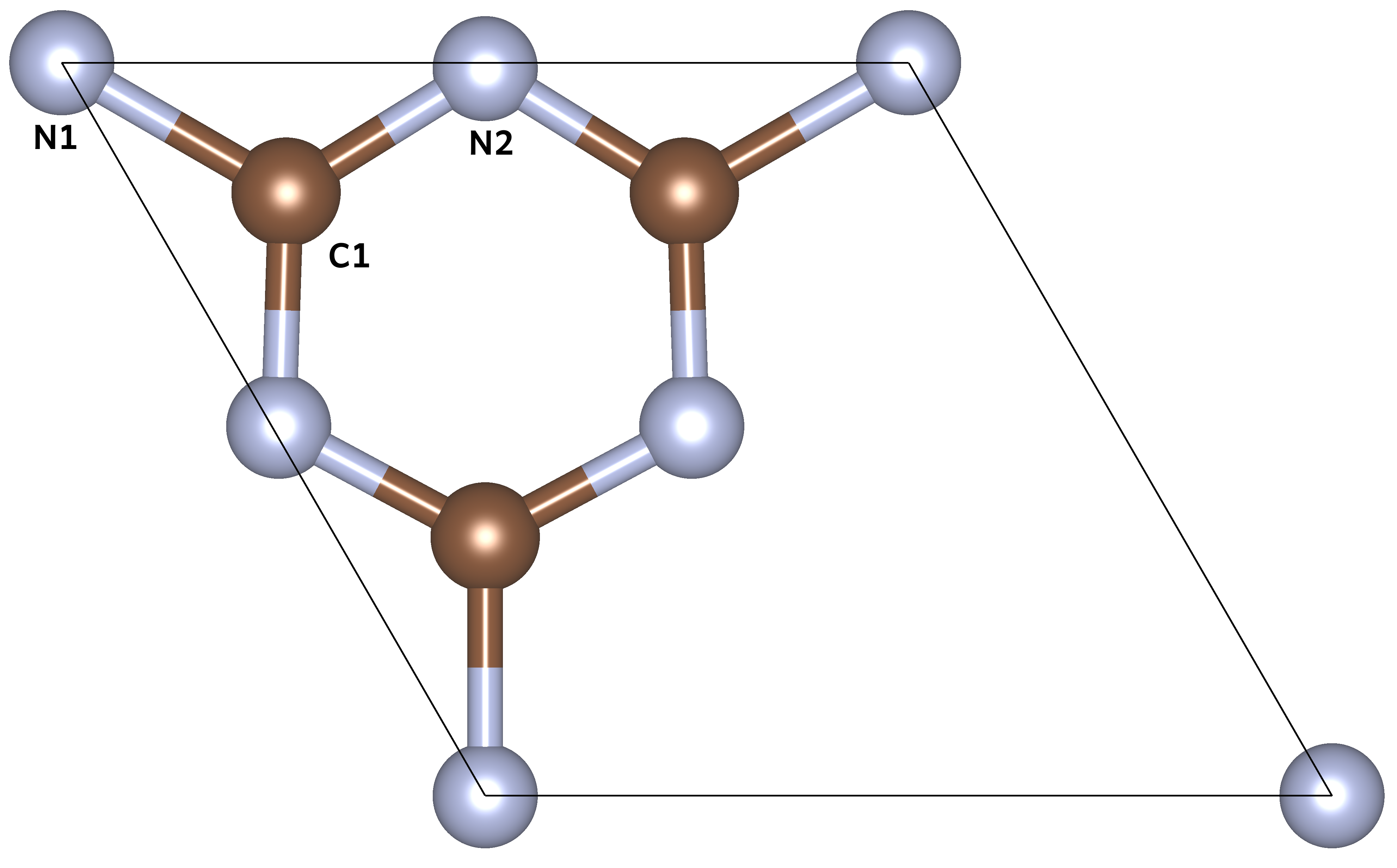, width=0.45\textwidth}} \quad
\subfigure[ ] {\epsfig{figure=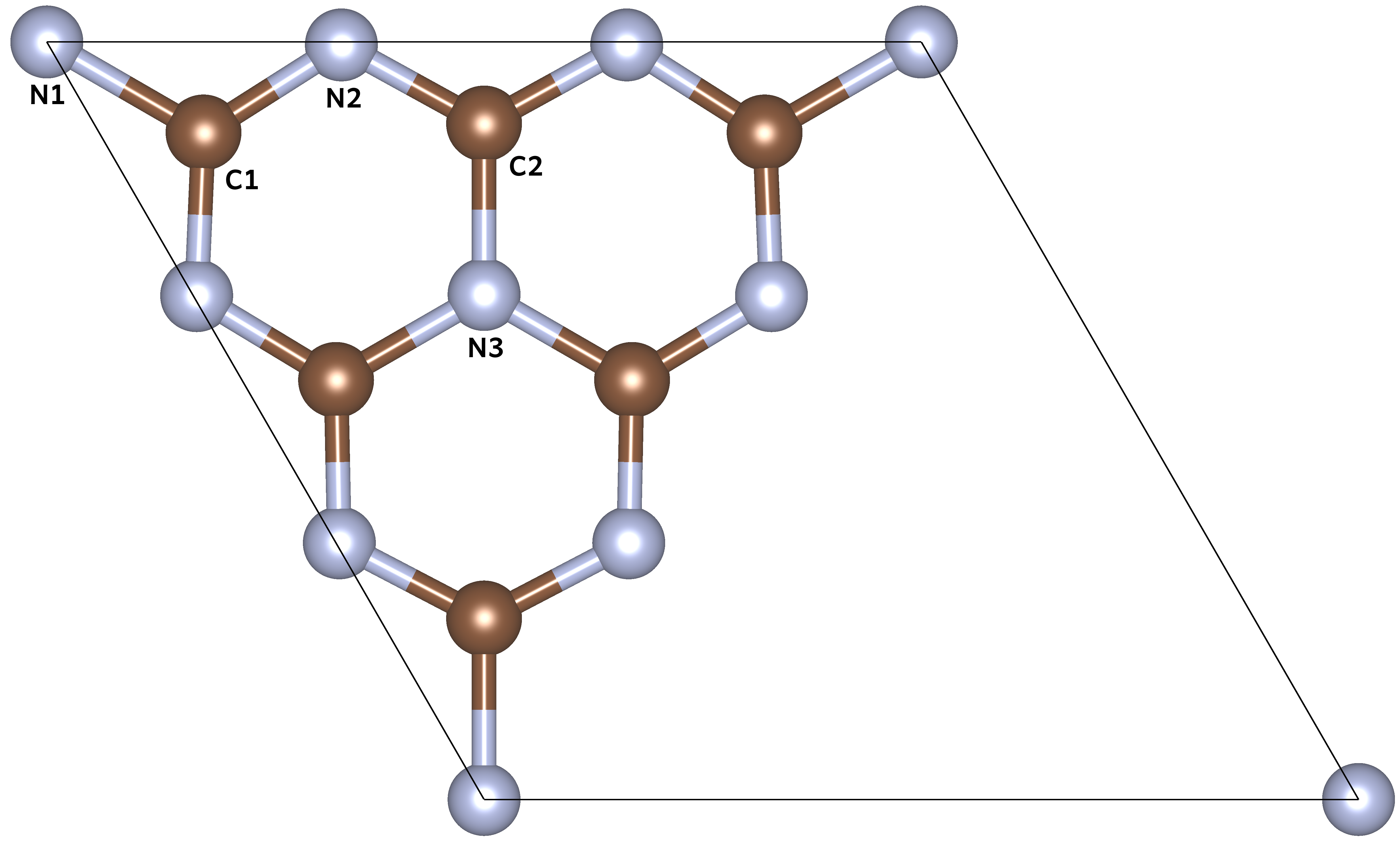, width=0.45\textwidth}} }
\caption{Models of g-C$_{{\rm3}}$N$_{{\rm4}}$: (a) gt-C$_{{\rm3}}$N$_{{\rm4}}$ and (b) gh-C$_{{\rm3}}$N$_{{\rm4}}$.}
\label{fig:fig1}
\end{figure}

Two forms were shown for the triazine-based g-C$_{\rm3}$N$_{\rm4}$, which represent two different atomic lattices, namely gt-C$_{\rm3}$N$_{\rm4}$ and gh-C$_{\rm3}$N$_{\rm4}$, made from s-triazine (C$_{\rm3}$N$_{\rm3}$) or tri-s-triazine (heptazine rings) building blocks connected by single N links. The g-C$_{\rm3}$N$_{\rm4}$ forms a 2D defect-rich and nitrogen-bridged honeycomb lattice. The models of the two forms of g-C$_{\rm3}$N$_{\rm4}$ are represented in Fig. \ref{fig:fig1}.

In the current study, calculations were performed using the DFTB method. Although, the method's capability in simulating carbon-based materials is well-documented, it is essential to start by validation its predictions of the lattice parameters, bond distances, and energy band gaps of gh-C$_{\rm3}$N$_{\rm4}$ and gt-C$_{\rm3}$N$_{\rm4}$. Results are listed in Table \ref{tab:geom}. In gh-C$_{\rm3}$N$_{\rm4}$, there are three and two nonequivalent N (N1, N2, and N3) and C (C1 and C2) atoms  respectively. Whereas, for the gt-C$_{\rm3}$N$_{\rm4}$ monolayer, there are one type of C atoms and two  nonequivalent N atoms (N1 and N2). Comparison of lattice parameters shows a small deviation from the reported DFT values. Moreover, we find that the xTB calculated bond lengths in both gh-C$_{\rm3}$N$_{\rm4}$ and gt-C$_{\rm3}$N$_{\rm4}$ are in excellent agreement with DFT. We hence conclude that our used method is suitable for the study of the graphitic carbon nitride.
Regarding structural stability, the experimental feasibility and/or the stability of a theoretically modeled 2D material is evaluated by its cohesive energy per atom with a higher absolute value corresponding to higher stability \cite{Zhang2018}. The cohesive energy, being the most important physical parameter in quantifying the stability of materials, is a measure of the strength of forces that bind atoms together in a crystal \cite{Zhang2018, Shein2012}. 
Here, we evaluated the energetic stability of both phases of the g-C$_{{\rm3}}$N$_{{\rm4}}$ by calculating their normalized cohesive energies ($\Bar{E}_{{\rm coh}}$) as follows: $\Bar{E}_{\rm coh}=[E{\rm (C_xN_y)}-x\times E({\rm C})-y\times E({\rm N})]/(x+y)$, \\
where $E({\rm C_xN_y})$, $E$(C) and $E$(N) are the total energies of g-C$_{{\rm3}}$N$_{{\rm4}}$, the energies of isolated C and N atoms, respectively. A more negative normalized cohesive energy indicates an energetically more stable structure. For the gt-C$_{{\rm3}}$N$_{{\rm4}}$ phase, $x$ and $y$ are equal to 3 and 4, respectively. However, for the gh-C$_{{\rm3}}$N$_{{\rm4}}$ phase, $x$=6 and $y$=8. $\Bar{E}_{{\rm coh}}$ are -6782.385 meV/atom and -6775.850 meV/atom for gt-C$_{{\rm3}}$N$_{{\rm4}}$ and gh-C$_{{\rm3}}$N$_{{\rm4}}$, respectively. The difference is a few meV only; therefore, we consider both structures to possess similar stability. In the remainder of the present work, we consider only the gt-C$_{{\rm3}}$N$_{{\rm4}}$ phase to investigate the electronic and magnetic properties and the effect of defects, hydrogen and oxygen passivation, and Fe adsorption on gt-C$_{{\rm3}}$N$_{{\rm4}}$. In gt-C$_{{\rm3}}$N$_{{\rm4}}$ 2D material, there are two inequivalent nitrogen atoms (N1 and N2) as shown in Fig. \ref{fig:fig1}. The calculated lattice parameter is found to be 4.74 \AA{} of the pure gt-C$_{{\rm3}}$N$_{{\rm4}}$, which compares well with previous theoretical reports \cite{Sakhraoui_2023, Xu_2019, Bafekry_2020, Rangraz_2021}. 
\subsection{Defected structures}
Due to its large surface area-to-volume ratio, gt-C$_{{\rm3}}$N$_{{\rm4}}$ monolayer has a high tendency to form defects that can significantly change its physicochemical properties \cite{Yuhan2023, Dong2020, Xue2019}. A 5$\times$5$\times$1 supercell containing 175 atoms of gt-C$_{{\rm3}}$N$_{{\rm4}}$ is employed for modeling the defects in gt-C$_{{\rm3}}$N$_{{\rm4}}$ system. 
\begin{figure}[H]
\centering
\subfigure[ ] {\epsfig{figure=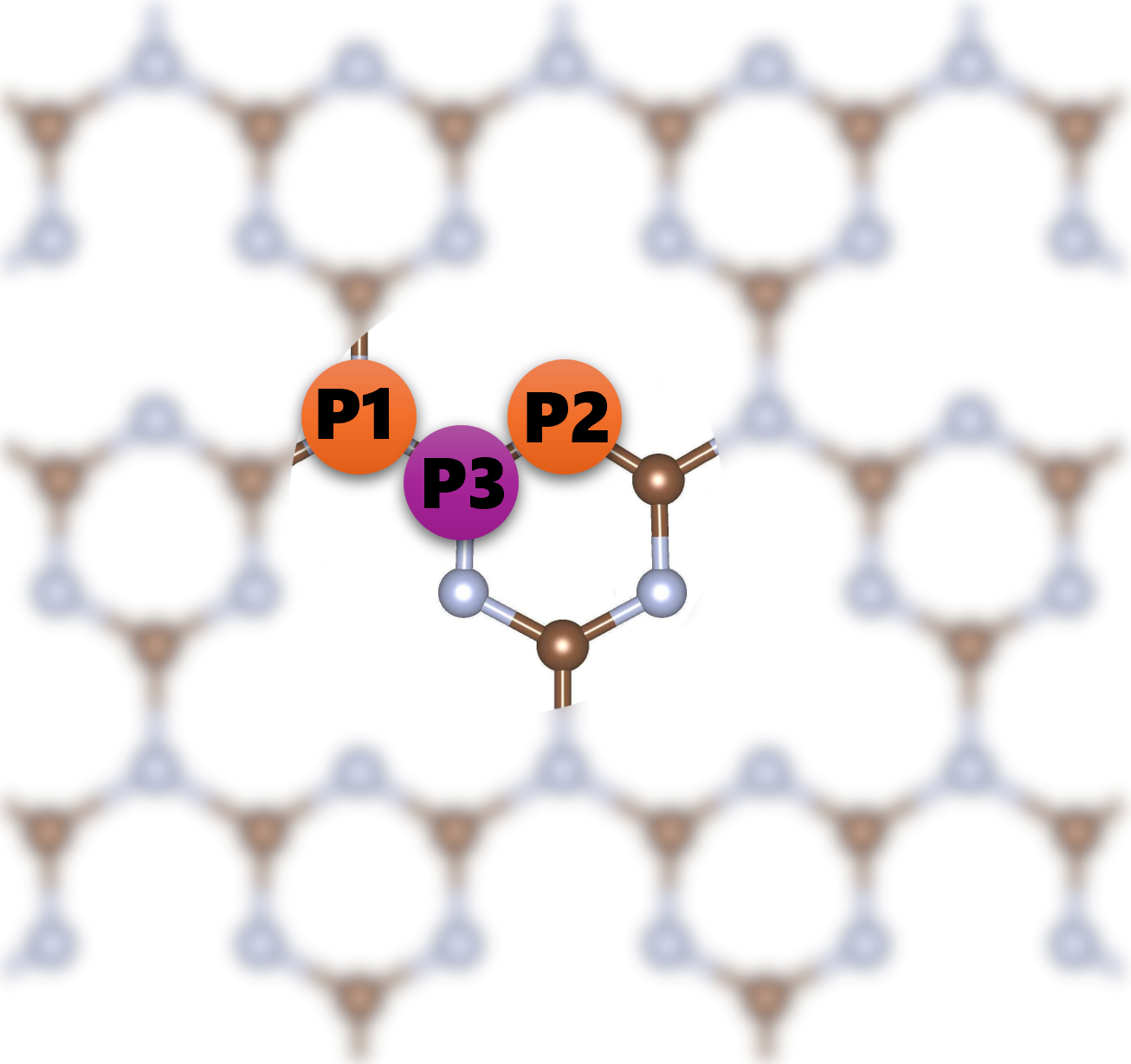, width=0.30\textwidth}} \quad
\subfigure[ ] {\epsfig{figure=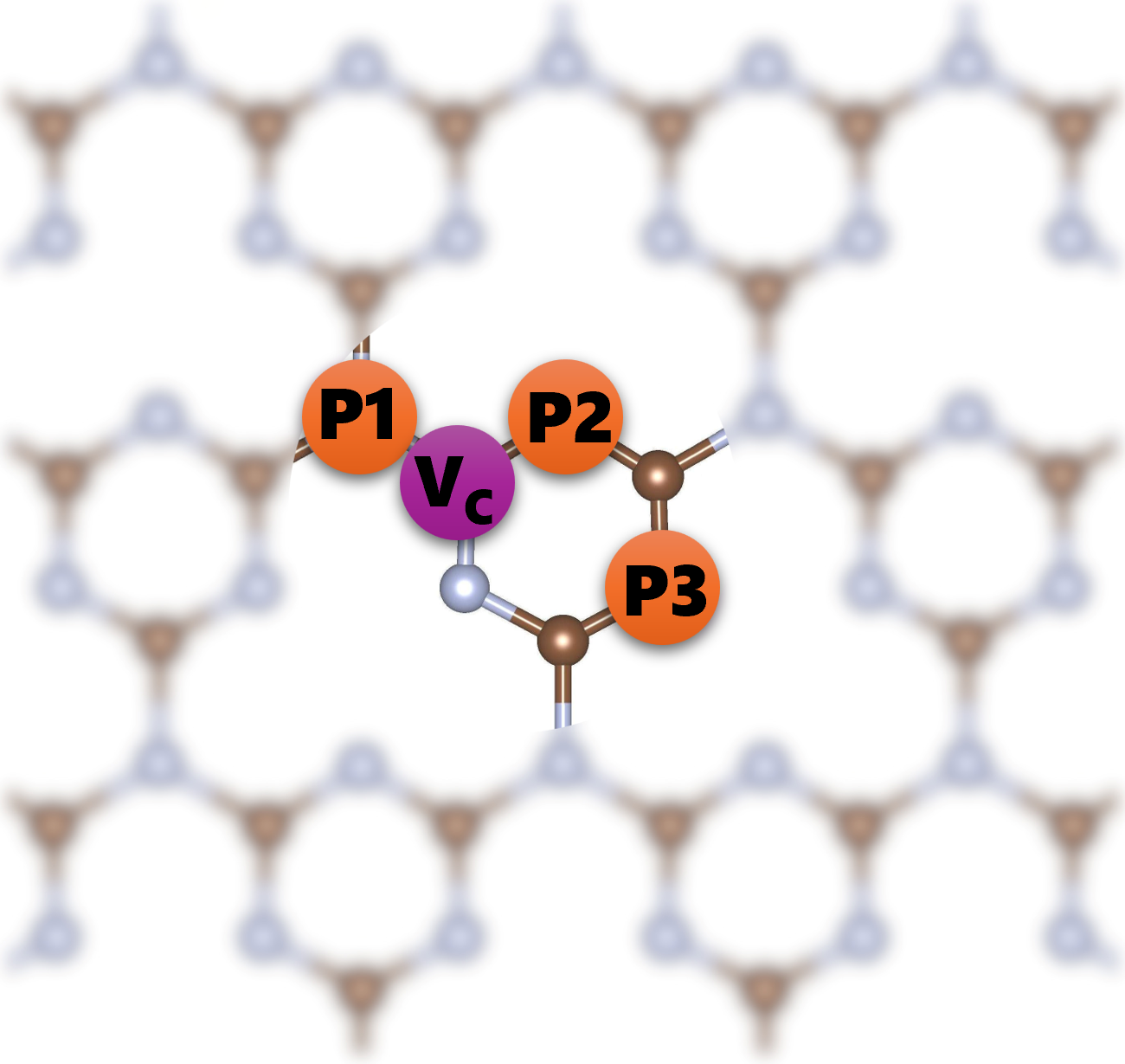, width=0.30\textwidth}} \quad
\subfigure[ ] {\epsfig{figure=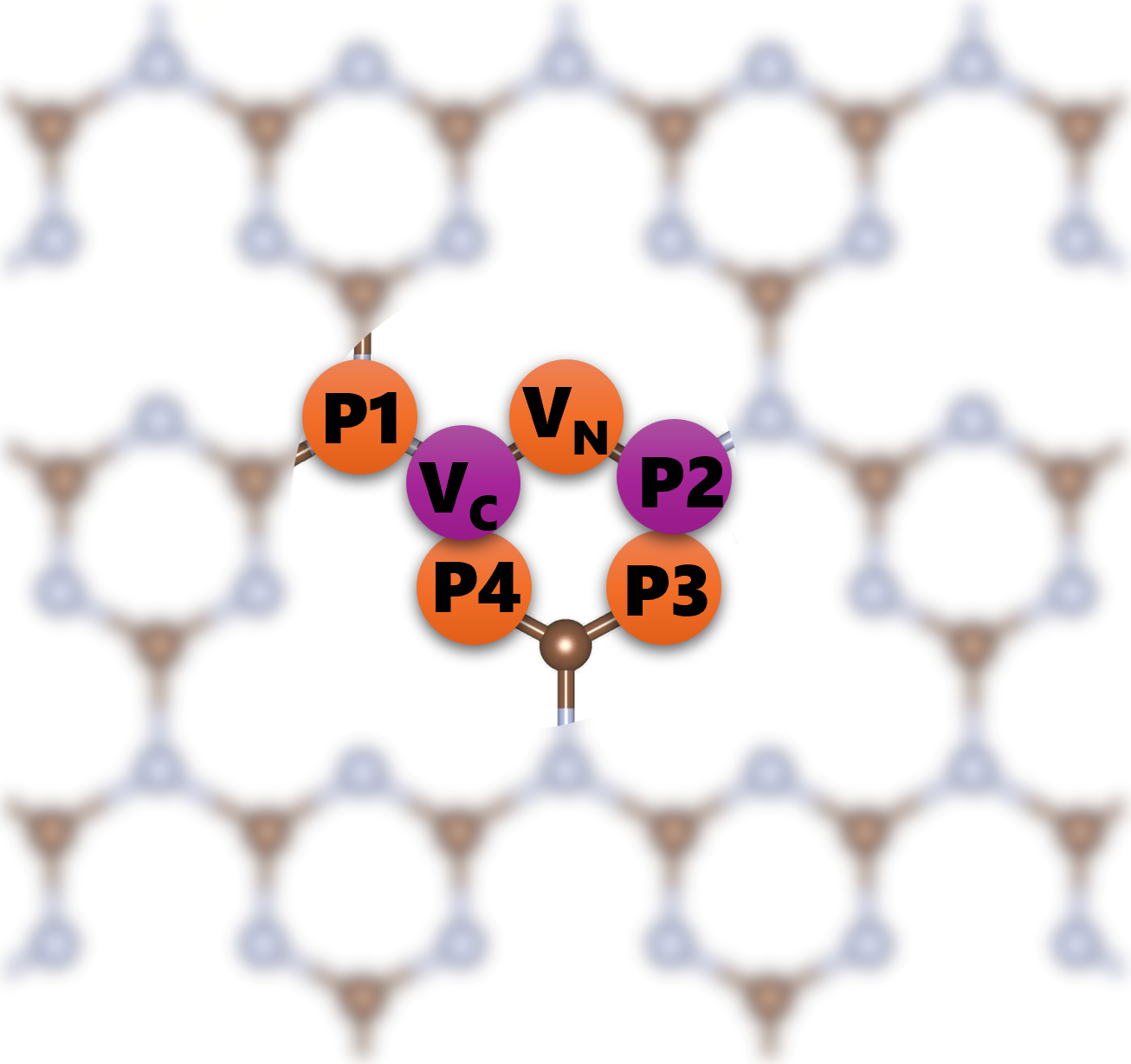, width=0.30\textwidth}} \\
\subfigure[ ] {\epsfig{figure=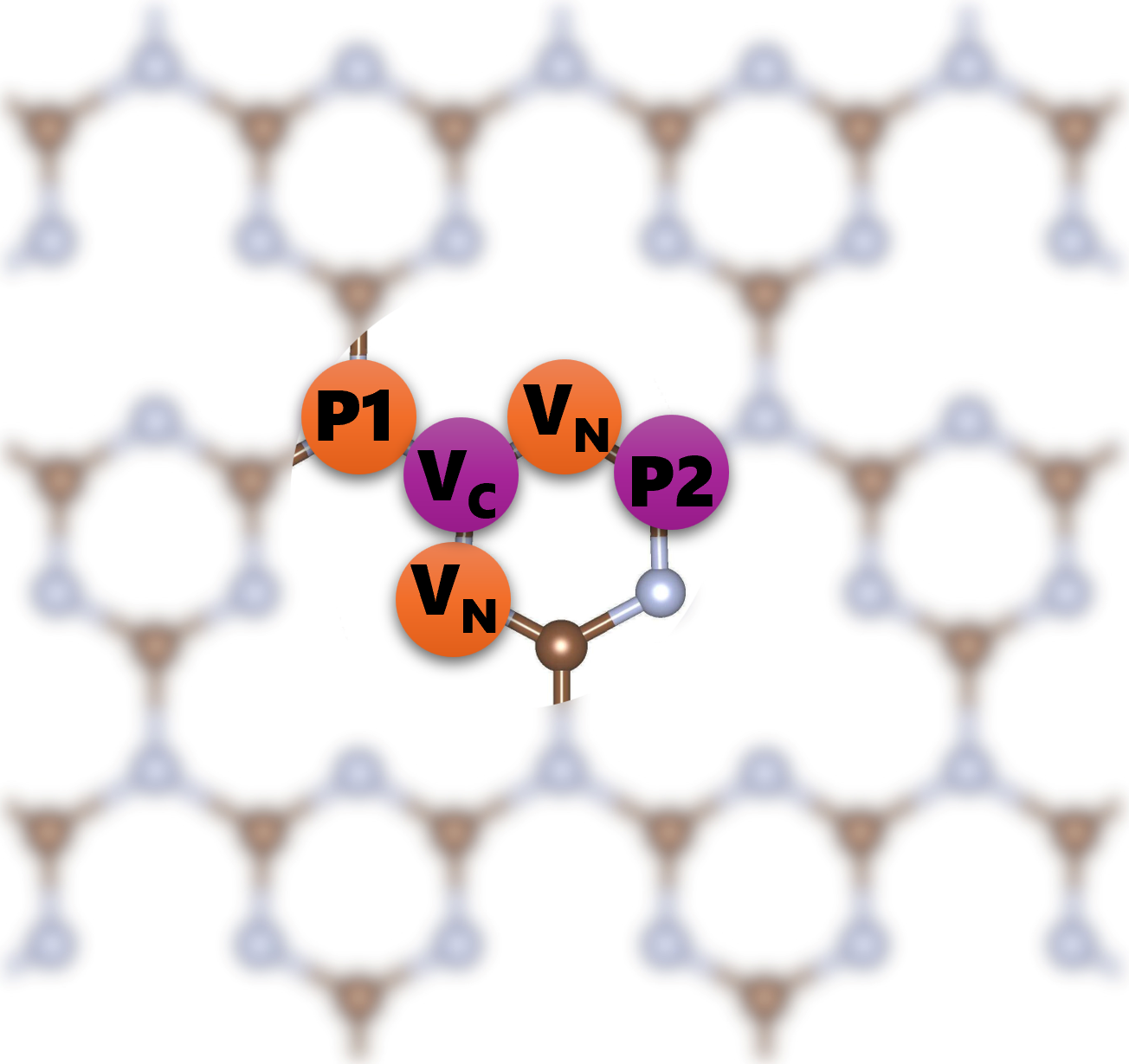, width=0.30\textwidth}} \quad
\subfigure[ ] {\epsfig{figure=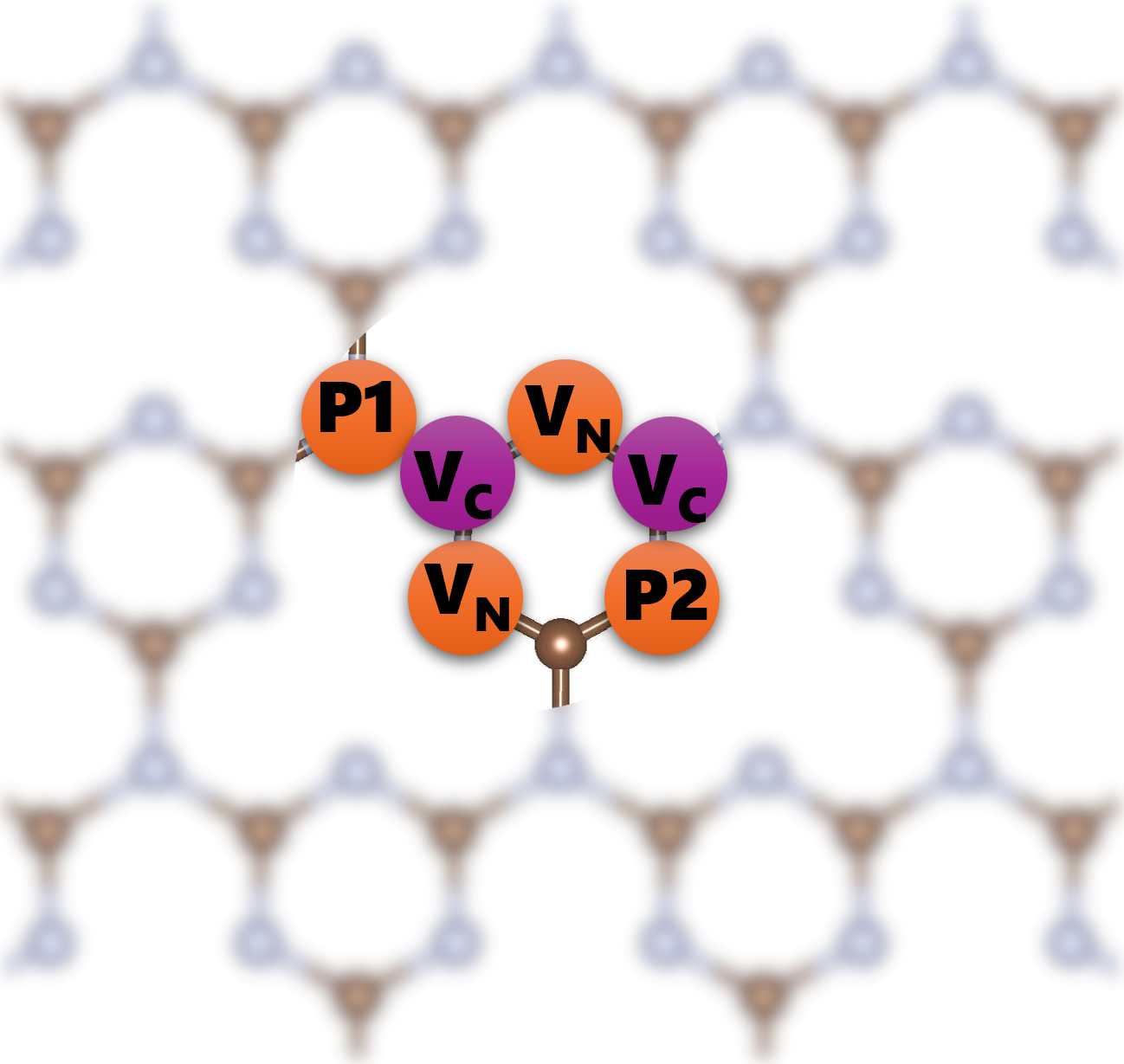, width=0.30\textwidth}} \quad
\subfigure[ ] {\epsfig{figure=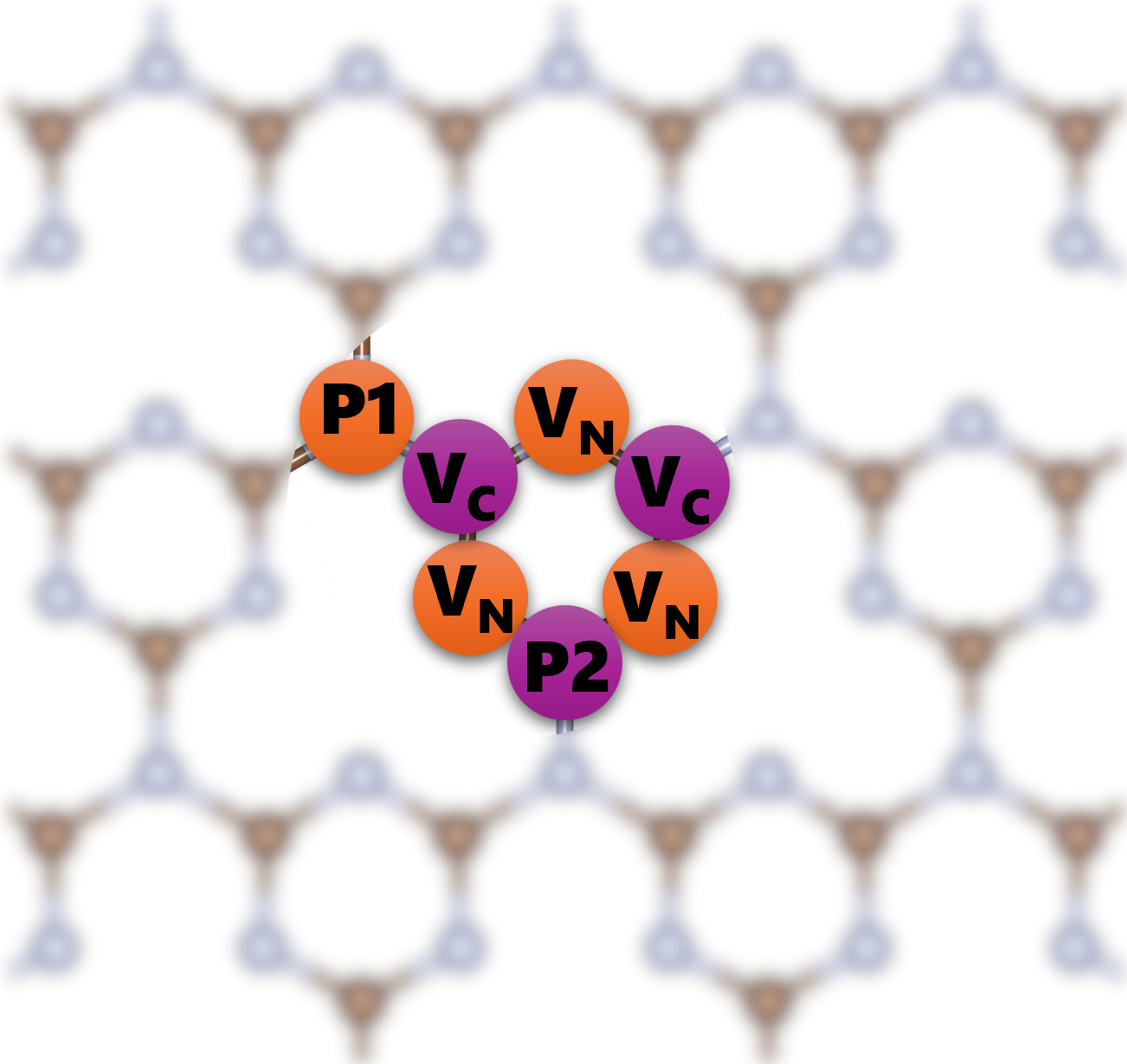, width=0.30\textwidth}}  
\caption{Defected C$_{{\rm3}}$N$_{{\rm4}}$, including (a) 1-, (b) 2-, (c) 3-, (d) 4-, (e) 5-, and (f) 6-vacancies. P$_{i}$ represents the different possibilities of the position of the defects, while V$_{\rm C}$ and V$_{\rm N}$ show established defects from the previous sub-panel (see also Table \ref{tab:form} for energetics).} 
\label{fig:mod-def}
\end{figure}
Fig. \ref{fig:mod-def} shows the optimized defect structures of the different types of defects considered in the present work, including a single defect (C or N1 or N2), 2-, 3-, 4-, 5-, 6-, and 7 defects. For multi-defects, we considered all possible combinations of atomic defects (P$_{i}$ in Fig. \ref{fig:mod-def}). The structural stability of the considered vacancy defects is investigated through their formation energies ($E_{{\rm f}}$) as follows: $E_{\rm f}=(-E_{{\rm tot}}^{{\rm perfect}}+E_{\rm tot}^{\rm defected}+n\times\mu_{C}+m\times\mu_{N})/(n+m)$, where $E_{\rm tot}^{\rm defected}$ and $E_{\rm tot}^{\rm perfect}$ are the total energies of 5$\times$5$\times$1-C$_{{\rm3}}$N$_{{\rm4}}$ supercell with and without defects, respectively. $\mu_{{\rm C}}$ and $\mu_{{\rm N}}$ are the chemical potentials of the C and N atoms, respectively. $\mu_{{\rm N}}$ and $\mu_{{\rm C}}$ were calculated from molecular nitrogen (gas phase) and graphite (solid carbon), respectively. $n$ and $m$ are the number of carbon and nitrogen atoms, that were removed from the perfect cell, respectively. We start by considering one vacancy defect, three possibilities are present (C-, N1- or N2-defect). By calculating the energy of vacancy formation, we found that the C-vacancy defect (referred to as P$_{3}$ in Fig. \ref{fig:mod-def}(a)) is the easiest to form. Then, to create $k$ defects, ($k$ = 2 to 7), we start by checking the probable vacancy positions in the structure of the n-1 defects. The different possible positions (P$_i$) of the defects for each number of defects in gt-C$_{{\rm3}}$N$_{{\rm4}}$ are given in Fig. \ref{fig:mod-def}. The path of defect creation is represented in Fig. \ref{fig:path}. We clearly observe that defects tend to form a cluster. Moreover, it is predicted that double vacancies are easier to create than single vacancies. In addition, it is found that 7 vacancies are more stable than single vacancies.
\begin{figure}[H]
\centering
\includegraphics[width=0.99\linewidth]{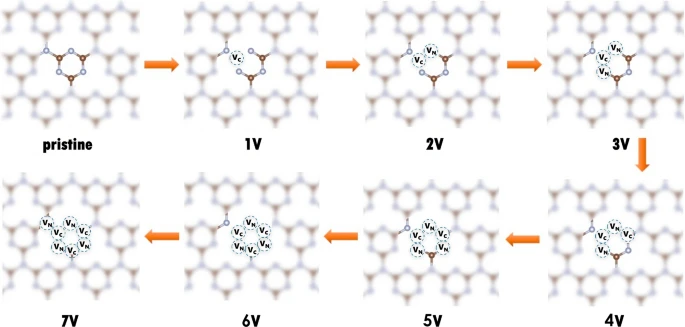}
\caption{Vacancy-defects creation path in gt-C$_{{\rm3}}$N$_{{\rm4}}$ 2D material. See also Table \ref{tab:form} and Figure \ref{fig:fig1} for details.}
\label{fig:path}
\end{figure}
For comparison, the authors in ref. \cite{Praus2022} showed by the analysis of positron annihilation spectroscopy the formation of carbon and nitrogen defects in gh-C$_{{\rm3}}$N$_{{\rm4}}$, including 
V$_{{\rm C}}$+3V$_{{\rm N}}$, and 2V$_{{\rm C}}$+2V$_{{\rm N}}$ (4-defects). Here, we found that the formation energy of the defects gradually decreased from 5.82 eV corresponding to V$_{{\rm N}}$ to the value of 1.45 eV calculated for 3V$_{{\rm C}}$+3V$_{{\rm N}}$, however with 2V$_{{\rm C}}$+2V$_{{\rm N}}$ energetically favorable during creation a path also (1.87 eV, cf. Table \ref{tab:form} and Figure \ref{fig:path}).  The presence of the 2V$_{{\rm C}}$+2V$_{{\rm N}}$ complexes in gh-C$_{{\rm3}}$N$_{{\rm4}}$ is assumed to affect its photocatalytic activity,\cite{Praus2022} and therefore a better understanding of the properties of g-C$_{{\rm3}}$N$_{{\rm4}}$ in the presence of defects is needed. 
\begin{table}[h]
\centering
\begin{tabular*}{\textwidth}{@{\extracolsep\fill}lcccc}
\toprule%
1 vacancy                       & P1   & P2   & {\bf P3} \\
${E_{\rm f}}$ (eV)              & 9.62 & 6.18 & 5.82     \\
Total magnetic moment ($\mu_B$) & 0.94 & 0.06 & NM       \\ [0.5em]
2 vacancies                     & P1   & {\bf P2}  & P3   \\
${E_{\rm f}}$ (eV)              & 4.70 & 4.08      & 4.14 \\
Total magnetic moment ($\mu_B$) & 0.81 & 1.00      & NM   \\ [0.5em]
3 vacancies                     & P1   & P2   & P3   & {\bf P4} \\
${E_{\rm f}}$ (eV)              & 3.89 & 3.63 & 3.62 & 2.61     \\
Total magnetic moment ($\mu_B$) & NM   & 2.76 & NM   & NM       \\ [0.5em]
4 vacancies                     & P1   & {\bf P2} \\
${E_{\rm f}}$ (eV)              & 2.98 & 1.87     \\
Total magnetic moment ($\mu_B$) & NM   & NM       \\ [0.5em]
5 vacancies                     & P1   & {\bf P2} \\
${E_{\rm f}}$ (eV)              & 2.28 & 2.13     \\
Total magnetic moment ($\mu_B$) & NM   & 0.50     \\ [0.5em]
6 vacancies                     & P1   & {\bf P2} \\
${E_{\rm f}}$ (eV)              & 2.47 & 1.45     \\
Total magnetic moment ($\mu_B$) & 0.60 & NM       \\ [0.5em]
7 vacancies                     & P1   \\
${E_{\rm f}}$ (eV)              & 1.71 \\
Total magnetic moment ($\mu_B$) & NM   \\
\botrule
\end{tabular*} 
\caption{Calculated formation energies ($E_{{\rm f}}$) and total magnetic moments of defected gt-C$_{{\rm3}}$N$_{{\rm4}}$ compound. The reader may refer to Fig. \ref{fig:mod-def} for more details regarding the positions P$_{i}$ (the most probable one is in bold).}
\label{tab:form}
\end{table}

\begin{figure}[H]
\centering
\includegraphics[width=0.8\textwidth]{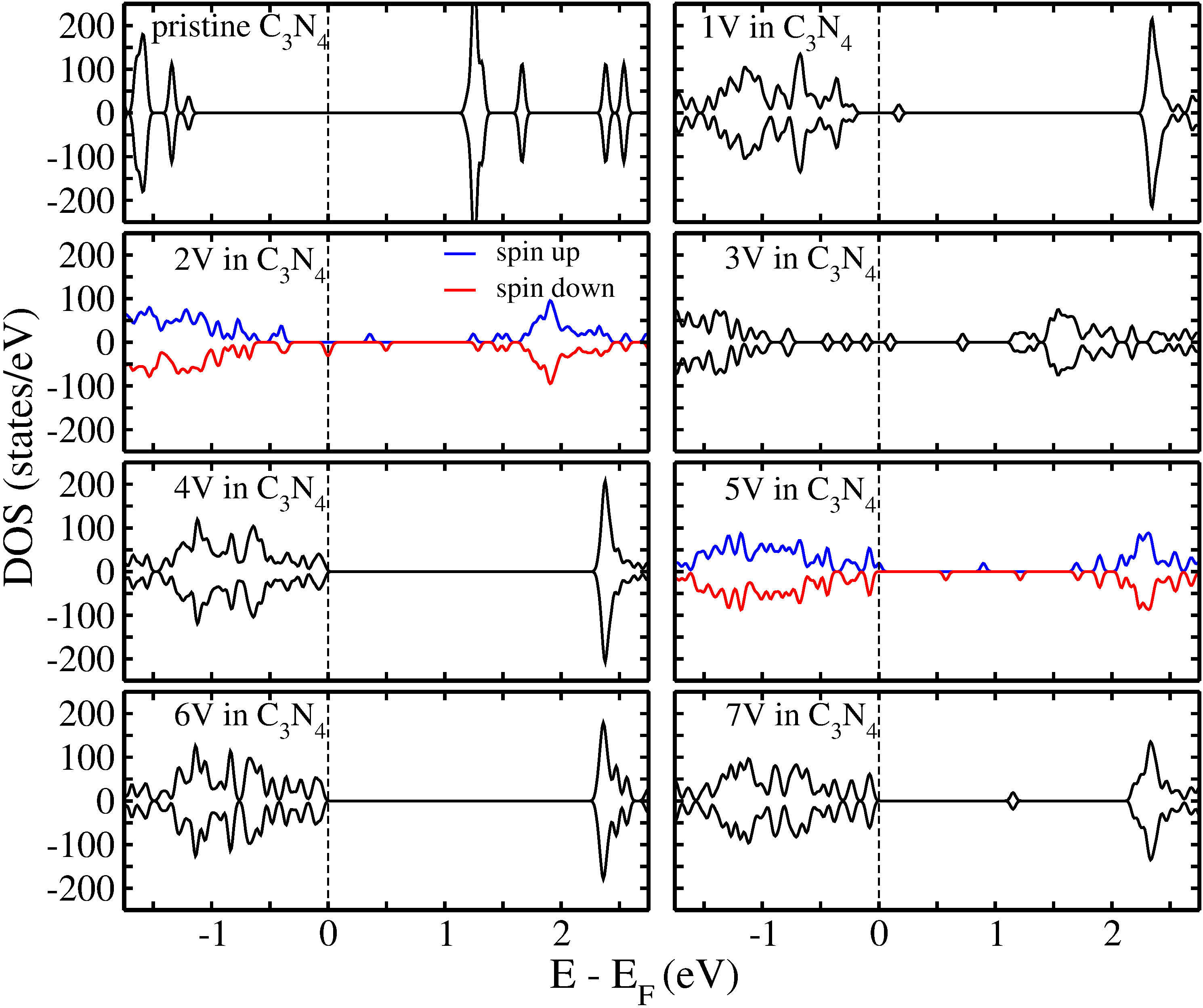} 
\caption{Density of states of (a) the pristine C$_{{\rm3}}$N$_{{\rm4}}$ and the defected C$_{{\rm3}}$N$_{{\rm4}}$. (b) 1 vacancy (V$_{{\rm C}}$), (c) 2 vacancies (V$_{{\rm N1-C}}$), (d) 3 vacancies (V$_{{\rm N1-C-N1}}$), (e) 4 vacancies (V$_{{\rm N1-C-N1-C}}$), (f) 5 vacancies (V$_{{\rm N1-C-N1-C-N1}}$), (g) 6 vacancies (V$_{{\rm N1-C-N1-C-N1-C}}$), and (h) 7 vacancies (V$_{{\rm N1-C-N1-C-N1-C-N2}}$). The zero energy (dashed black line) was set to the Fermi energy E$_{\rm F}$. Corresponding geometric structures are depicted in Figure \ref{fig:path}.} 
\label{fig:dos-defect} 
\end{figure}

The defect-free gt-C$_{{\rm3}}$N$_{{\rm4}}$ 2D material was explored very briefly to understand its general electronic and magnetic properties to be compared with the defected structures. We found that the pure gt-C$_{{\rm3}}$N$_{{\rm4}}$ appears as a non-magnetic semiconductor with a band gap of 2.28 eV. Moreover, we found that the considered vacancies in C$_{\rm3}$N$_{\rm4}$ 2D material do not considerably alter the semiconductor characteristics considerably; however, they affect the magnetic properties. \\
We plot the density of states of the pristine and the defected gt-C$_{{\rm3}}$N$_{{\rm4}}$ in Fig. \ref{fig:dos-defect}. It is revealed that vacant gt-C$_{{\rm3}}$N$_{{\rm4}}$ material keeps the semiconducting characteristics with slightly smaller bandgaps compared to perfect gt-C$_{{\rm3}}$N$_{{\rm4}}$, except single defect, which shows an electronic band gap of 2.31 eV. Moreover, we found that in gt-C$_{{\rm3}}$N$_{{\rm4}}$ with 5-vacancy defects the bandgap reduction is more severe with respect to its counterpart without defect. As can be seen, there are some defect states in the band gap near the Fermi level observed in defective structures. Interestingly, the creation of 2 and 5 vacancy defects turns on magnetism on gt-C$_{{\rm3}}$N$_{{\rm4}}$. Ferromagnetic order at room temperature has been reported in monolayers of FePS$_{{\rm3}}$ \cite{Lee2016}, CrI$_{{\rm3}}$ \cite{Huang2017}, VSe$_{{\rm2}}$ \cite{Zhong-Liu2018}, Cr$_{{\rm2}}$Ge$_{{\rm2}}$Te$_{{\rm6}}$ \cite{Gong2017}, Fe$_{{\rm3}}$GeTe$_{{\rm2}}$ \cite{Fei2018}, etc. ... The magnetic defected C$_{{\rm3}}$N$_{{\rm4}}$ show a magnetic moments in the same order of NiCl$_{{\rm3}}$ (0.94 $\mu_{{\rm B}}$) \cite{Shalini2019} and Ti$_{{\rm2}}$N (0.5$\mu_{{\rm B}}$) \cite{Gao2016}. To compare to the available magnetic 2D material, the reader may refer to the following review \cite{Jiang2021}.
\subsection{Hydrogen and oxygen passivation of the defects}
\begin{figure}[H]
\centering
\includegraphics[width=0.8\textwidth]{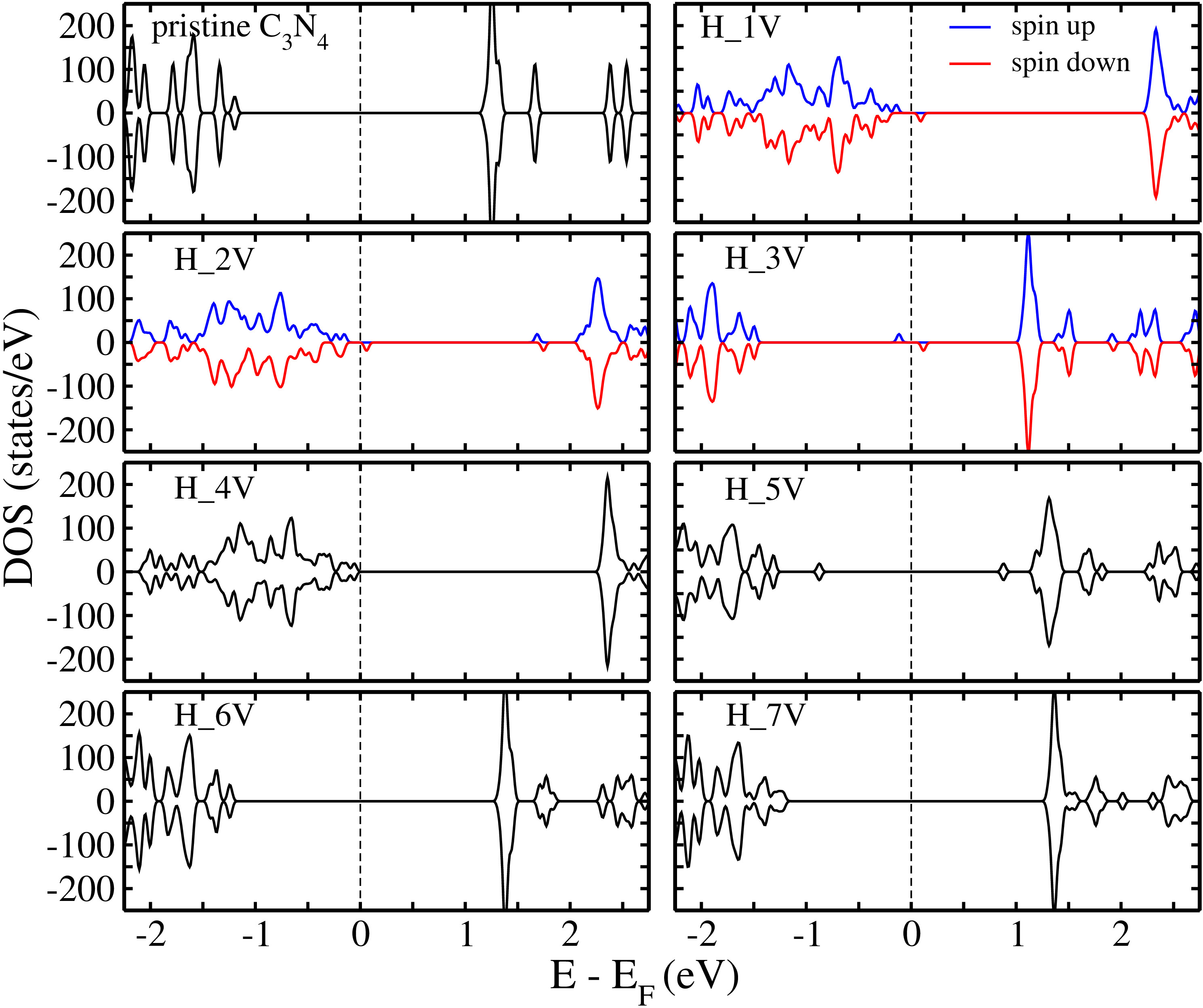} 
\caption{Density of states of (a) the pristine C$_{{\rm3}}$N$_{{\rm4}}$ and H-passivated defects in C$_{{\rm3}}$N$_{{\rm4}}$; (b) single defect, (c) 2, (d) 3, (e) 4, (f) 5, (g) 6, and (h) 7 defects. The zero energy (dashed black line) was set to the Fermi energy E$_{\rm F}$.} 
\label{fig:dos-H-in-defect} 
\end{figure}

\begin{figure}[H]
\centering
\includegraphics[width=0.8\textwidth]{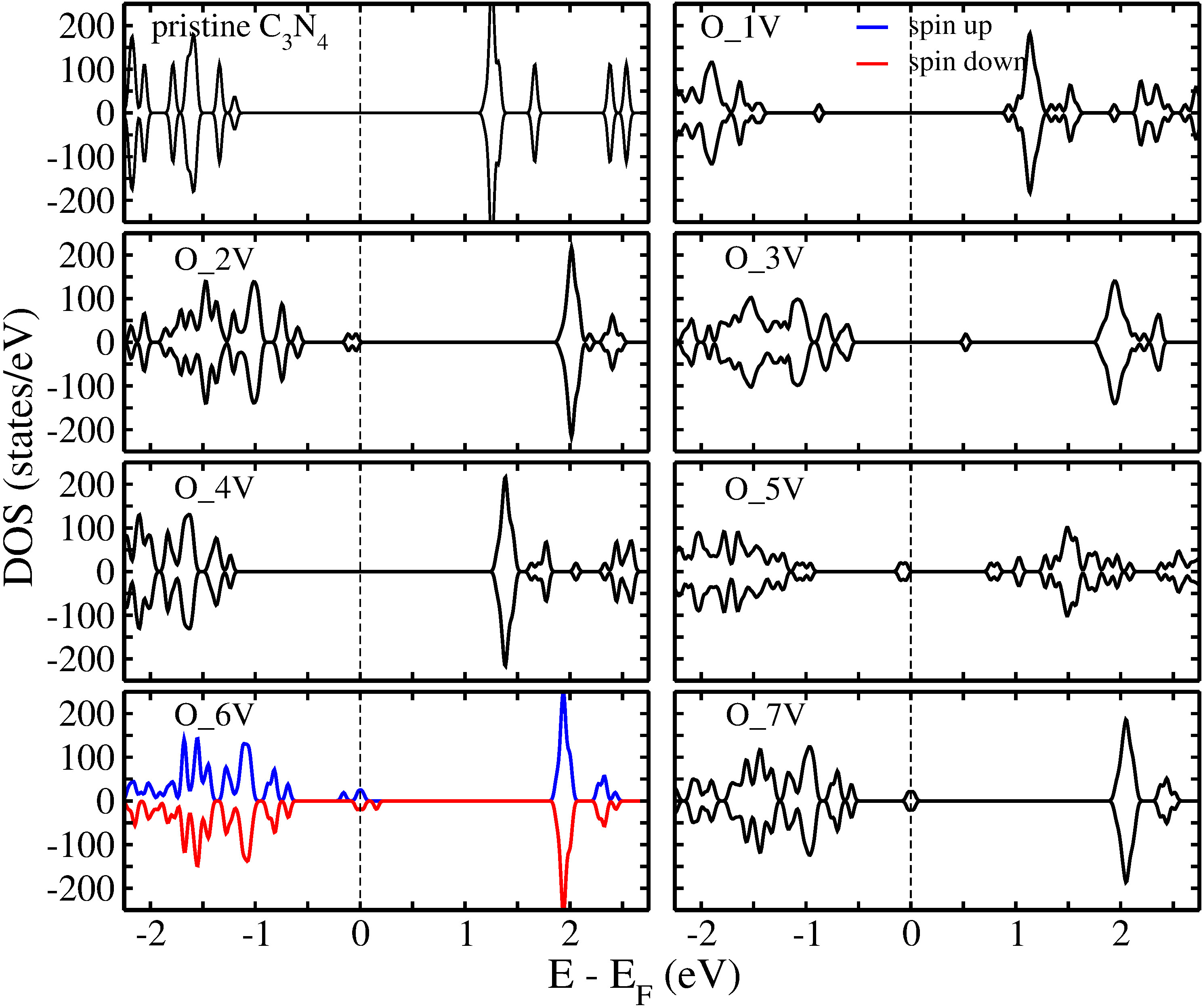} 
\caption{Density of states of (a) the pristine C$_{{\rm3}}$N$_{{\rm4}}$ and O-passivated defects in C$_{{\rm3}}$N$_{{\rm4}}$; (b) single defect, (c) 2, (d) 3, (e) 4, (f) 5, (g) 6, and (h) 7 defects. The zero energy (dashed black line) was set to the Fermi energy E$_{\rm F}$.} 
\label{fig:dos-O-in-defect} 
\end{figure}

\begin{figure}[H]
{\centering
\captionsetup{justification=centering} 
\subfigure[ ] {\epsfig{figure=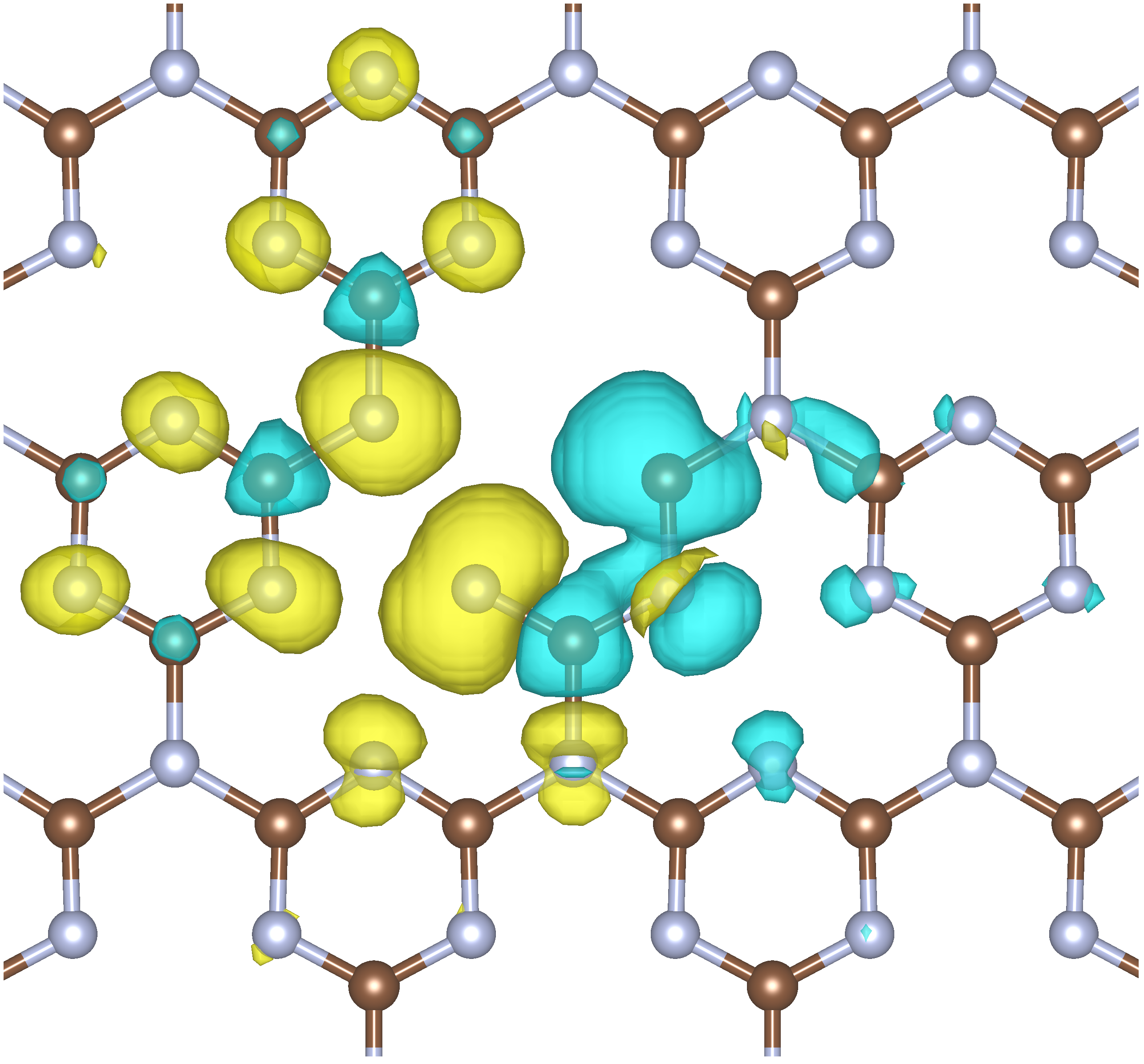, width=0.26\textwidth}} \quad
\subfigure[ ] {\epsfig{figure=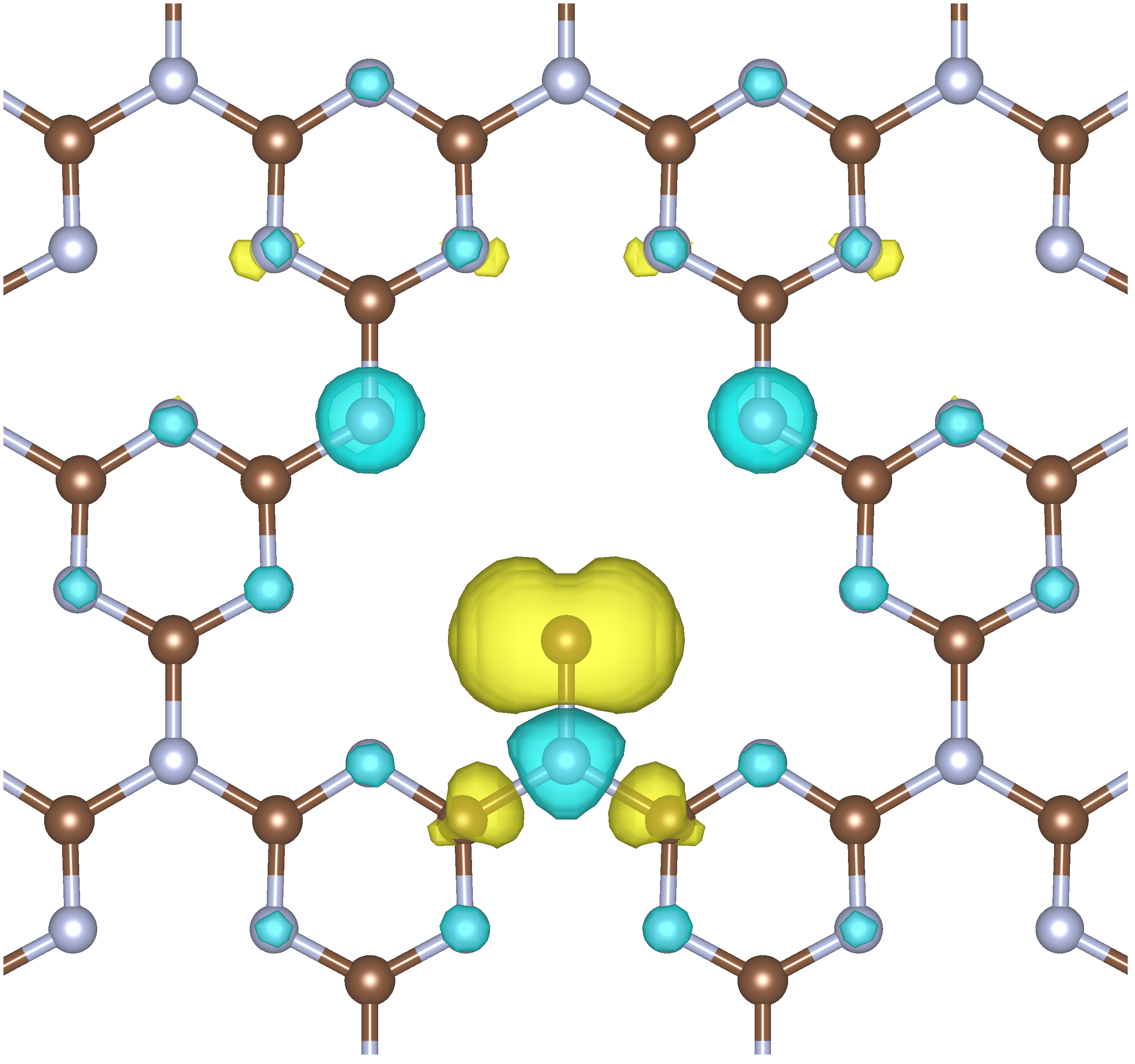, width=0.26\textwidth}} \quad
\subfigure[ ] {\epsfig{figure=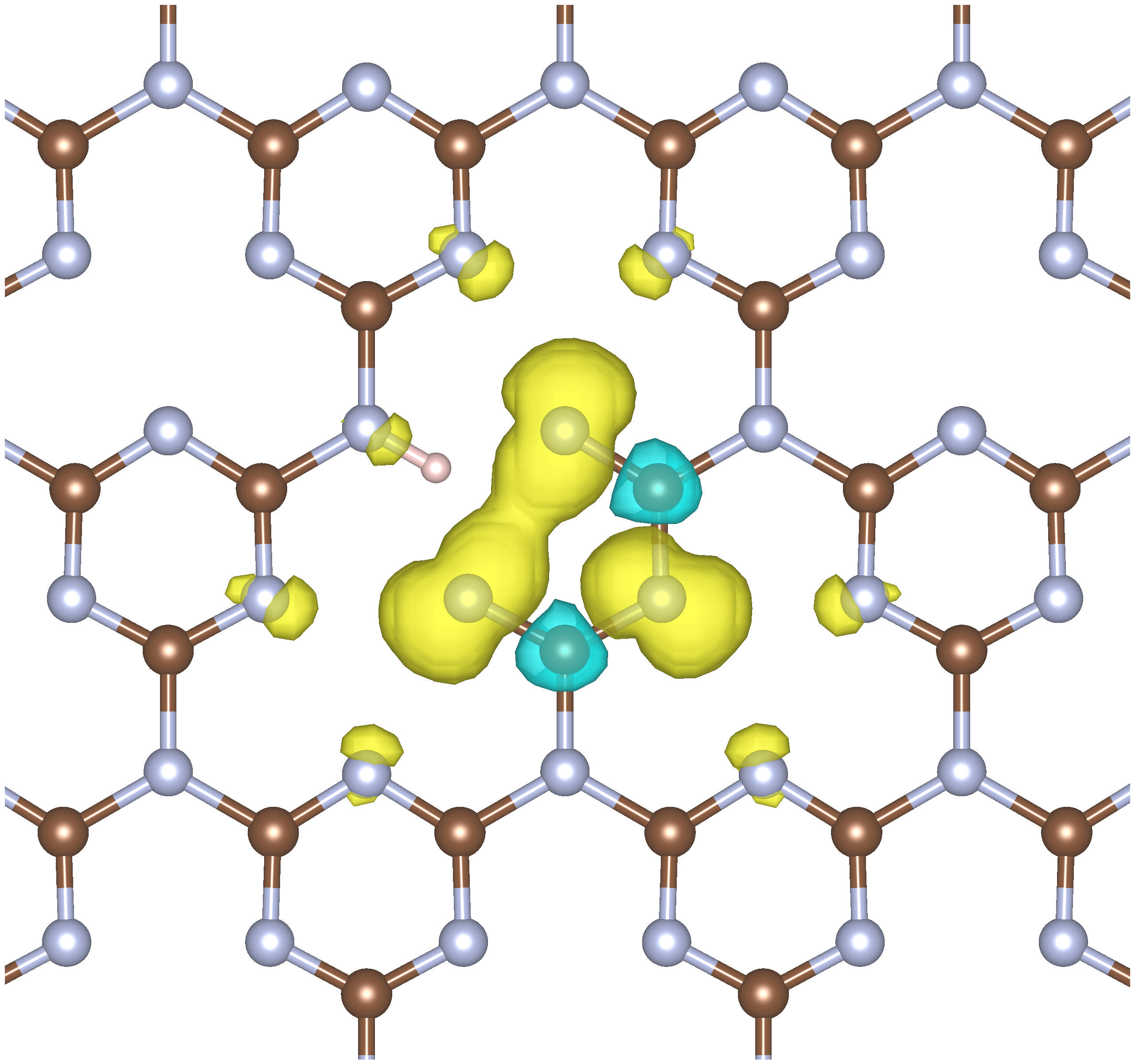, width=0.26\textwidth}} \\
\subfigure[ ] {\epsfig{figure=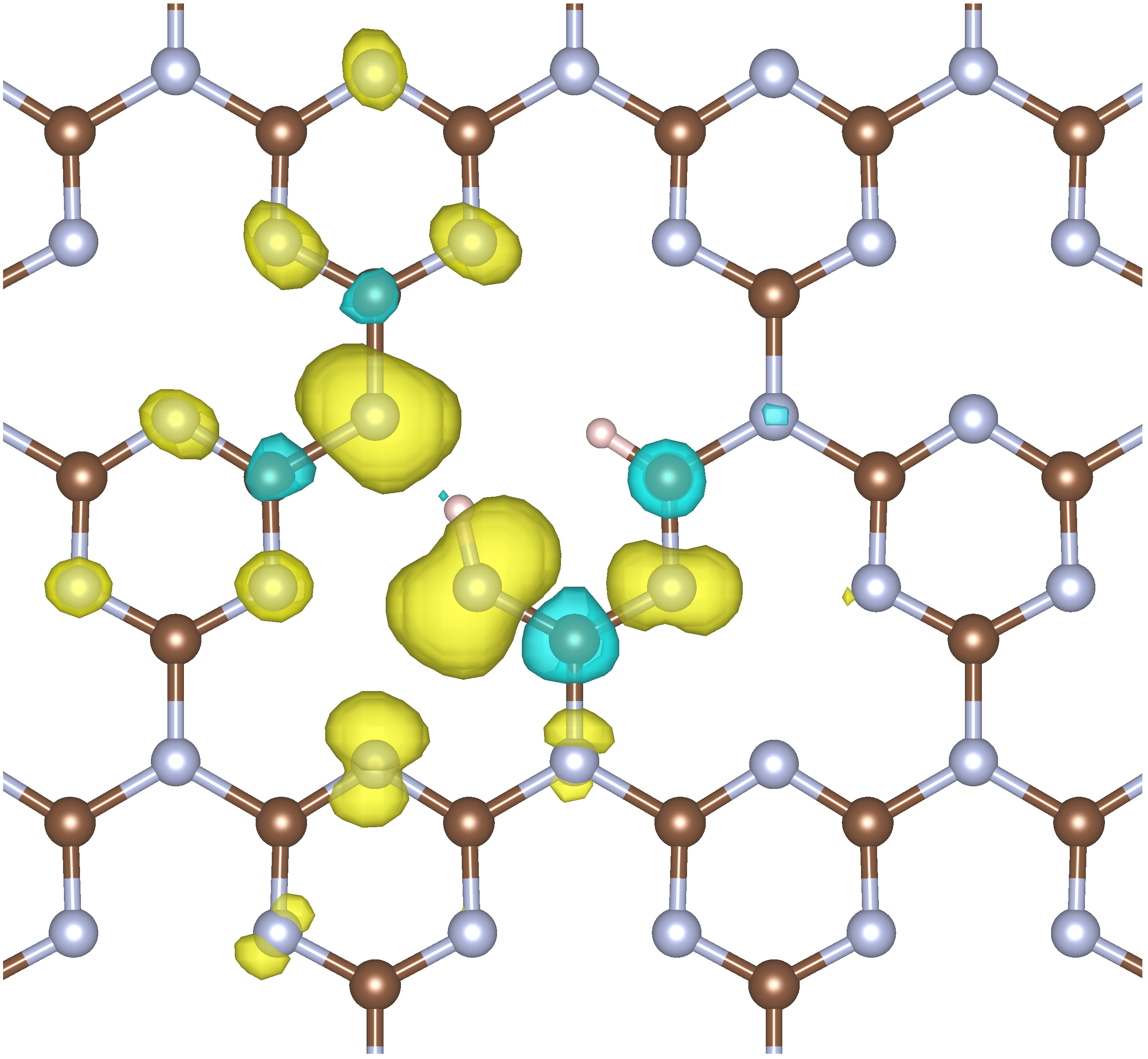, width=0.26\textwidth}} \quad
\subfigure[ ] {\epsfig{figure=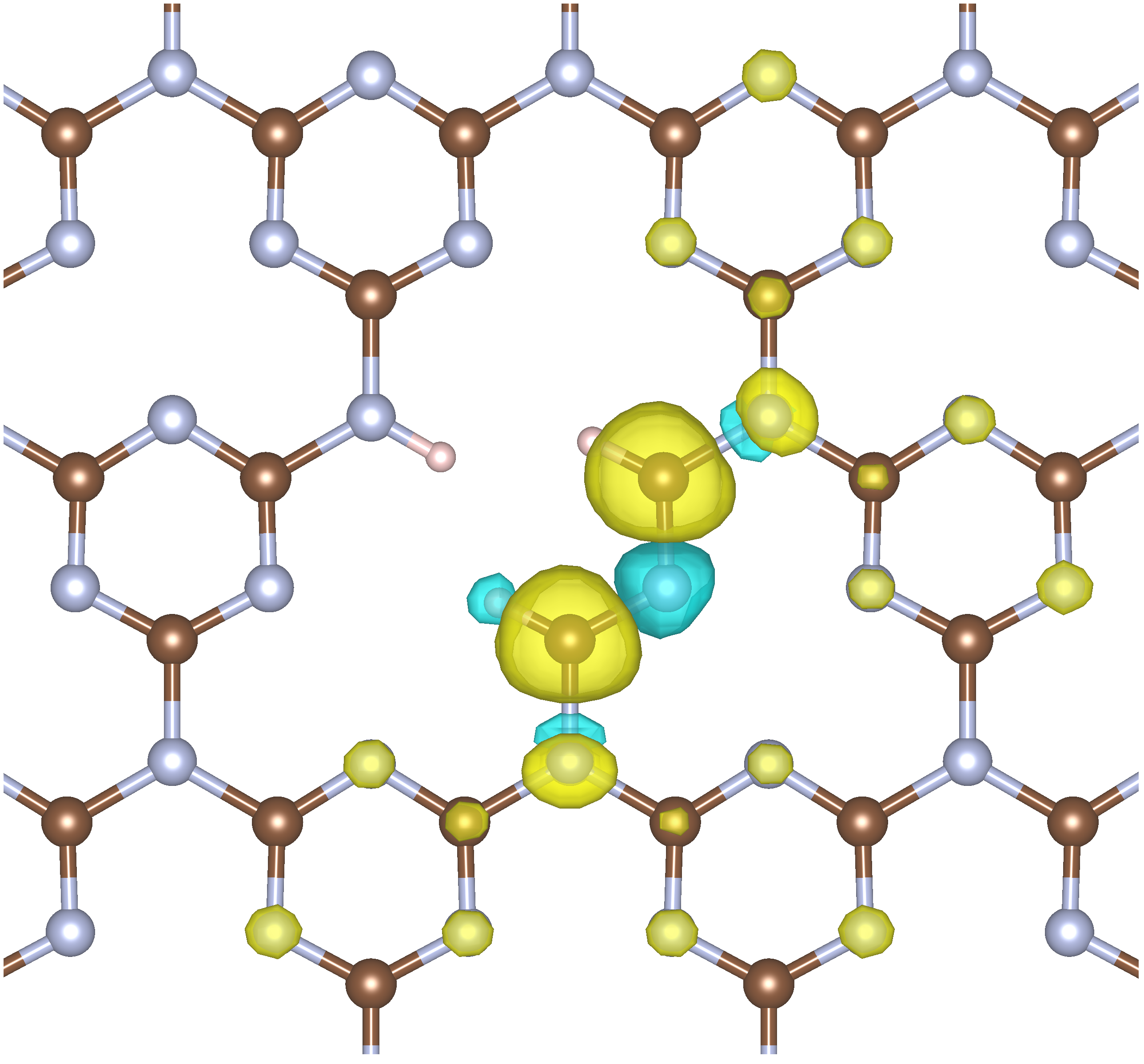, width=0.26\textwidth}} \quad
\subfigure[ ] {\epsfig{figure=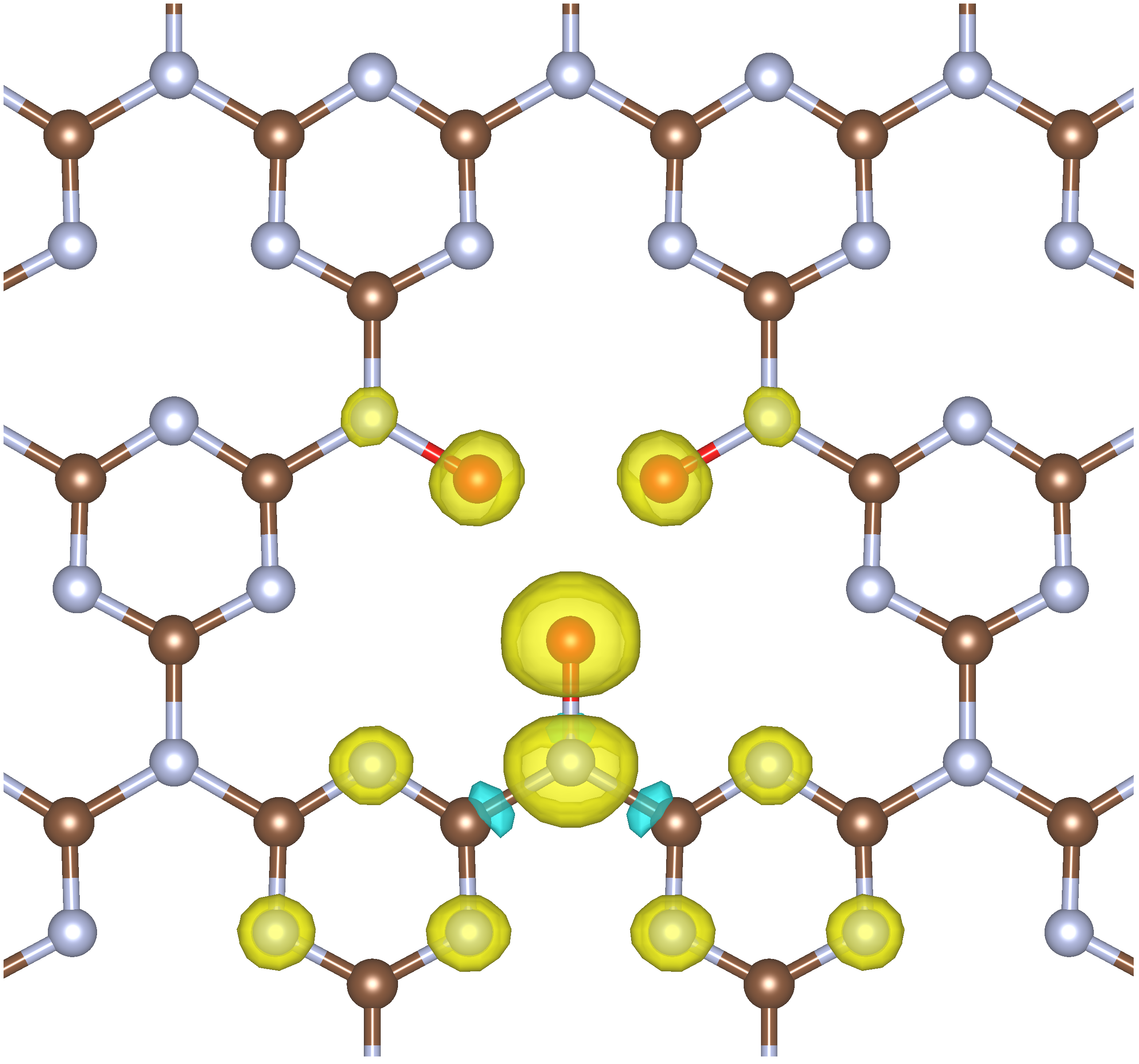, width=0.26\textwidth}} \\ }
\caption{Spin density distribution of vacancy defects and passivated defects in gt-C$_{{\rm3}}$N$_{{\rm4}}$. (a) 2 vacancies, (b) 5 vacancies, H-passivated (c) 1 defect, (d) 2, (e) 3 defects, and (f) O-passivated 6 defects. Isosurface is set to 0.0015 e/a.u.$^3$. Yellow and cyan represent positive and negative values, respectively.}
\label{spall}
\end{figure}

\begin{table}[h]
\centering
\begin{tabular*}{\textwidth}{@{\extracolsep\fill}lccccccc}
\toprule%
               &  1V  &  2V  &  3V  &  4V  &  5V  &  6V  &  7V  \\
\midrule
& \multicolumn{6}{c}{Magnetic moment} & \\
\cmidrule{4-5}
non passivated & NM   & 1.00 & NM   & NM   & 0.50 & NM   & NM   \\
H-passivation  & 0.71 & 0.74 & 0.72 & NM   & NM   & NM   & NM \\
O-passivation  & NM   & NM   & NM   & NM   & NM   & 0.82 & NM   \\[0.25em]
& \multicolumn{6}{c}{Band gap} & \\
\cmidrule{4-5}
non passivated & 2.31 & 2.23 & 2.14 & 2.26 & 2.04 & 2.24 & 2.13 \\
H-passivation  & 2.33 & 2.13 & 2.45 & 2.27 & 2.36 & 2.47 & 2.39 \\
O-passivation  & 2.33 & 2.37 & 2.32 & 2.43 & 2.14 & 2.42 & 2.40 \\
\botrule
\end{tabular*} 
\caption{Calculated total magnetic moments (in $\mu_B$/cell) and energy band gaps (in eV) of H-passivated defects in gt-C$_{{\rm3}}$N$_{{\rm4}}$. NM stands for nonmagnetic.}
\label{tab:passiv}
\end{table}
Further, we performed the spin-polarized density of states calculations for hydrogen and oxygen passivated defects, results are shown in Fig. \ref{fig:dos-H-in-defect} and Fig. \ref{fig:dos-O-in-defect}. We found an asymmetry between spin-up and spin-down densities in some of the H/O-passivated defects, which induces an obvious magnetism, including H-passivated 1, 2, and 3 defects and O-passivated 6-defects. Interestingly, it is clear that the semiconducting characteristics are preserved and the band gap is larger than that for defective structures. \\
Table \ref{tab:passiv} lists the band gap and magnetic moment values. Passivating the 1-, 2-, 3-, 4-, 5-, 6-, and 7-vacancy defect structures by hydrogen modulates the band gaps to 2.33, 2.13, 2.45, 2.27, 2.36, 2.47 and 2.39 eV, respectively. For both H and O passivation, the band gaps of the passivated structures are higher than the band gap of perfect gt-C$_{{\rm3}}$N$_{{\rm4}}$. 
To further explore the effect of the creation of vacancy defects and the passivation of defects on magnetism, we plot the spin density distributions for 2-, 5-vacancies, 1-, 2-, and 3- H-passivated defects, and 6- O-passivated defects in Fig. \ref{spall}. When a single C and N2 atom or five atoms (3 nitrogen and 2 carbon atoms) are simultaneously removed from the C$_{{\rm3}}$N$_{{\rm4}}$, the structure optimization leads to the formation of dangling bonds or an under-coordinated atom around the local defect region. Thus, the magnetism is due to the dangling bonds which have a localized unpaired spin, and the defect levels have a significant contribution on the C and N atoms around the vacancy site, whereas the rest of the C and N atoms do not carry any magnetic moments. 
Fig. \ref{spall} (c-g) illustrates the views of the spin isosurface of hydrogen-saturated defects. It can be clearly seen that the magnetism is predominantly located around the hydrogenated C and N atoms. The positive spin density (in yellow) is seen more over the C and N atoms close to the H atom. Very low negative spin density (blue) is seen in some cases. It should be noted that the same behavior was also observed in the case of O passivation of 6 defects.

\subsection{Adsorption of transition metal atom on C$_{\rm 3}$N$_{\rm 4}$}
\begin{figure}[H]
\centering
\includegraphics[width=0.7\textwidth]{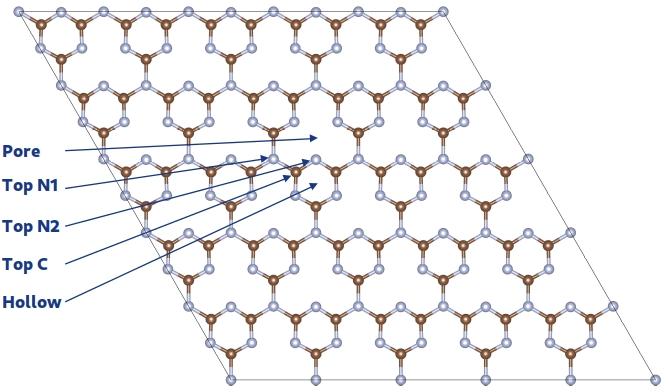} 
\caption{Possible adsorption sites for transition metal (TM) atom (i.e. V, Cr, Mn, Fe, Co, Ni) in gt-C$_{{\rm3}}$N$_{{\rm4}}$.} 
\label{fig:fe-pos} 
\end{figure} 
Next, we explore the electronic and magnetic properties of the TM atom (i.e. V, Cr, Mn, Fe, Co, Ni) adsorbed on C$_{{\rm3}}$N$_{{\rm4}}$ monolayer. We considered structures where a single TM is adsorbed in top N1, top N2, top C, hollow, and pore sites using a 5$\times$5$\times$1 super-cell as shown in Fig. \ref{fig:fe-pos}. The adsorption energy of the TM atom adsorbed on the gt-C$_{\rm 3}$N$_{\rm 4}$ monolayer is defined as $E_{\rm ads.}$ = $E_{{\rm tot.}}$(TM-adsorbed)-$E$(TM)-$E_{{\rm tot.}}$(free), where $E_{{\rm tot.}}$(TM-adsorbed) is the total energy of the TM-adsorbed monolayer C$_{\rm 3}$N$_{\rm 4}$, $E_{{\rm tot.}}$(free) is the total energy of the non-adsorbed C$_{\rm 3}$N$_{\rm 4}$, and $E$(TM) represents the energy of an isolated TM atom. The more negative the values of $E_{{\rm ads.}}$, the more stable the structures. We found that the top N2 position is the most stable for Fe atom, followed by the pore, hollow, top C, and top N1 with energy differences of 0.13 eV, 0.75 eV, 0.86 eV, and 1.30 eV, respectively, with respect to the top N2 position. The V, Co, and Ni atoms prefer the hollow site, whereas the Cr and Mn atoms favor the pore position. Here, the TM atom forms strong covalent bonds with carbon/nitrogen atoms in the gt-C$_{{\rm3}}$N$_{{\rm4}}$ surface. Interestingly, we show that TM adsorption in gt-C$_{{\rm3}}$N$_{{\rm4}}$ sheet does not change the structures drastically, which is already experimentally found by X-ray diffraction study of TM-gt-C$_{{\rm3}}$N$_{{\rm4}}$ by X. Wang, et al. \cite{XWang_2009}.

\begin{figure}[H] 
\centering
\includegraphics[width=0.8\textwidth]{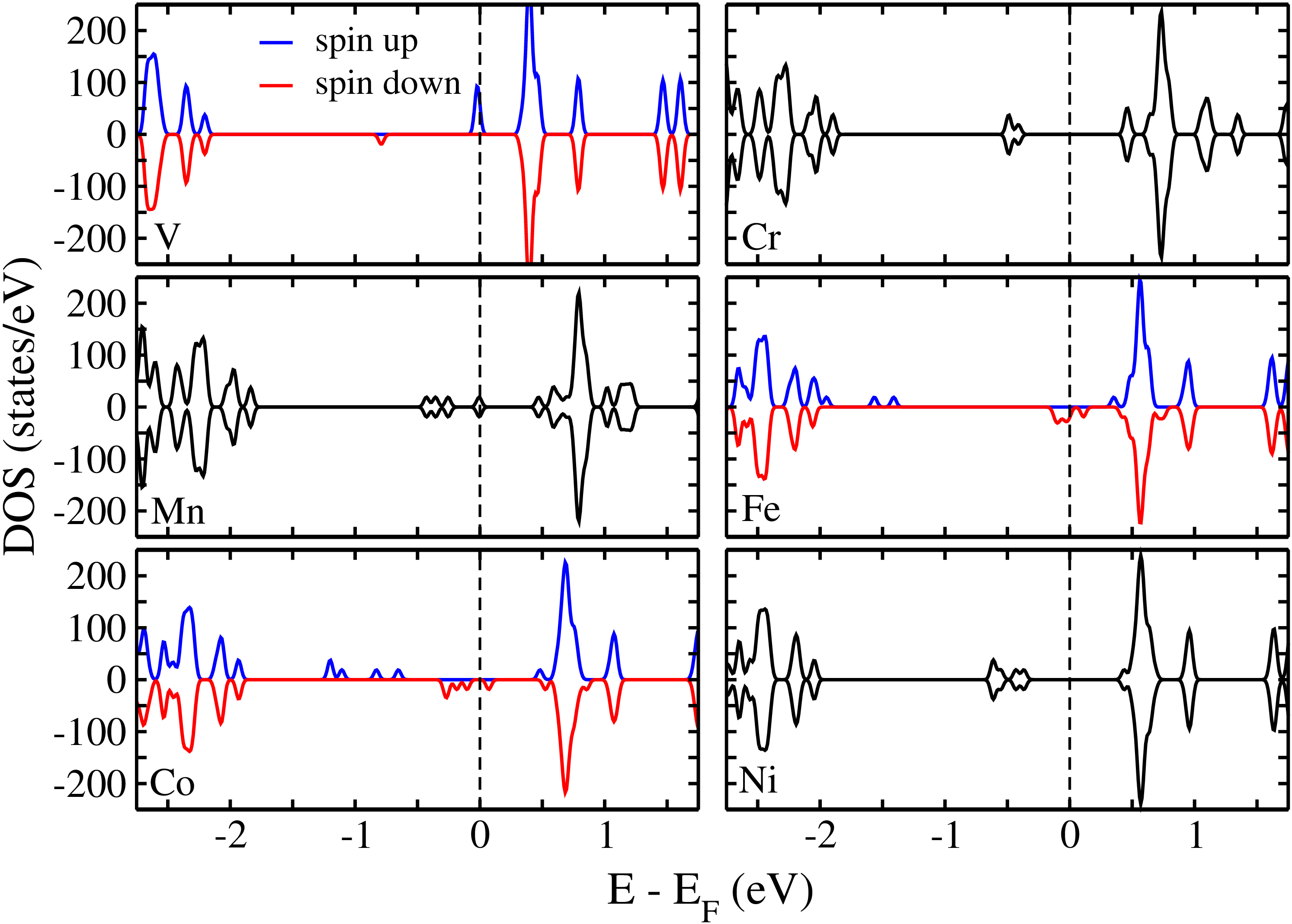}
\caption{Density of states of the TM adsorbed on C$_{{\rm3}}$N$_{{\rm4}}$ (a) V in hollow site, (b) Cr in pore position, (c) Mn in pore position, (d) Fe in top N2 site, (e) Co in hollow site, and (f) Ni in hollow site. The zero energy (dashed black line) was set to the Fermi energy E$_{\rm F}$. Corresponding adsorption positions are depicted in Figure \ref{fig:fe-pos}. }
\label{fig:dos-fecn}
\end{figure}

\begin{figure}[H]
{\centering
\subfigure[ ] {\epsfig{figure=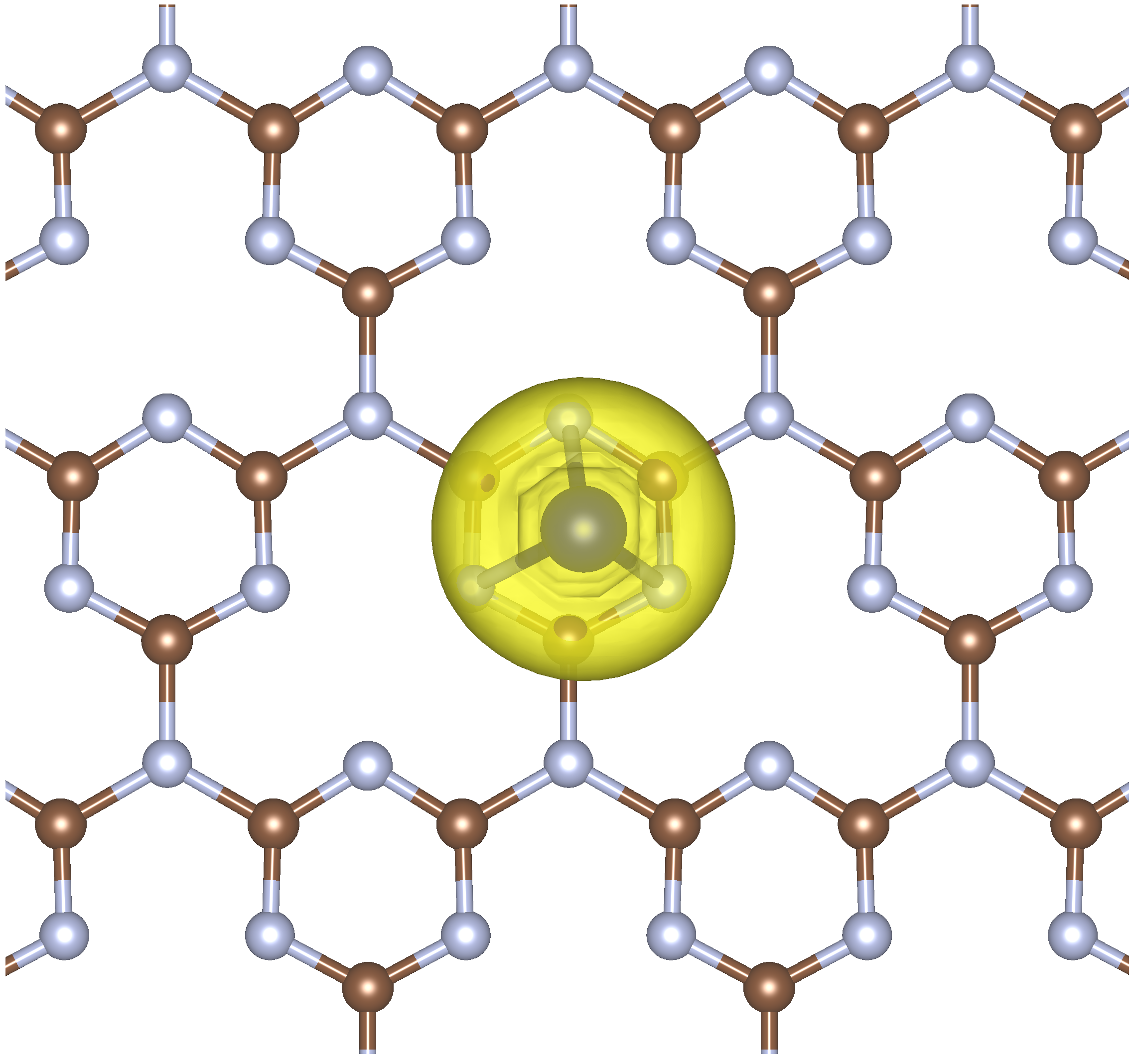, width=0.30\textwidth}} \quad
\subfigure[ ] {\epsfig{figure=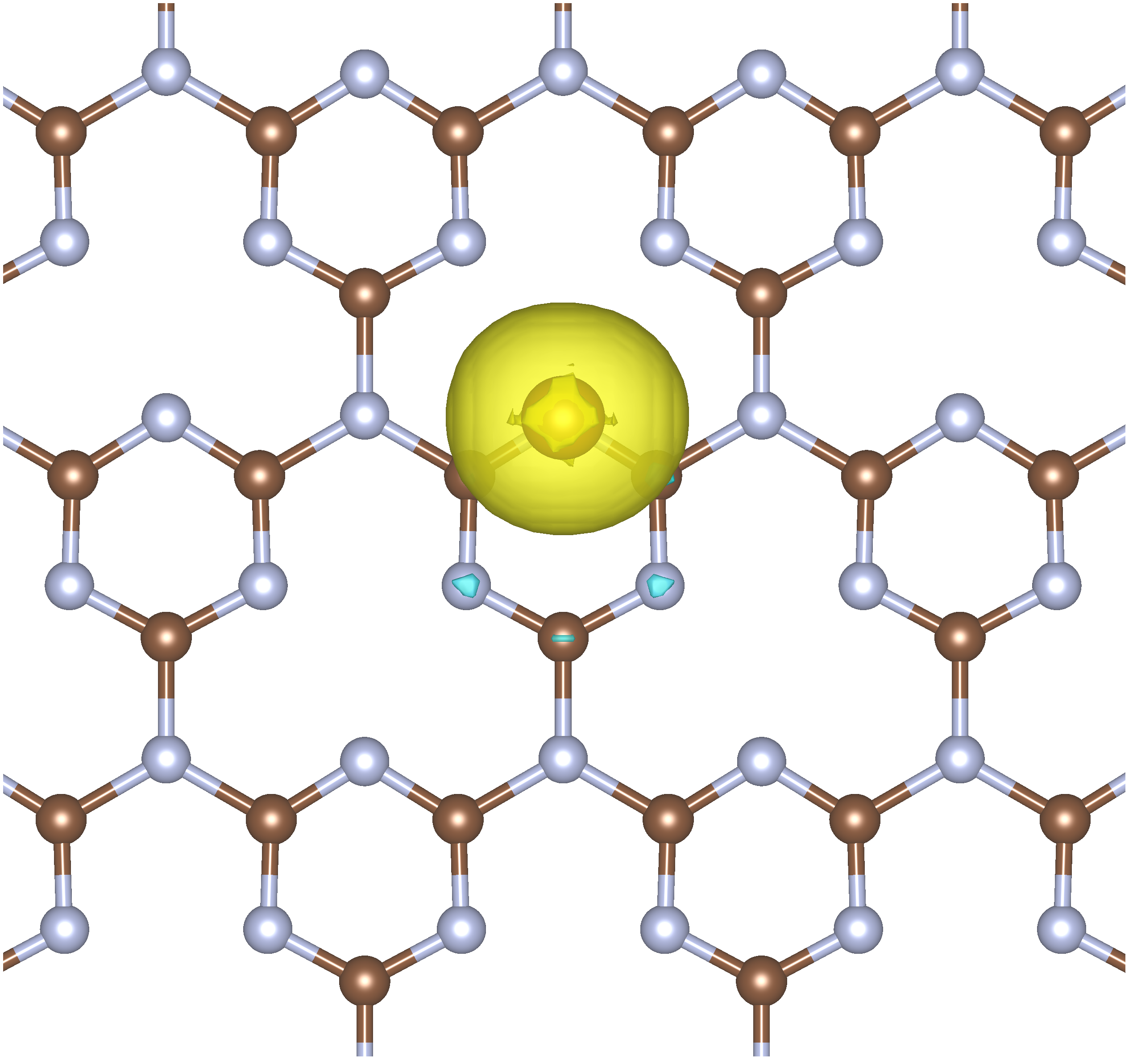, width=0.30\textwidth}} \quad
\subfigure[ ] {\epsfig{figure=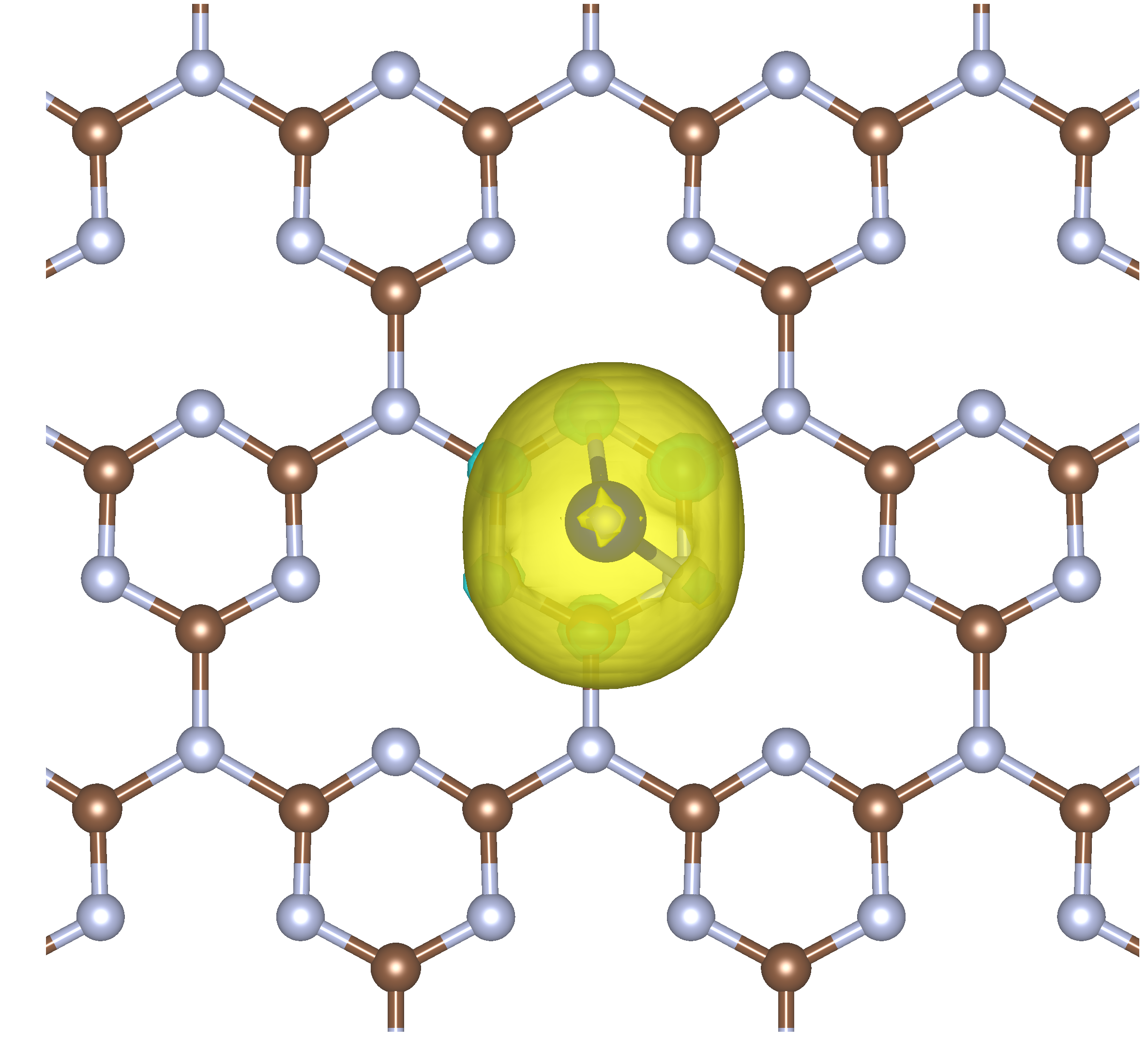, width=0.30\textwidth}} }
\caption{Spin density distribution of transition metal adsorbed in gt-C$_{{\rm3}}$N$_{{\rm4}}$ (a) V in hollow site, (b) Fe in top N2 position, (c) Co in hollow site. Isosurface is set to 0.008 e/a.u.$^3$. Yellow and cyan represent positive and negative values, respectively. Here, we plot the most stable adsorption site only.}
\label{sp-fecn}
\end{figure}

Fig. \ref{fig:dos-fecn} shows the DOS of TM adsorbed on C$_{\rm3}$N$_{\rm4}$ in the most favorable sites (i.e. top N2 position for Fe atom, hollow site for V, Co, and Ni atoms, and pore site for Cr and Mn). Obviously, the TM induces significant variations in DOS around the Fermi level, characterized mainly by defect states. The induced local magnetic moment is most likely due to the different amounts of hybridization between the d orbitals of the TM atom and the p orbitals of C and N atoms. Interestingly, we found that Fe and V atoms adsorbed on C$_{\rm3}$N$_{\rm4}$ induces higher magnetic moments compared to Fe and Co adsorption in antimonene \cite{Yungang_2018}, Fe, Co, Ni adsorption in arsenene \cite{Liu_2018}, Co, Ni, Cu adsorption in germanene \cite{Kaloni_2014}, and Rh and Pd adsorption in graphene \cite{Minglei_2017}. However, adsorption of the Fe and Mn atoms in SbAs material induces a spin polarization higher than 3.5 $\mu_B$ \cite{SAKHRAOUI2022}. \\
The corresponding spin densities of these magnetic states of TM in gt-C$_{{\rm3}}$N$_{{\rm4}}$ can be seen in Fig. \ref{sp-fecn}. Our calculations show that V, Fe, and Co in the C$_{{\rm3}}$N$_{{\rm4}}$ material is magnetic with a total spin moment of 2.89 $\mu_B$/cell, 2 $\mu_B$/cell, and 1 $\mu_B$/cell respectively. We show that the TM is the origin of magnetism induction in the nonmagnetic bare C$_{{\rm3}}$N$_{{\rm4}}$. Moreover, the neighboring N and C atoms of the TM atom are magnetized. So the above results suggest that TM-C$_{{\rm3}}$N$_{{\rm4}}$ can also regulate the magnetic behaviors of C$_{{\rm3}}$N$_{{\rm4}}$. And TM-C$_{{\rm3}}$N$_{{\rm4}}$ systems will lay the foundation for designing nano-spintronics devices. Introducing a magnetic moment into nonmagnetic 2D semiconducting compounds is significant for them in spintronic application. On the other hand, no induced magnetism is found by the adsorption of Cr, Mn, and Ni atoms. Similar behavior was found in the case of the adsorption of  Ni in B$_{{\rm4}}$C$_{{\rm3}}$ 2D material \cite{Muhammad2024}, Ni in MgAl$_{{\rm2}}$S$_{{\rm4}}$ monolayer \cite{Wenyu2024} and Cr in WSSe compound \cite{Gebredingle2024}. 

\begin{figure}[H]
\centering
\includegraphics[width=0.5\textwidth]{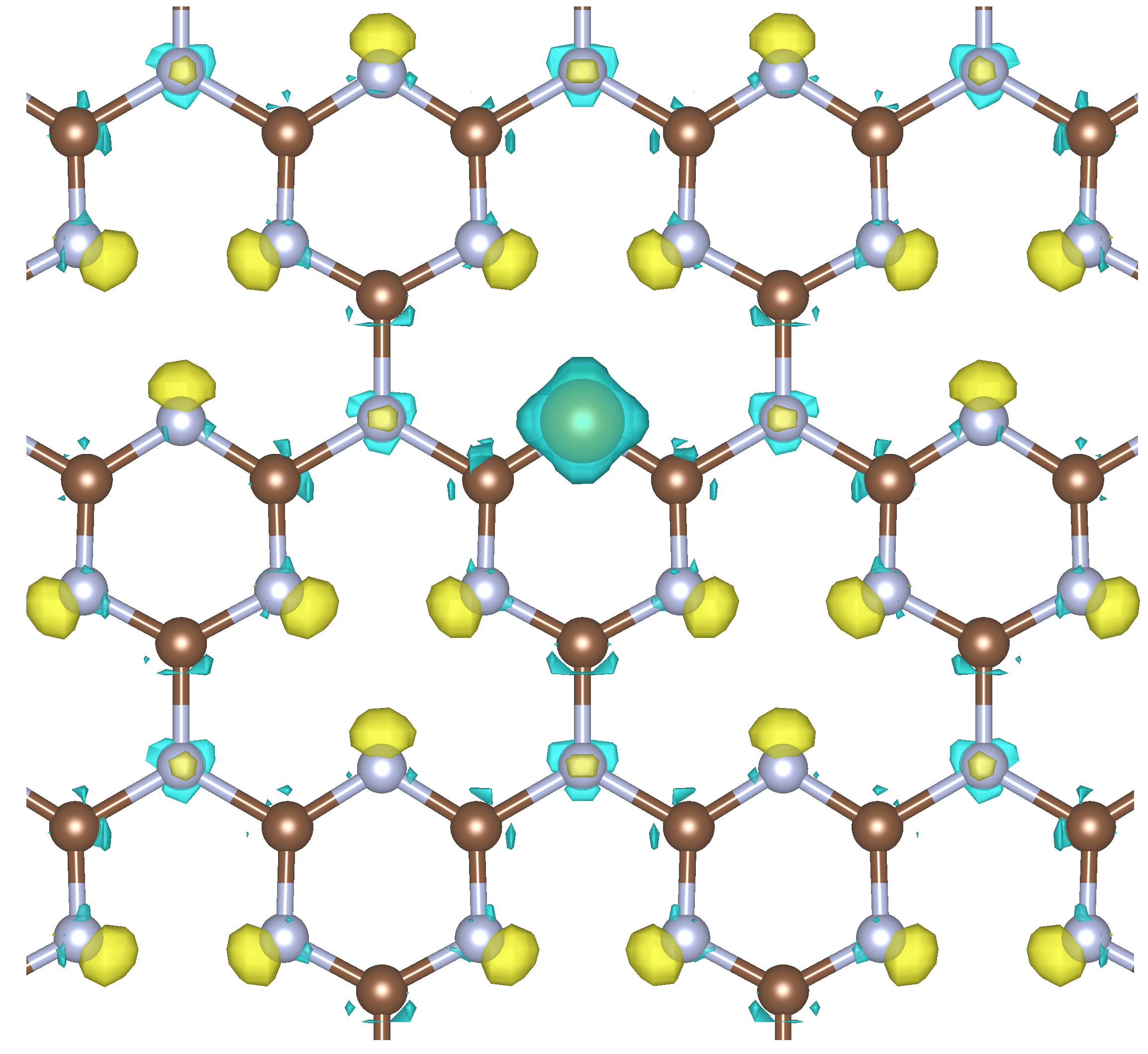} 
\caption{Charge density difference of iron metal adsorbed in gt-C$_{{\rm3}}$N$_{{\rm4}}$. Isosurface is set to 0.006 e/a.u.$^3$. Yellow and cyan colors represent electron accumulation and electron depletion regions respectively. }
\label{ch-fecn}
\end{figure}

The transition metal impurities, which have a much bigger radius, have gained great interest due to their ability to introduce charge into the electron arrangement of C$_{{\rm 3}}$N$_{{\rm 4}}$ material. The charge transfer effect at the transition metal C$_{{\rm3}}$N$_{{\rm4}}$ interface is shown in the Fig. \ref{ch-fecn}, we show the case of Fe-adsorption as an example. The induced magnetic properties are arising due to the charge transfer effect at the  interface. Ferromagnetic ordering is observed as a result of this induced magnetism in the presence of V, Fe, and Co atoms on the C$_{{\rm3}}$N$_{{\rm4}}$ surface. Induced magnetism due to charge transfer between transition metal and graphene was already validated by C. Majumder et al. \cite{Majumder2019, Liu2011}. For transition-metal adsorption, electrons are shared between the adatom and the atoms of gt-C$_{{\rm3}}$N$_{{\rm4}}$ material due to the formation of covalent bonds. 

\section{Conclusion}
In summary, we investigated the effect of vacancy defect creation, hydrogen and oxygen passivation, and TM adsorption on the electronic and magnetic properties of the gt-C$_{\rm3}$N$_{\rm4}$. The vacancy creation pathway leads to the creation of clusters, i.e., six-tuple (6-defects) empty rings. During its formation, the four-fold semicircle C-N-C-N vacancy is significantly stable. The vacancies induce defect states in the band gap that can be effectively passivated by hydrogen or oxygen. We find that defects and H/O passivation behave in a different way, the defects tend to slightly decrease the electronic band gap and keep the nonmagnetic behavior, except the 2- and 5-defects, which are magnetic. However, the passivation of the defects increases the band gaps and turns on magnetism in the system. Furthermore, our results show that the Fe, V, and Co TM atoms adsorbed in gt-C$_{\rm3}$N$_{\rm4}$ induces magnetism on the order of 2 $\mu_B$, 2.89 $\mu_B$, and 1 $\mu_B$, respectively and the TM atom contains most of the magnetic moments of TM-gt-C$_{\rm3}$N$_{\rm4}$, and results in an FM alignment. No magnetism was induced by adsorption of Cr, Mn, and Ni atoms. Interestingly, the introduction of TM maintains the semiconductor nature of gt-C$_{\rm3}$N$_{\rm4}$. 


\backmatter

\bmhead{Acknowledgements}
This article was produced with the financial support of the European Union under the LERCO project (number ${\rm CZ.10.03.01/00/22\_003/0000003}$) {\it via} the Operational Programme Just Transition. The calculations were performed at IT4Innovations National Supercomputing Center through the e-INFRA CZ (grant ID:90254).
\bmhead{Conflicts of interest or competing interests}
There is no conflict to declare.
\bmhead{Data and code availability} 
Not Applicable
\bmhead{Supplementary information}
Not Applicable
\bmhead{Ethical approval}
Not Applicable

\bibliography{a-refs}


\begin{thebibliography}{99}
\ifx \bisbn   \undefined \def \bisbn  #1{ISBN #1}\fi
\ifx \binits  \undefined \def \binits#1{#1}\fi
\ifx \bauthor  \undefined \def \bauthor#1{#1}\fi
\ifx \batitle  \undefined \def \batitle#1{#1}\fi
\ifx \bjtitle  \undefined \def \bjtitle#1{#1}\fi
\ifx \bvolume  \undefined \def \bvolume#1{\textbf{#1}}\fi
\ifx \byear  \undefined \def \byear#1{#1}\fi
\ifx \bissue  \undefined \def \bissue#1{#1}\fi
\ifx \bfpage  \undefined \def \bfpage#1{#1}\fi
\ifx \blpage  \undefined \def \blpage #1{#1}\fi
\ifx \burl  \undefined \def \burl#1{\textsf{#1}}\fi
\ifx \doiurl  \undefined \def \doiurl#1{\url{https://doi.org/#1}}\fi
\ifx \betal  \undefined \def \betal{\textit{et al.}}\fi
\ifx \binstitute  \undefined \def \binstitute#1{#1}\fi
\ifx \binstitutionaled  \undefined \def \binstitutionaled#1{#1}\fi
\ifx \bctitle  \undefined \def \bctitle#1{#1}\fi
\ifx \beditor  \undefined \def \beditor#1{#1}\fi
\ifx \bpublisher  \undefined \def \bpublisher#1{#1}\fi
\ifx \bbtitle  \undefined \def \bbtitle#1{#1}\fi
\ifx \bedition  \undefined \def \bedition#1{#1}\fi
\ifx \bseriesno  \undefined \def \bseriesno#1{#1}\fi
\ifx \blocation  \undefined \def \blocation#1{#1}\fi
\ifx \bsertitle  \undefined \def \bsertitle#1{#1}\fi
\ifx \bsnm \undefined \def \bsnm#1{#1}\fi
\ifx \bsuffix \undefined \def \bsuffix#1{#1}\fi
\ifx \bparticle \undefined \def \bparticle#1{#1}\fi
\ifx \barticle \undefined \def \barticle#1{#1}\fi
\bibcommenthead
\ifx \bconfdate \undefined \def \bconfdate #1{#1}\fi
\ifx \botherref \undefined \def \botherref #1{#1}\fi
\ifx \url \undefined \def \url#1{\textsf{#1}}\fi
\ifx \bchapter \undefined \def \bchapter#1{#1}\fi
\ifx \bbook \undefined \def \bbook#1{#1}\fi
\ifx \bcomment \undefined \def \bcomment#1{#1}\fi
\ifx \oauthor \undefined \def \oauthor#1{#1}\fi
\ifx \citeauthoryear \undefined \def \citeauthoryear#1{#1}\fi
\ifx \endbibitem  \undefined \def \endbibitem {}\fi
\ifx \bconflocation  \undefined \def \bconflocation#1{#1}\fi
\ifx \arxivurl  \undefined \def \arxivurl#1{\textsf{#1}}\fi
\csname PreBibitemsHook\endcsname

\bibitem[\protect\citeauthoryear{Novoselov et~al.}{2004}]{Novoselov2004}
\begin{barticle}
\bauthor{\bsnm{Novoselov}, \binits{K.S.}},
\bauthor{\bsnm{Geim}, \binits{A.K.}},
\bauthor{\bsnm{Morozov}, \binits{S.V.}},
\bauthor{\bsnm{Jiang}, \binits{D.}},
\bauthor{\bsnm{Zhang}, \binits{Y.}},
\bauthor{\bsnm{Dubonos}, \binits{S.V.}},
\bauthor{\bsnm{Grigorieva}, \binits{I.V.}},
\bauthor{\bsnm{Firsov}, \binits{A.A.}}:
\batitle{Electric field effect in atomically thin carbon films}.
\bjtitle{Science}
\bvolume{306}(\bissue{5696}),
\bfpage{666}--\blpage{669}
(\byear{2004})
\doiurl{10.1126/science.1102896}
\end{barticle}
\endbibitem

\bibitem[\protect\citeauthoryear{Geim and Novoselov}{2007}]{Geim2007}
\begin{barticle}
\bauthor{\bsnm{Geim}, \binits{A.K.}},
\bauthor{\bsnm{Novoselov}, \binits{K.S.}}:
\batitle{The rise of graphene}.
\bjtitle{Nature Materials}
\bvolume{6},
\bfpage{183}--\blpage{191}
(\byear{2007})
\doiurl{10.1038/nmat1849}
\end{barticle}
\endbibitem

\bibitem[\protect\citeauthoryear{Mortazavi et~al.}{2020}]{Mortzavi2020}
\begin{barticle}
\bauthor{\bsnm{Mortazavi}, \binits{B.}},
\bauthor{\bsnm{Shojaei}, \binits{F.}},
\bauthor{\bsnm{Shahrokhi}, \binits{M.}},
\bauthor{\bsnm{Azizi}, \binits{M.}},
\bauthor{\bsnm{Rabczuk}, \binits{T.}},
\bauthor{\bsnm{Shapeev}, \binits{A.V.}},
\bauthor{\bsnm{Zhuang}, \binits{X.}}:
\batitle{Nanoporous c3n4, c3n5 and c3n6 nanosheets; novel strong semiconductors
  with low thermal conductivities and appealing optical/electronic properties}.
\bjtitle{Carbon}
\bvolume{167},
\bfpage{40}--\blpage{50}
(\byear{2020})
\doiurl{10.1016/j.carbon.2020.05.105}
\end{barticle}
\endbibitem

\bibitem[\protect\citeauthoryear{Algara-Siller
  et~al.}{2014}]{Algara-Siller2014}
\begin{barticle}
\bauthor{\bsnm{Algara-Siller}, \binits{G.}},
\bauthor{\bsnm{Severin}, \binits{N.}},
\bauthor{\bsnm{Chong}, \binits{S.Y.}},
\bauthor{\bsnm{Björkman}, \binits{T.}},
\bauthor{\bsnm{Palgrave}, \binits{R.G.}},
\bauthor{\bsnm{Laybourn}, \binits{A.}},
\bauthor{\bsnm{Antonietti}, \binits{M.}},
\bauthor{\bsnm{Khimyak}, \binits{Y.Z.}},
\bauthor{\bsnm{Krasheninnikov}, \binits{A.V.}},
\bauthor{\bsnm{Rabe}, \binits{J.P.}},
\bauthor{\bsnm{Kaiser}, \binits{U.}},
\bauthor{\bsnm{Cooper}, \binits{A.I.}},
\bauthor{\bsnm{Thomas}, \binits{A.}},
\bauthor{\bsnm{Bojdys}, \binits{M.J.}}:
\batitle{Triazine-based graphitic carbon nitride: a two-dimensional
  semiconductor}.
\bjtitle{Angewandte Chemie International Edition}
\bvolume{53},
\bfpage{7450}--\blpage{7455}
(\byear{2014})
\doiurl{10.1002/anie.201402191}
\end{barticle}
\endbibitem

\bibitem[\protect\citeauthoryear{Sakhraoui}{2023}]{Sakhraoui_2023}
\begin{barticle}
\bauthor{\bsnm{Sakhraoui}, \binits{T.}}:
\batitle{Effect of vacancy defect and strain on the structural, electronic and
  magnetic properties of carbon nitride 2d monolayers by dftb method}.
\bjtitle{Journal of Physics: Condensed Matter}
\bvolume{35},
\bfpage{324003}
(\byear{2023})
\doiurl{10.1088/1361-648X/acd293}
\end{barticle}
\endbibitem

\bibitem[\protect\citeauthoryear{Luis~Francisco
  et~al.}{2020}]{LuisFrancisco2020}
\begin{barticle}
\bauthor{\bsnm{Luis~Francisco}, \binits{V.}},
\bauthor{\bsnm{Mohammad~Tohidi}, \binits{V.}},
\bauthor{\bsnm{Mostapha}, \binits{D.}},
\bauthor{\bsnm{Zahra}, \binits{N.}},
\bauthor{\bsnm{Mounir}, \binits{M.}},
\bauthor{\bsnm{Emad}, \binits{O.}},
\bauthor{\bsnm{Davide}, \binits{C.}},
\bauthor{\bsnm{Nicola}, \binits{M.}},
\bauthor{\bsnm{Kumar~Varoon}, \binits{A.}}:
\batitle{Large-scale synthesis of crystalline g-c3n4 nanosheets and
  high-temperature h2 sieving from assembled films}.
\bjtitle{Science Advances}
\bvolume{6},
\bfpage{9851}
(\byear{2020})
\doiurl{10.1126/sciadv.aay9851}
\end{barticle}
\endbibitem

\bibitem[\protect\citeauthoryear{Zheng et~al.}{2011}]{Zheng2011}
\begin{barticle}
\bauthor{\bsnm{Zheng}, \binits{Y.}},
\bauthor{\bsnm{Jiao}, \binits{Y.}},
\bauthor{\bsnm{Chen}, \binits{J.}},
\bauthor{\bsnm{Liu}, \binits{J.}},
\bauthor{\bsnm{Liang}, \binits{J.}},
\bauthor{\bsnm{Du}, \binits{A.}},
\bauthor{\bsnm{Zhang}, \binits{W.}},
\bauthor{\bsnm{Zhu}, \binits{Z.}},
\bauthor{\bsnm{Smith}, \binits{S.C.}},
\bauthor{\bsnm{Jaroniec}, \binits{M.}},
\bauthor{\bsnm{Lu}, \binits{G.Q.M.}},
\bauthor{\bsnm{Qiao}, \binits{S.Z.}}:
\batitle{Nanoporous graphitic-c3n4@carbon metal-free electrocatalysts for
  highly efficient oxygen reduction}.
\bjtitle{Journal of the American Chemical Society}
\bvolume{133},
\bfpage{20116}--\blpage{20119}
(\byear{2011})
\doiurl{10.1021/ja209206c}
\end{barticle}
\endbibitem

\bibitem[\protect\citeauthoryear{Thomas et~al.}{2008}]{Thomas2008}
\begin{barticle}
\bauthor{\bsnm{Thomas}, \binits{A.}},
\bauthor{\bsnm{Fischer}, \binits{A.}},
\bauthor{\bsnm{Goettmann}, \binits{F.}},
\bauthor{\bsnm{Antonietti}, \binits{M.}},
\bauthor{\bsnm{Müller}, \binits{J.-O.}},
\bauthor{\bsnm{Schlögl}, \binits{R.}},
\bauthor{\bsnm{Carlsson}, \binits{J.M.}}:
\batitle{Graphitic carbon nitride materials: variation of structure and
  morphology and their use as metal-free catalysts}.
\bjtitle{J. Mater. Chem.}
\bvolume{18},
\bfpage{4893}--\blpage{4908}
(\byear{2008})
\doiurl{10.1039/B800274F}
\end{barticle}
\endbibitem

\bibitem[\protect\citeauthoryear{Zhu et~al.}{2014}]{Zhu2014}
\begin{barticle}
\bauthor{\bsnm{Zhu}, \binits{J.}},
\bauthor{\bsnm{Xiao}, \binits{P.}},
\bauthor{\bsnm{Li}, \binits{H.}},
\bauthor{\bsnm{Carabineiro}, \binits{S.A.C.}}:
\batitle{Graphitic carbon nitride: Synthesis, properties, and applications in
  catalysis}.
\bjtitle{ACS Applied Materials \& Interfaces}
\bvolume{6},
\bfpage{16449}--\blpage{16465}
(\byear{2014})
\doiurl{10.1021/am502925j}
\end{barticle}
\endbibitem

\bibitem[\protect\citeauthoryear{Botari et~al.}{2017}]{Botari2017}
\begin{barticle}
\bauthor{\bsnm{Botari}, \binits{T.}},
\bauthor{\bsnm{Huhn}, \binits{W.P.}},
\bauthor{\bsnm{Lau}, \binits{V.W.-h.}},
\bauthor{\bsnm{Lotsch}, \binits{B.V.}},
\bauthor{\bsnm{Blum}, \binits{V.}}:
\batitle{Thermodynamic equilibria in carbon nitride photocatalyst materials and
  conditions for the existence of graphitic carbon nitride g-c3n4}.
\bjtitle{Chemistry of Materials}
\bvolume{29},
\bfpage{4445}--\blpage{4453}
(\byear{2017})
\doiurl{10.1021/acs.chemmater.7b00965}
\end{barticle}
\endbibitem

\bibitem[\protect\citeauthoryear{Miller et~al.}{2017}]{Miller2017}
\begin{barticle}
\bauthor{\bsnm{Miller}, \binits{T.S.}},
\bauthor{\bsnm{Jorge}, \binits{A.B.}},
\bauthor{\bsnm{Suter}, \binits{T.M.}},
\bauthor{\bsnm{Sella}, \binits{A.}},
\bauthor{\bsnm{Corà}, \binits{F.}},
\bauthor{\bsnm{McMillan}, \binits{P.F.}}:
\batitle{Carbon nitrides: synthesis and characterization of a new class of
  functional materials}.
\bjtitle{Phys. Chem. Chem. Phys.}
\bvolume{19},
\bfpage{15613}--\blpage{15638}
(\byear{2017})
\doiurl{10.1039/C7CP02711G}
\end{barticle}
\endbibitem

\bibitem[\protect\citeauthoryear{Banhart et~al.}{2011}]{Banhart2011}
\begin{barticle}
\bauthor{\bsnm{Banhart}, \binits{F.}},
\bauthor{\bsnm{Kotakoski}, \binits{J.}},
\bauthor{\bsnm{Krasheninnikov}, \binits{A.V.}}:
\batitle{Structural defects in graphene}.
\bjtitle{ACS Nano}
\bvolume{5},
\bfpage{26}--\blpage{41}
(\byear{2011})
\doiurl{10.1021/nn102598m}
\end{barticle}
\endbibitem

\bibitem[\protect\citeauthoryear{Vicarelli et~al.}{2015}]{Vicarelli2015}
\begin{barticle}
\bauthor{\bsnm{Vicarelli}, \binits{L.}},
\bauthor{\bsnm{Heerema}, \binits{S.J.}},
\bauthor{\bsnm{Dekker}, \binits{C.}},
\bauthor{\bsnm{Zandbergen}, \binits{H.W.}}:
\batitle{Controlling defects in graphene for optimizing the electrical
  properties of graphene nanodevices}.
\bjtitle{ACS Nano}
\bvolume{9},
\bfpage{3428}--\blpage{3435}
(\byear{2015})
\doiurl{10.1021/acsnano.5b01762}
\end{barticle}
\endbibitem

\bibitem[\protect\citeauthoryear{López-Polín et~al.}{2015}]{Lopez-Polin2015}
\begin{barticle}
\bauthor{\bsnm{López-Polín}, \binits{G.}},
\bauthor{\bsnm{Gómez-Navarro}, \binits{C.}},
\bauthor{\bsnm{Parente}, \binits{V.}},
\bauthor{\bsnm{Guinea}, \binits{F.}},
\bauthor{\bsnm{Katsnelson}, \binits{M.I.}},
\bauthor{\bsnm{Pérez-Murano}, \binits{F.}},
\bauthor{\bsnm{Gómez-Herrero}, \binits{J.}}:
\batitle{Increasing the elastic modulus of graphene by controlled defect
  creation}.
\bjtitle{Nature Physics}
\bvolume{11},
\bfpage{26}--\blpage{31}
(\byear{2015})
\doiurl{10.1038/nphys3183}
\end{barticle}
\endbibitem

\bibitem[\protect\citeauthoryear{Wong et~al.}{2015}]{Wong2015}
\begin{barticle}
\bauthor{\bsnm{Wong}, \binits{D.}},
\bauthor{\bsnm{Velasco}, \binits{J.}},
\bauthor{\bsnm{Ju}, \binits{L.}},
\bauthor{\bsnm{Lee}, \binits{J.}},
\bauthor{\bsnm{Kahn}, \binits{S.}},
\bauthor{\bsnm{Tsai}, \binits{H.-Z.}},
\bauthor{\bsnm{Germany}, \binits{C.}},
\bauthor{\bsnm{Taniguchi}, \binits{T.}},
\bauthor{\bsnm{Watanabe}, \binits{K.}},
\bauthor{\bsnm{Zettl}, \binits{A.}},
\bauthor{\bsnm{Wang}, \binits{F.}},
\bauthor{\bsnm{Crommie}, \binits{M.F.}}:
\batitle{Characterization and manipulation of individual defects in insulating
  hexagonal boron nitride using scanning tunnelling microscopy}.
\bjtitle{Nature Nanotechnology}
\bvolume{10},
\bfpage{949}--\blpage{953}
(\byear{2015})
\doiurl{10.1038/nnano.2015.188}
\end{barticle}
\endbibitem

\bibitem[\protect\citeauthoryear{Liu et~al.}{2012}]{Liu2012}
\begin{barticle}
\bauthor{\bsnm{Liu}, \binits{Y.}},
\bauthor{\bsnm{Zou}, \binits{X.}},
\bauthor{\bsnm{Yakobson}, \binits{B.I.}}:
\batitle{Dislocations and grain boundaries in two-dimensional boron nitride}.
\bjtitle{ACS Nano}
\bvolume{6},
\bfpage{7053}--\blpage{7058}
(\byear{2012})
\doiurl{10.1021/nn302099q}
\end{barticle}
\endbibitem

\bibitem[\protect\citeauthoryear{Bourrellier et~al.}{2016}]{Bourrellier2016}
\begin{barticle}
\bauthor{\bsnm{Bourrellier}, \binits{R.}},
\bauthor{\bsnm{Meuret}, \binits{S.}},
\bauthor{\bsnm{Tararan}, \binits{A.}},
\bauthor{\bsnm{Stéphan}, \binits{O.}},
\bauthor{\bsnm{Kociak}, \binits{M.}},
\bauthor{\bsnm{Tizei}, \binits{L.H.G.}},
\bauthor{\bsnm{Zobelli}, \binits{A.}}:
\batitle{Bright uv single photon emission at point defects in h-bn}.
\bjtitle{Nano Letters}
\bvolume{16},
\bfpage{4317}--\blpage{4321}
(\byear{2016})
\doiurl{10.1021/acs.nanolett.6b01368}
\end{barticle}
\endbibitem

\bibitem[\protect\citeauthoryear{Sang et~al.}{2016}]{Sang2016}
\begin{barticle}
\bauthor{\bsnm{Sang}, \binits{X.}},
\bauthor{\bsnm{Xie}, \binits{Y.}},
\bauthor{\bsnm{Lin}, \binits{M.-W.}},
\bauthor{\bsnm{Alhabeb}, \binits{M.}},
\bauthor{\bsnm{Van~Aken}, \binits{K.L.}},
\bauthor{\bsnm{Gogotsi}, \binits{Y.}},
\bauthor{\bsnm{Kent}, \binits{P.R.C.}},
\bauthor{\bsnm{Xiao}, \binits{K.}},
\bauthor{\bsnm{Unocic}, \binits{R.R.}}:
\batitle{Atomic defects in monolayer titanium carbide (ti3c2tx) mxene}.
\bjtitle{ACS Nano}
\bvolume{10},
\bfpage{9193}--\blpage{9200}
(\byear{2016})
\doiurl{10.1021/acsnano.6b05240}
\end{barticle}
\endbibitem

\bibitem[\protect\citeauthoryear{Karlsson et~al.}{2015}]{Karlsson2015}
\begin{barticle}
\bauthor{\bsnm{Karlsson}, \binits{L.H.}},
\bauthor{\bsnm{Birch}, \binits{J.}},
\bauthor{\bsnm{Halim}, \binits{J.}},
\bauthor{\bsnm{Barsoum}, \binits{M.W.}},
\bauthor{\bsnm{Persson}, \binits{P.O.A.}}:
\batitle{Atomically resolved structural and chemical investigation of single
  mxene sheets}.
\bjtitle{Nano Letters}
\bvolume{15},
\bfpage{4955}--\blpage{4960}
(\byear{2015})
\doiurl{10.1021/acs.nanolett.5b00737}
\end{barticle}
\endbibitem

\bibitem[\protect\citeauthoryear{Sakhraoui and
  Karlický}{2022}]{Sakhraoui_2022}
\begin{barticle}
\bauthor{\bsnm{Sakhraoui}, \binits{T.}},
\bauthor{\bsnm{Karlický}, \binits{F.}}:
\batitle{Electronic nature transition and magnetism creation in
  vacancy-defected ti2co2 mxene under biaxial strain: A dftb + u study}.
\bjtitle{ACS Omega}
\bvolume{7},
\bfpage{42221}--\blpage{42232}
(\byear{2022})
\doiurl{10.1021/acsomega.2c05037}
\end{barticle}
\endbibitem

\bibitem[\protect\citeauthoryear{Halim et~al.}{2016}]{Halim2016}
\begin{barticle}
\bauthor{\bsnm{Halim}, \binits{J.}},
\bauthor{\bsnm{Kota}, \binits{S.}},
\bauthor{\bsnm{Lukatskaya}, \binits{M.R.}},
\bauthor{\bsnm{Naguib}, \binits{M.}},
\bauthor{\bsnm{Zhao}, \binits{M.-Q.}},
\bauthor{\bsnm{Moon}, \binits{E.J.}},
\bauthor{\bsnm{Pitock}, \binits{J.}},
\bauthor{\bsnm{Nanda}, \binits{J.}},
\bauthor{\bsnm{May}, \binits{S.J.}},
\bauthor{\bsnm{Gogotsi}, \binits{Y.}},
\bauthor{\bsnm{Barsoum}, \binits{M.W.}}:
\batitle{Synthesis and characterization of 2d molybdenum carbide (mxene)}.
\bjtitle{Advanced Functional Materials}
\bvolume{26},
\bfpage{3118}--\blpage{3127}
(\byear{2016})
\doiurl{10.1002/adfm.201505328}
\end{barticle}
\endbibitem

\bibitem[\protect\citeauthoryear{Lipatov et~al.}{2016}]{Lipatov2016}
\begin{barticle}
\bauthor{\bsnm{Lipatov}, \binits{A.}},
\bauthor{\bsnm{Alhabeb}, \binits{M.}},
\bauthor{\bsnm{Lukatskaya}, \binits{M.R.}},
\bauthor{\bsnm{Boson}, \binits{A.}},
\bauthor{\bsnm{Gogotsi}, \binits{Y.}},
\bauthor{\bsnm{Sinitskii}, \binits{A.}}:
\batitle{Effect of synthesis on quality, electronic properties and
  environmental stability of individual monolayer ti3c2 mxene flakes}.
\bjtitle{Advanced Electronic Materials}
\bvolume{2},
\bfpage{1600255}
(\byear{2016})
\doiurl{10.1002/aelm.201600255}
\end{barticle}
\endbibitem

\bibitem[\protect\citeauthoryear{Zhou et~al.}{2013}]{Zhou2013}
\begin{barticle}
\bauthor{\bsnm{Zhou}, \binits{W.}},
\bauthor{\bsnm{Zou}, \binits{X.}},
\bauthor{\bsnm{Najmaei}, \binits{S.}},
\bauthor{\bsnm{Liu}, \binits{Z.}},
\bauthor{\bsnm{Shi}, \binits{Y.}},
\bauthor{\bsnm{Kong}, \binits{J.}},
\bauthor{\bsnm{Lou}, \binits{J.}},
\bauthor{\bsnm{Ajayan}, \binits{P.M.}},
\bauthor{\bsnm{Yakobson}, \binits{B.I.}},
\bauthor{\bsnm{Idrobo}, \binits{J.-C.}}:
\batitle{Intrinsic structural defects in monolayer molybdenum disulfide}.
\bjtitle{Nano Letters}
\bvolume{13},
\bfpage{2615}--\blpage{2622}
(\byear{2013})
\doiurl{10.1021/nl4007479}
\end{barticle}
\endbibitem

\bibitem[\protect\citeauthoryear{Amani et~al.}{2015}]{Martin2015}
\begin{barticle}
\bauthor{\bsnm{Amani}, \binits{M.}},
\bauthor{\bsnm{Lien}, \binits{D.-H.}},
\bauthor{\bsnm{Kiriya}, \binits{D.}},
\bauthor{\bsnm{Xiao}, \binits{J.}},
\bauthor{\bsnm{Azcatl}, \binits{A.}},
\bauthor{\bsnm{Noh}, \binits{J.}},
\bauthor{\bsnm{Madhvapathy}, \binits{S.R.}},
\bauthor{\bsnm{Addou}, \binits{R.}},
\bauthor{\bsnm{KC}, \binits{S.}},
\bauthor{\bsnm{Dubey}, \binits{M.}},
\bauthor{\bsnm{Cho}, \binits{K.}},
\bauthor{\bsnm{Wallace}, \binits{R.M.}},
\bauthor{\bsnm{Lee}, \binits{S.-C.}},
\bauthor{\bsnm{He}, \binits{J.-H.}},
\bauthor{\bsnm{Ager}, \binits{J.W.}},
\bauthor{\bsnm{Zhang}, \binits{X.}},
\bauthor{\bsnm{Yablonovitch}, \binits{E.}},
\bauthor{\bsnm{Javey}, \binits{A.}}:
\batitle{Near-unity photoluminescence quantum yield in mos<sub>2</sub>}.
\bjtitle{Science}
\bvolume{350},
\bfpage{1065}--\blpage{1068}
(\byear{2015})
\doiurl{10.1126/science.aad2114}
\end{barticle}
\endbibitem

\bibitem[\protect\citeauthoryear{Wei et~al.}{2012}]{Wei2012}
\begin{barticle}
\bauthor{\bsnm{Wei}, \binits{Y.}},
\bauthor{\bsnm{Wu}, \binits{J.}},
\bauthor{\bsnm{Yin}, \binits{H.}},
\bauthor{\bsnm{Shi}, \binits{X.}},
\bauthor{\bsnm{Yang}, \binits{R.}},
\bauthor{\bsnm{Dresselhaus}, \binits{M.}}:
\batitle{The nature of strength enhancement and weakening by
  pentagon–heptagon defects in graphene}.
\bjtitle{Nature Materials}
\bvolume{11},
\bfpage{759}--\blpage{763}
(\byear{2012})
\doiurl{10.1038/nmat3370}
\end{barticle}
\endbibitem

\bibitem[\protect\citeauthoryear{Yazyev and Louie}{2010}]{Yazyev2010}
\begin{barticle}
\bauthor{\bsnm{Yazyev}, \binits{O.V.}},
\bauthor{\bsnm{Louie}, \binits{S.G.}}:
\batitle{Electronic transport in polycrystalline graphene}.
\bjtitle{Nature Materials}
\bvolume{9},
\bfpage{806}--\blpage{809}
(\byear{2010})
\doiurl{10.1038/nmat2830}
\end{barticle}
\endbibitem

\bibitem[\protect\citeauthoryear{Li et~al.}{2023}]{Yuhan2023}
\begin{barticle}
\bauthor{\bsnm{Li}, \binits{Y.}},
\bauthor{\bsnm{He}, \binits{Z.}},
\bauthor{\bsnm{Liu}, \binits{L.}},
\bauthor{\bsnm{Jiang}, \binits{Y.}},
\bauthor{\bsnm{Ong}, \binits{W.-J.}},
\bauthor{\bsnm{Duan}, \binits{Y.}},
\bauthor{\bsnm{Ho}, \binits{W.}},
\bauthor{\bsnm{Dong}, \binits{F.}}:
\batitle{Inside-and-out modification of graphitic carbon nitride (g-c3n4)
  photocatalysts via defect engineering for energy and environmental science}.
\bjtitle{Nano Energy}
\bvolume{105},
\bfpage{108032}
(\byear{2023})
\doiurl{10.1016/j.nanoen.2022.108032}
\end{barticle}
\endbibitem

\bibitem[\protect\citeauthoryear{Dong et~al.}{2020}]{Dong2020}
\begin{barticle}
\bauthor{\bsnm{Dong}, \binits{G.}},
\bauthor{\bsnm{Wen}, \binits{Y.}},
\bauthor{\bsnm{Fan}, \binits{H.}},
\bauthor{\bsnm{Wang}, \binits{C.}},
\bauthor{\bsnm{Cheng}, \binits{Z.}},
\bauthor{\bsnm{Zhang}, \binits{M.}},
\bauthor{\bsnm{Ma}, \binits{J.}},
\bauthor{\bsnm{Zhang}, \binits{S.}}:
\batitle{Graphitic carbon nitride with thermally-induced nitrogen defects: an
  efficient process to enhance photocatalytic h2 production performance}.
\bjtitle{RSC Advances}
\bvolume{10},
\bfpage{18632}--\blpage{18638}
(\byear{2020})
\doiurl{10.1039/D0RA01425G}
\end{barticle}
\endbibitem

\bibitem[\protect\citeauthoryear{Xue et~al.}{2019}]{Xue2019}
\begin{barticle}
\bauthor{\bsnm{Xue}, \binits{J.}},
\bauthor{\bsnm{Fujitsuka}, \binits{M.}},
\bauthor{\bsnm{Majima}, \binits{T.}}:
\batitle{The role of nitrogen defects in graphitic carbon nitride for
  visible-light-driven hydrogen evolution}.
\bjtitle{Phys. Chem. Chem. Phys.}
\bvolume{21},
\bfpage{2318}--\blpage{2324}
(\byear{2019})
\doiurl{10.1039/C8CP06922K}
\end{barticle}
\endbibitem

\bibitem[\protect\citeauthoryear{Liu et~al.}{2024}]{Liu2024}
\begin{barticle}
\bauthor{\bsnm{Liu}, \binits{Y.}},
\bauthor{\bsnm{Chen}, \binits{X.}},
\bauthor{\bsnm{Kamali}, \binits{M.}},
\bauthor{\bsnm{Rossi}, \binits{B.}},
\bauthor{\bsnm{Appels}, \binits{L.}},
\bauthor{\bsnm{Dewil}, \binits{R.}}:
\batitle{Unraveling the presence and positions of nitrogen defects in defective
  g-c3n4 for improved organic photocatalytic degradation: Insights from
  experiments and theoretical calculations}.
\bjtitle{Advanced Functional Materials}
\bvolume{34},
\bfpage{2405741}
(\byear{2024})
\doiurl{10.1002/adfm.202405741}
\end{barticle}
\endbibitem

\bibitem[\protect\citeauthoryear{Li et~al.}{2018}]{YuhanLi2018}
\begin{barticle}
\bauthor{\bsnm{Li}, \binits{Y.}},
\bauthor{\bsnm{Ho}, \binits{W.}},
\bauthor{\bsnm{Lv}, \binits{K.}},
\bauthor{\bsnm{Zhu}, \binits{B.}},
\bauthor{\bsnm{Lee}, \binits{S.C.}}:
\batitle{Carbon vacancy-induced enhancement of the visible light-driven
  photocatalytic oxidation of no over g-c3n4 nanosheets}.
\bjtitle{Applied Surface Science}
\bvolume{430},
\bfpage{380}--\blpage{389}
(\byear{2018})
\doiurl{10.1016/j.apsusc.2017.06.054}
\end{barticle}
\endbibitem

\bibitem[\protect\citeauthoryear{Li et~al.}{2020}]{YuhanLi2020}
\begin{barticle}
\bauthor{\bsnm{Li}, \binits{Y.}},
\bauthor{\bsnm{Gu}, \binits{M.}},
\bauthor{\bsnm{Zhang}, \binits{X.}},
\bauthor{\bsnm{Fan}, \binits{J.}},
\bauthor{\bsnm{Lv}, \binits{K.}},
\bauthor{\bsnm{Carabineiro}, \binits{S.A.C.}},
\bauthor{\bsnm{Dong}, \binits{F.}}:
\batitle{2d g-c3n4 for advancement of photo-generated carrier dynamics: Status
  and challenges}.
\bjtitle{Materials Today}
\bvolume{41},
\bfpage{270}--\blpage{303}
(\byear{2020})
\doiurl{10.1016/j.mattod.2020.09.004}
\end{barticle}
\endbibitem

\bibitem[\protect\citeauthoryear{Song et~al.}{2021}]{Peng2021}
\begin{barticle}
\bauthor{\bsnm{Song}, \binits{P.}},
\bauthor{\bsnm{Sun}, \binits{S.}},
\bauthor{\bsnm{Cui}, \binits{J.}},
\bauthor{\bsnm{Zheng}, \binits{X.}},
\bauthor{\bsnm{Liang}, \binits{S.}}:
\batitle{Organic dye-reformed construction of porous-defect g-c3n4 nanosheet
  for improved visible-light-driven photocatalytic activity}.
\bjtitle{Applied Surface Science}
\bvolume{568},
\bfpage{150986}
(\byear{2021})
\doiurl{10.1016/j.apsusc.2021.150986}
\end{barticle}
\endbibitem

\bibitem[\protect\citeauthoryear{Wang et~al.}{2025}]{Wang2025}
\begin{barticle}
\bauthor{\bsnm{Wang}, \binits{S.}},
\bauthor{\bsnm{Wang}, \binits{L.}},
\bauthor{\bsnm{Liu}, \binits{W.}},
\bauthor{\bsnm{Ke}, \binits{C.}},
\bauthor{\bsnm{Li}, \binits{M.}},
\bauthor{\bsnm{Hui}, \binits{J.}}:
\batitle{Recent progress in intrinsic defect modulation of g-c3n4 based
  materials and their photocatalytic properties}.
\bjtitle{Nano Research}
\bvolume{18},
\bfpage{94907125}
(\byear{2025})
\doiurl{10.26599/NR.2025.94907125}
\end{barticle}
\endbibitem

\bibitem[\protect\citeauthoryear{Kumar et~al.}{2021}]{Kumar2021}
\begin{barticle}
\bauthor{\bsnm{Kumar}, \binits{A.}},
\bauthor{\bsnm{Raizada}, \binits{P.}},
\bauthor{\bsnm{Hosseini-Bandegharaei}, \binits{A.}},
\bauthor{\bsnm{Thakur}, \binits{V.K.}},
\bauthor{\bsnm{Nguyen}, \binits{V.-H.}},
\bauthor{\bsnm{Singh}, \binits{P.}}:
\batitle{C-{,} n-vacancy defect engineered polymeric carbon nitride towards
  photocatalysis: viewpoints and challenges}.
\bjtitle{J. Mater. Chem. A}
\bvolume{9},
\bfpage{111}--\blpage{153}
(\byear{2021})
\doiurl{10.1039/D0TA08384D}
\end{barticle}
\endbibitem

\bibitem[\protect\citeauthoryear{Tay et~al.}{2015}]{Tay2015}
\begin{barticle}
\bauthor{\bsnm{Tay}, \binits{Q.}},
\bauthor{\bsnm{Kanhere}, \binits{P.}},
\bauthor{\bsnm{Ng}, \binits{C.F.}},
\bauthor{\bsnm{Chen}, \binits{S.}},
\bauthor{\bsnm{Chakraborty}, \binits{S.}},
\bauthor{\bsnm{Huan}, \binits{A.C.H.}},
\bauthor{\bsnm{Sum}, \binits{T.C.}},
\bauthor{\bsnm{Ahuja}, \binits{R.}},
\bauthor{\bsnm{Chen}, \binits{Z.}}:
\batitle{Defect engineered g-c3n4 for efficient visible light photocatalytic
  hydrogen production}.
\bjtitle{Chemistry of Materials}
\bvolume{27},
\bfpage{4930}--\blpage{4933}
(\byear{2015})
\doiurl{10.1021/acs.chemmater.5b02344}
\end{barticle}
\endbibitem

\bibitem[\protect\citeauthoryear{Zhong et~al.}{2024}]{Daopeng2024}
\begin{barticle}
\bauthor{\bsnm{Zhong}, \binits{D.}},
\bauthor{\bsnm{Jia}, \binits{X.}},
\bauthor{\bsnm{Zhang}, \binits{X.}},
\bauthor{\bsnm{Zhao}, \binits{J.}},
\bauthor{\bsnm{Meng}, \binits{F.}},
\bauthor{\bsnm{Wang}, \binits{D.}},
\bauthor{\bsnm{Fang}, \binits{Y.}},
\bauthor{\bsnm{Zhang}, \binits{Z.}}:
\batitle{Facile synthesis of distinctive nitrogen defect-regulated g-c3n4 for
  efficient photocatalytic hydrogen evolution}.
\bjtitle{Diamond and Related Materials}
\bvolume{142},
\bfpage{110816}
(\byear{2024})
\doiurl{10.1016/j.diamond.2024.110816}
\end{barticle}
\endbibitem

\bibitem[\protect\citeauthoryear{Wang et~al.}{2023}]{Wang2023}
\begin{barticle}
\bauthor{\bsnm{Wang}, \binits{Q.}},
\bauthor{\bsnm{Li}, \binits{Y.}},
\bauthor{\bsnm{Huang}, \binits{F.}},
\bauthor{\bsnm{Song}, \binits{S.}},
\bauthor{\bsnm{Ai}, \binits{G.}},
\bauthor{\bsnm{Xin}, \binits{X.}},
\bauthor{\bsnm{Zhao}, \binits{B.}},
\bauthor{\bsnm{Zheng}, \binits{Y.}},
\bauthor{\bsnm{Zhang}, \binits{Z.}}:
\batitle{Recent advances in g-c3n4-based materials and their application in
  energy and environmental sustainability}.
\bjtitle{Molecules}
\bvolume{28},
\bfpage{432}
(\byear{2023})
\doiurl{10.3390/molecules28010432}
\end{barticle}
\endbibitem

\bibitem[\protect\citeauthoryear{Nor et~al.}{2021}]{Nur2021}
\begin{barticle}
\bauthor{\bsnm{Nor}, \binits{N.U.M.}},
\bauthor{\bsnm{Mazalan}, \binits{E.}},
\bauthor{\bsnm{Amin}, \binits{N.A.S.}}:
\batitle{Insights into enhancing photocatalytic reduction of co2:
  Substitutional defect strategy of modified g-c3n4 by experimental and
  theoretical calculation approaches}.
\bjtitle{Journal of Alloys and Compounds}
\bvolume{871},
\bfpage{159464}
(\byear{2021})
\doiurl{10.1016/j.jallcom.2021.159464}
\end{barticle}
\endbibitem

\bibitem[\protect\citeauthoryear{Zhu et~al.}{2017}]{Zhu2017}
\begin{barticle}
\bauthor{\bsnm{Zhu}, \binits{B.}},
\bauthor{\bsnm{Zhang}, \binits{J.}},
\bauthor{\bsnm{Jiang}, \binits{C.}},
\bauthor{\bsnm{Cheng}, \binits{B.}},
\bauthor{\bsnm{Yu}, \binits{J.}}:
\batitle{First principle investigation of halogen-doped monolayer g-c3n4
  photocatalyst}.
\bjtitle{Applied Catalysis B: Environmental}
\bvolume{207},
\bfpage{27}--\blpage{34}
(\byear{2017})
\doiurl{10.1016/j.apcatb.2017.02.020}
\end{barticle}
\endbibitem

\bibitem[\protect\citeauthoryear{Shen et~al.}{2020}]{Quanhao2020}
\begin{barticle}
\bauthor{\bsnm{Shen}, \binits{Q.}},
\bauthor{\bsnm{Li}, \binits{N.}},
\bauthor{\bsnm{Bibi}, \binits{R.}},
\bauthor{\bsnm{Richard}, \binits{N.}},
\bauthor{\bsnm{Liu}, \binits{M.}},
\bauthor{\bsnm{Zhou}, \binits{J.}},
\bauthor{\bsnm{Jing}, \binits{D.}}:
\batitle{Incorporating nitrogen defects into novel few-layer carbon nitride
  nanosheets for enhanced photocatalytic h2 production}.
\bjtitle{Applied Surface Science}
\bvolume{529},
\bfpage{147104}
(\byear{2020})
\doiurl{10.1016/j.apsusc.2020.147104}
\end{barticle}
\endbibitem

\bibitem[\protect\citeauthoryear{Hou et~al.}{2024}]{Hou2024}
\begin{barticle}
\bauthor{\bsnm{Hou}, \binits{S.}},
\bauthor{\bsnm{Gao}, \binits{X.}},
\bauthor{\bsnm{Lv}, \binits{X.}},
\bauthor{\bsnm{Zhao}, \binits{Y.}},
\bauthor{\bsnm{Yin}, \binits{X.}},
\bauthor{\bsnm{Liu}, \binits{Y.}},
\bauthor{\bsnm{Fang}, \binits{J.}},
\bauthor{\bsnm{Yu}, \binits{X.}},
\bauthor{\bsnm{Ma}, \binits{X.}},
\bauthor{\bsnm{Ma}, \binits{T.}},
\bauthor{\bsnm{Su}, \binits{D.}}:
\batitle{Decade milestone advancement of defect-engineered g-c3n4 for solar
  catalytic applications}.
\bjtitle{Nano-Micro Letters}
\bvolume{16},
\bfpage{70}
(\byear{2024})
\doiurl{10.1007/s40820-023-01297-x}
\end{barticle}
\endbibitem

\bibitem[\protect\citeauthoryear{Jiao et~al.}{2020}]{Shilong2020}
\begin{barticle}
\bauthor{\bsnm{Jiao}, \binits{S.}},
\bauthor{\bsnm{Fu}, \binits{X.}},
\bauthor{\bsnm{Zhang}, \binits{L.}},
\bauthor{\bsnm{Zeng}, \binits{Y.-J.}},
\bauthor{\bsnm{Huang}, \binits{H.}}:
\batitle{Point-defect-optimized electron distribution for enhanced
  electrocatalysis: Towards the perfection of the imperfections}.
\bjtitle{Nano Today}
\bvolume{31},
\bfpage{100833}
(\byear{2020})
\doiurl{10.1016/j.nantod.2019.100833}
\end{barticle}
\endbibitem

\bibitem[\protect\citeauthoryear{Sun et~al.}{2022}]{Sun2022}
\begin{barticle}
\bauthor{\bsnm{Sun}, \binits{S.P.}},
\bauthor{\bsnm{Wang}, \binits{Y.R.}},
\bauthor{\bsnm{Gu}, \binits{S.}},
\bauthor{\bsnm{Wang}, \binits{B.}},
\bauthor{\bsnm{Sun}, \binits{J.H.}},
\bauthor{\bsnm{Jiang}, \binits{Y.}}:
\batitle{Effects of vacancies on the electronic structures and photocatalytic
  properties of g-c3n4}.
\bjtitle{Vacuum}
\bvolume{206},
\bfpage{111483}
(\byear{2022})
\doiurl{10.1016/j.vacuum.2022.111483}
\end{barticle}
\endbibitem

\bibitem[\protect\citeauthoryear{Wu et~al.}{2020}]{Aiqing2020}
\begin{barticle}
\bauthor{\bsnm{Wu}, \binits{A.}},
\bauthor{\bsnm{Song}, \binits{Q.}},
\bauthor{\bsnm{Liu}, \binits{H.}}:
\batitle{Oxygen atom adsorbed on the sulphur vacancy of monolayer mos2: A
  promising method for the passivation of the vacancy defect}.
\bjtitle{Computational and Theoretical Chemistry}
\bvolume{1187},
\bfpage{112906}
(\byear{2020})
\doiurl{10.1016/j.comptc.2020.112906}
\end{barticle}
\endbibitem

\bibitem[\protect\citeauthoryear{Nan et~al.}{2014}]{Nan2014}
\begin{barticle}
\bauthor{\bsnm{Nan}, \binits{H.}},
\bauthor{\bsnm{Wang}, \binits{Z.}},
\bauthor{\bsnm{Wang}, \binits{W.}},
\bauthor{\bsnm{Liang}, \binits{Z.}},
\bauthor{\bsnm{Lu}, \binits{Y.}},
\bauthor{\bsnm{Chen}, \binits{Q.}},
\bauthor{\bsnm{He}, \binits{D.}},
\bauthor{\bsnm{Tan}, \binits{P.}},
\bauthor{\bsnm{Miao}, \binits{F.}},
\bauthor{\bsnm{Wang}, \binits{X.}},
\bauthor{\bsnm{Wang}, \binits{J.}},
\bauthor{\bsnm{Ni}, \binits{Z.}}:
\batitle{Strong photoluminescence enhancement of mos2 through defect
  engineering and oxygen bonding}.
\bjtitle{ACS Nano}
\bvolume{8},
\bfpage{5738}--\blpage{5745}
(\byear{2014})
\doiurl{10.1021/nn500532f}
\end{barticle}
\endbibitem

\bibitem[\protect\citeauthoryear{Tongay et~al.}{2013}]{Tongay2013}
\begin{barticle}
\bauthor{\bsnm{Tongay}, \binits{S.}},
\bauthor{\bsnm{Zhou}, \binits{J.}},
\bauthor{\bsnm{Ataca}, \binits{C.}},
\bauthor{\bsnm{Liu}, \binits{J.}},
\bauthor{\bsnm{Kang}, \binits{J.S.}},
\bauthor{\bsnm{Matthews}, \binits{T.S.}},
\bauthor{\bsnm{You}, \binits{L.}},
\bauthor{\bsnm{Li}, \binits{J.}},
\bauthor{\bsnm{Grossman}, \binits{J.C.}},
\bauthor{\bsnm{Wu}, \binits{J.}}:
\batitle{Broad-range modulation of light emission in two-dimensional
  semiconductors by molecular physisorption gating}.
\bjtitle{Nano Letters}
\bvolume{13},
\bfpage{2831}--\blpage{2836}
(\byear{2013})
\doiurl{10.1021/nl4011172}
\end{barticle}
\endbibitem

\bibitem[\protect\citeauthoryear{Kim et~al.}{2018}]{Kim2018}
\begin{barticle}
\bauthor{\bsnm{Kim}, \binits{Y.}},
\bauthor{\bsnm{Lee}, \binits{Y.}},
\bauthor{\bsnm{Kim}, \binits{H.}},
\bauthor{\bsnm{Roy}, \binits{S.}},
\bauthor{\bsnm{Kim}, \binits{J.}}:
\batitle{Near-field exciton imaging of chemically treated mos2 monolayers}.
\bjtitle{Nanoscale}
\bvolume{10},
\bfpage{8851}--\blpage{8858}
(\byear{2018})
\doiurl{10.1039/C8NR00606G}
\end{barticle}
\endbibitem

\bibitem[\protect\citeauthoryear{Atallah et~al.}{2017}]{Atallah2017}
\begin{barticle}
\bauthor{\bsnm{Atallah}, \binits{T.L.}},
\bauthor{\bsnm{Wang}, \binits{J.}},
\bauthor{\bsnm{Bosch}, \binits{M.}},
\bauthor{\bsnm{Seo}, \binits{D.}},
\bauthor{\bsnm{Burke}, \binits{R.A.}},
\bauthor{\bsnm{Moneer}, \binits{O.}},
\bauthor{\bsnm{Zhu}, \binits{J.}},
\bauthor{\bsnm{Theibault}, \binits{M.}},
\bauthor{\bsnm{Brus}, \binits{L.E.}},
\bauthor{\bsnm{Hone}, \binits{J.}},
\bauthor{\bsnm{Zhu}, \binits{X.-Y.}}:
\batitle{Electrostatic screening of charged defects in monolayer mos2}.
\bjtitle{The Journal of Physical Chemistry Letters}
\bvolume{8},
\bfpage{2148}--\blpage{2152}
(\byear{2017})
\doiurl{10.1021/acs.jpclett.7b00710}
\end{barticle}
\endbibitem

\bibitem[\protect\citeauthoryear{Zhang et~al.}{2018}]{ZhangSiyuan2018}
\begin{barticle}
\bauthor{\bsnm{Zhang}, \binits{S.}},
\bauthor{\bsnm{Hill}, \binits{H.M.}},
\bauthor{\bsnm{Moudgil}, \binits{K.}},
\bauthor{\bsnm{Richter}, \binits{C.A.}},
\bauthor{\bsnm{Hight~Walker}, \binits{A.R.}},
\bauthor{\bsnm{Barlow}, \binits{S.}},
\bauthor{\bsnm{Marder}, \binits{S.R.}},
\bauthor{\bsnm{Hacker}, \binits{C.A.}},
\bauthor{\bsnm{Pookpanratana}, \binits{S.J.}}:
\batitle{Controllable, wide-ranging n-doping and p-doping of monolayer group 6
  transition-metal disulfides and diselenides}.
\bjtitle{Advanced Materials}
\bvolume{30},
\bfpage{1802991}
(\byear{2018})
\doiurl{10.1002/adma.201802991}
\end{barticle}
\endbibitem

\bibitem[\protect\citeauthoryear{Zhou et~al.}{2024}]{Xiang2024}
\begin{barticle}
\bauthor{\bsnm{Zhou}, \binits{X.}},
\bauthor{\bsnm{He}, \binits{W.}},
\bauthor{\bsnm{Liu}, \binits{C.}},
\bauthor{\bsnm{Zhang}, \binits{H.}},
\bauthor{\bsnm{Sun}, \binits{J.}},
\bauthor{\bsnm{Wang}, \binits{W.}}:
\batitle{Effects of defect creation and passivation on graphite friction under
  ultra-high vacuum conditions}.
\bjtitle{Carbon}
\bvolume{225},
\bfpage{119103}
(\byear{2024})
\doiurl{10.1016/j.carbon.2024.119103}
\end{barticle}
\endbibitem

\bibitem[\protect\citeauthoryear{Xu et~al.}{2023}]{Keliang_2023}
\begin{barticle}
\bauthor{\bsnm{Xu}, \binits{K.}},
\bauthor{\bsnm{Li}, \binits{P.}},
\bauthor{\bsnm{Wang}, \binits{S.}},
\bauthor{\bsnm{Ma}, \binits{J.}},
\bauthor{\bsnm{Xu}, \binits{H.}},
\bauthor{\bsnm{Liu}, \binits{Y.}}:
\batitle{Passivation of oxygen vacancy defects in conductive zno nanoparticles
  via low-temperature annealing in nf3}.
\bjtitle{Journal of Physics D: Applied Physics}
\bvolume{56},
\bfpage{085301}
(\byear{2023})
\doiurl{10.1088/1361-6463/acb4a5}
\end{barticle}
\endbibitem

\bibitem[\protect\citeauthoryear{Aradi et~al.}{2007}]{dftb_2007}
\begin{barticle}
\bauthor{\bsnm{Aradi}, \binits{B.}},
\bauthor{\bsnm{Hourahine}, \binits{B.}},
\bauthor{\bsnm{Frauenheim}, \binits{T.}}:
\batitle{Dftb+, a sparse matrix-based implementation of the dftb method}.
\bjtitle{The Journal of Physical Chemistry A}
\bvolume{111},
\bfpage{5678}--\blpage{5684}
(\byear{2007})
\doiurl{10.1021/jp070186p}
\end{barticle}
\endbibitem

\bibitem[\protect\citeauthoryear{Hourahine et~al.}{2020}]{dftb_2020}
\begin{barticle}
\bauthor{\bsnm{Hourahine}, \binits{B.}},
\bauthor{\bsnm{Aradi}, \binits{B.}},
\bauthor{\bsnm{Blum}, \binits{V.}},
\bauthor{\bsnm{Bonafé}, \binits{F.}},
\bauthor{\bsnm{Buccheri}, \binits{A.}},
\bauthor{\bsnm{Camacho}, \binits{C.}},
\bauthor{\bsnm{Cevallos}, \binits{C.}},
\bauthor{\bsnm{Deshaye}, \binits{M.Y.}},
\bauthor{\bsnm{Dumitrică}, \binits{T.}},
\bauthor{\bsnm{Dominguez}, \binits{A.}},
\bauthor{\bsnm{Ehlert}, \binits{S.}},
\bauthor{\bsnm{Elstner}, \binits{M.}},
\bauthor{\bsnm{Heide}, \binits{T.}},
\bauthor{\bsnm{Hermann}, \binits{J.}},
\bauthor{\bsnm{Irle}, \binits{S.}},
\bauthor{\bsnm{Kranz}, \binits{J.J.}},
\bauthor{\bsnm{Köhler}, \binits{C.}},
\bauthor{\bsnm{Kowalczyk}, \binits{T.}},
\bauthor{\bsnm{Kubař}, \binits{T.}},
\bauthor{\bsnm{Lee}, \binits{I.S.}},
\bauthor{\bsnm{Lutsker}, \binits{V.}},
\bauthor{\bsnm{Maurer}, \binits{R.J.}},
\bauthor{\bsnm{Min}, \binits{S.K.}},
\bauthor{\bsnm{Mitchell}, \binits{I.}},
\bauthor{\bsnm{Negre}, \binits{C.}},
\bauthor{\bsnm{Niehaus}, \binits{T.A.}},
\bauthor{\bsnm{Niklasson}, \binits{A.M.N.}},
\bauthor{\bsnm{Page}, \binits{A.J.}},
\bauthor{\bsnm{Pecchia}, \binits{A.}},
\bauthor{\bsnm{Penazzi}, \binits{G.}},
\bauthor{\bsnm{Persson}, \binits{M.P.}},
\bauthor{\bsnm{Řezáč}, \binits{J.}},
\bauthor{\bsnm{Sánchez}, \binits{C.G.}},
\bauthor{\bsnm{Sternberg}, \binits{M.}},
\bauthor{\bsnm{Stöhr}, \binits{M.}},
\bauthor{\bsnm{Stuckenberg}, \binits{F.}},
\bauthor{\bsnm{Tkatchenko}, \binits{A.}},
\bauthor{\bsnm{Yu}, \binits{V.W.-z.}},
\bauthor{\bsnm{Frauenheim}, \binits{T.}}:
\batitle{Dftb+, a software package for efficient approximate density functional
  theory based atomistic simulations}.
\bjtitle{The Journal of Chemical Physics}
\bvolume{152},
\bfpage{124101}
(\byear{2020})
\doiurl{10.1063/1.5143190}
\end{barticle}
\endbibitem

\bibitem[\protect\citeauthoryear{Grimme et~al.}{2017}]{grimme-xTB1_2017}
\begin{barticle}
\bauthor{\bsnm{Grimme}, \binits{S.}},
\bauthor{\bsnm{Bannwarth}, \binits{C.}},
\bauthor{\bsnm{Shushkov}, \binits{P.}}:
\batitle{A robust and accurate tight-binding quantum chemical method for
  structures, vibrational frequencies, and noncovalent interactions of large
  molecular systems parametrized for all spd-block elements (z = 1–86)}.
\bjtitle{Journal of Chemical Theory and Computation}
\bvolume{13},
\bfpage{1989}--\blpage{2009}
(\byear{2017})
\doiurl{10.1021/acs.jctc.7b00118}
\end{barticle}
\endbibitem

\bibitem[\protect\citeauthoryear{Bannwarth et~al.}{2021}]{grimme-xTB2_2021}
\begin{barticle}
\bauthor{\bsnm{Bannwarth}, \binits{C.}},
\bauthor{\bsnm{Caldeweyher}, \binits{E.}},
\bauthor{\bsnm{Ehlert}, \binits{S.}},
\bauthor{\bsnm{Hansen}, \binits{A.}},
\bauthor{\bsnm{Pracht}, \binits{P.}},
\bauthor{\bsnm{Seibert}, \binits{J.}},
\bauthor{\bsnm{Spicher}, \binits{S.}},
\bauthor{\bsnm{Grimme}, \binits{S.}}:
\batitle{Extended tight-binding quantum chemistry methods}.
\bjtitle{WIREs Computational Molecular Science}
\bvolume{11},
\bfpage{1493}
(\byear{2021})
\doiurl{10.1002/wcms.1493}
\end{barticle}
\endbibitem

\bibitem[\protect\citeauthoryear{Sakhraoui and Karlický}{2024}]{Taoufik2024}
\begin{barticle}
\bauthor{\bsnm{Sakhraoui}, \binits{T.}},
\bauthor{\bsnm{Karlický}, \binits{F.}}:
\batitle{Prediction of induced magnetism in 2d ti2c based mxenes by
  manipulating the mixed surface functionalization and metal substitution
  computed by xtb model hamiltonian of the dftb method}.
\bjtitle{Phys. Chem. Chem. Phys.}
\bvolume{26},
\bfpage{12862}--\blpage{12868}
(\byear{2024})
\doiurl{10.1039/D3CP05665A}
\end{barticle}
\endbibitem

\bibitem[\protect\citeauthoryear{Vicent-Luna et~al.}{2021}]{Vincent2021}
\begin{barticle}
\bauthor{\bsnm{Vicent-Luna}, \binits{J.M.}},
\bauthor{\bsnm{Apergi}, \binits{S.}},
\bauthor{\bsnm{Tao}, \binits{S.}}:
\batitle{Efficient computation of structural and electronic properties of
  halide perovskites using density functional tight binding: Gfn1-xtb method.}
\bjtitle{Journal of chemical information and modeling}
\bvolume{61},
\bfpage{4415}--\blpage{4424}
(\byear{2021})
\doiurl{10.1021/acs.jcim.1c00432}
\end{barticle}
\endbibitem

\bibitem[\protect\citeauthoryear{Nurhuda et~al.}{2022}]{Maryam2022}
\begin{barticle}
\bauthor{\bsnm{Nurhuda}, \binits{M.}},
\bauthor{\bsnm{Perry}, \binits{C.C.}},
\bauthor{\bsnm{Addicoat}, \binits{M.A.}}:
\batitle{Performance of gfn1-xtb for periodic optimization of metal organic
  frameworks}.
\bjtitle{Phys. Chem. Chem. Phys.}
\bvolume{24},
\bfpage{10906}--\blpage{10914}
(\byear{2022})
\doiurl{10.1039/D2CP00184E}
\end{barticle}
\endbibitem

\bibitem[\protect\citeauthoryear{Wei et~al.}{2024}]{Xinru_2024}
\begin{barticle}
\bauthor{\bsnm{Wei}, \binits{X.}},
\bauthor{\bsnm{Zhang}, \binits{G.}},
\bauthor{\bsnm{Lv}, \binits{Y.}},
\bauthor{\bsnm{Ma}, \binits{R.}},
\bauthor{\bsnm{Yang}, \binits{Y.}},
\bauthor{\bsnm{Wang}, \binits{F.}},
\bauthor{\bsnm{Hang}, \binits{L.}},
\bauthor{\bsnm{Kang}, \binits{B.}}:
\batitle{Unveiling bandgap behavior of chiral carbon nitride nanotubes}.
\bjtitle{Diamond and Related Materials}
\bvolume{146},
\bfpage{111244}
(\byear{2024})
\doiurl{10.1016/j.diamond.2024.111244}
\end{barticle}
\endbibitem

\bibitem[\protect\citeauthoryear{Zhang et~al.}{2023}]{Zhang_2023}
\begin{barticle}
\bauthor{\bsnm{Zhang}, \binits{G.}},
\bauthor{\bsnm{Lv}, \binits{Y.}},
\bauthor{\bsnm{Wei}, \binits{X.}},
\bauthor{\bsnm{Yuan}, \binits{Y.}},
\bauthor{\bsnm{Wang}, \binits{F.}},
\bauthor{\bsnm{Lee}, \binits{J.Y.}},
\bauthor{\bsnm{Kang}, \binits{B.}}:
\batitle{Band-gap behaviors of graphyne-/graphdiyne-derived chiral nanotubes}.
\bjtitle{The Journal of Physical Chemistry C}
\bvolume{127},
\bfpage{14982}--\blpage{14990}
(\byear{2023})
\doiurl{10.1021/acs.jpcc.3c03182}
\end{barticle}
\endbibitem

\bibitem[\protect\citeauthoryear{Sakhraoui and
  Karlický}{2022}]{Sakhraoui_2021}
\begin{barticle}
\bauthor{\bsnm{Sakhraoui}, \binits{T.}},
\bauthor{\bsnm{Karlický}, \binits{F.}}:
\batitle{Dftb investigations of the electronic and magnetic properties of
  fluorographene with vacancies and with adsorbed chemical groups}.
\bjtitle{Phys. Chem. Chem. Phys.}
\bvolume{24},
\bfpage{3312}--\blpage{3321}
(\byear{2022})
\doiurl{10.1039/D1CP00995H}
\end{barticle}
\endbibitem

\bibitem[\protect\citeauthoryear{Kang et~al.}{2017}]{Kang_2017}
\begin{barticle}
\bauthor{\bsnm{Kang}, \binits{B.}},
\bauthor{\bsnm{Cho}, \binits{D.}},
\bauthor{\bsnm{Lee}, \binits{J.Y.}}:
\batitle{Periodicity of band gaps of chiral $\alpha$-graphyne nanotubes}.
\bjtitle{Phys. Chem. Chem. Phys.}
\bvolume{19},
\bfpage{7919}--\blpage{7922}
(\byear{2017})
\doiurl{10.1039/C7CP00137A}
\end{barticle}
\endbibitem

\bibitem[\protect\citeauthoryear{Monkhorst and
  Pack}{1976}]{Monkhorst_Pack_1976}
\begin{barticle}
\bauthor{\bsnm{Monkhorst}, \binits{H.J.}},
\bauthor{\bsnm{Pack}, \binits{J.D.}}:
\batitle{Special points for brillouin-zone integrations}.
\bjtitle{Phys. Rev. B}
\bvolume{13},
\bfpage{5188}--\blpage{5192}
(\byear{1976})
\doiurl{10.1103/PhysRevB.13.5188}
\end{barticle}
\endbibitem

\bibitem[\protect\citeauthoryear{Aspera et~al.}{2010}]{Aspera_2010}
\begin{barticle}
\bauthor{\bsnm{Aspera}, \binits{S.M.}},
\bauthor{\bsnm{David}, \binits{M.}},
\bauthor{\bsnm{Kasai}, \binits{H.}}:
\batitle{First-principles study of the adsorption of water on
  tri-s-triazine-based graphitic carbon nitride}.
\bjtitle{Japanese Journal of Applied Physics}
\bvolume{49},
\bfpage{115703}
(\byear{2010})
\doiurl{10.1143/JJAP.49.115703}
\end{barticle}
\endbibitem

\bibitem[\protect\citeauthoryear{Bafekry et~al.}{2019}]{Bafekry2019}
\begin{barticle}
\bauthor{\bsnm{Bafekry}, \binits{A.}},
\bauthor{\bsnm{Shayesteh}, \binits{S.F.}},
\bauthor{\bsnm{Peeters}, \binits{F.M.}}:
\batitle{Two-dimensional carbon nitride (2dcn) nanosheets: Tuning of novel
  electronic and magnetic properties by hydrogenation, atom substitution and
  defect engineering}.
\bjtitle{Journal of Applied Physics}
\bvolume{126},
\bfpage{215104}
(\byear{2019})
\doiurl{10.1063/1.5120525}
\end{barticle}
\endbibitem

\bibitem[\protect\citeauthoryear{Wang et~al.}{2014}]{Wang2014}
\begin{barticle}
\bauthor{\bsnm{Wang}, \binits{J.}},
\bauthor{\bsnm{Guan}, \binits{Z.}},
\bauthor{\bsnm{Huang}, \binits{J.}},
\bauthor{\bsnm{Li}, \binits{Q.}},
\bauthor{\bsnm{Yang}, \binits{J.}}:
\batitle{Enhanced photocatalytic mechanism for the hybrid g-c3n4/mos2
  nanocomposite}.
\bjtitle{J. Mater. Chem. A}
\bvolume{2},
\bfpage{7960}--\blpage{7966}
(\byear{2014})
\doiurl{10.1039/C4TA00275J}
\end{barticle}
\endbibitem

\bibitem[\protect\citeauthoryear{Zhao et~al.}{2019}]{Yali2019}
\begin{barticle}
\bauthor{\bsnm{Zhao}, \binits{Y.}},
\bauthor{\bsnm{Lin}, \binits{Y.}},
\bauthor{\bsnm{Wang}, \binits{G.}},
\bauthor{\bsnm{Jiang}, \binits{Z.}},
\bauthor{\bsnm{Zhang}, \binits{R.}},
\bauthor{\bsnm{Zhu}, \binits{C.}}:
\batitle{Photocatalytic water splitting of (f, ti) codoped heptazine/triazine
  based g-c3n4 heterostructure: A hybrid dft study}.
\bjtitle{Applied Surface Science}
\bvolume{463},
\bfpage{809}--\blpage{819}
(\byear{2019})
\doiurl{10.1016/j.apsusc.2018.08.013}
\end{barticle}
\endbibitem

\bibitem[\protect\citeauthoryear{Liu}{2015}]{Liu2015}
\begin{barticle}
\bauthor{\bsnm{Liu}, \binits{J.}}:
\batitle{Origin of high photocatalytic efficiency in monolayer g-c3n4/cds
  heterostructure: A hybrid dft study}.
\bjtitle{J. Phys. Chem. C}
\bvolume{119},
\bfpage{28417}--\blpage{28423}
(\byear{2015})
\doiurl{10.1021/acs.jpcc.5b09092}
\end{barticle}
\endbibitem

\bibitem[\protect\citeauthoryear{Mane et~al.}{2022}]{Pratap2022}
\begin{barticle}
\bauthor{\bsnm{Mane}, \binits{P.}},
\bauthor{\bsnm{Vaidyanathan}, \binits{A.}},
\bauthor{\bsnm{Chakraborty}, \binits{B.}}:
\batitle{Graphitic carbon nitride (g-c3n4) decorated with yttrium as potential
  hydrogen storage material: Acumen from quantum simulations}.
\bjtitle{International Journal of Hydrogen Energy}
\bvolume{47},
\bfpage{41898}--\blpage{41910}
(\byear{2022})
\doiurl{10.1016/j.ijhydene.2022.04.184}
\end{barticle}
\endbibitem

\bibitem[\protect\citeauthoryear{Wang et~al.}{2009}]{Wang2009}
\begin{barticle}
\bauthor{\bsnm{Wang}, \binits{X.}},
\bauthor{\bsnm{Maeda}, \binits{K.}},
\bauthor{\bsnm{Thomas}, \binits{A.}},
\bauthor{\bsnm{Takanabe}, \binits{K.}},
\bauthor{\bsnm{Xin}, \binits{G.}},
\bauthor{\bsnm{Carlsson}, \binits{J.M.}},
\bauthor{\bsnm{Domen}, \binits{K.}},
\bauthor{\bsnm{Antonietti}, \binits{M.}}:
\batitle{A metal-free polymeric photocatalyst for hydrogen production from
  water under visible light}.
\bjtitle{Nature Materials}
\bvolume{8},
\bfpage{76}--\blpage{80}
(\byear{2009})
\doiurl{10.1038/nmat2317}
\end{barticle}
\endbibitem

\bibitem[\protect\citeauthoryear{Re~Fiorentin et~al.}{2021}]{Fiorentin2021}
\begin{barticle}
\bauthor{\bsnm{Re~Fiorentin}, \binits{M.}},
\bauthor{\bsnm{Risplendi}, \binits{F.}},
\bauthor{\bsnm{Palummo}, \binits{M.}},
\bauthor{\bsnm{Cicero}, \binits{G.}}:
\batitle{First-principles calculations of exciton radiative lifetimes in
  monolayer graphitic carbon nitride nanosheets: Implications for
  photocatalysis}.
\bjtitle{ACS Applied Nano Materials}
\bvolume{4},
\bfpage{1985}--\blpage{1993}
(\byear{2021})
\doiurl{10.1021/acsanm.0c03317}
\end{barticle}
\endbibitem

\bibitem[\protect\citeauthoryear{Liu et~al.}{2016}]{Liu2016}
\begin{barticle}
\bauthor{\bsnm{Liu}, \binits{J.}},
\bauthor{\bsnm{Cheng}, \binits{B.}},
\bauthor{\bsnm{Yu}, \binits{J.}}:
\batitle{A new understanding of the photocatalytic mechanism of the direct
  z-scheme g-c3n4/tio2 heterostructure}.
\bjtitle{Phys. Chem. Chem. Phys.}
\bvolume{18},
\bfpage{31175}--\blpage{31183}
(\byear{2016})
\doiurl{10.1039/C6CP06147H}
\end{barticle}
\endbibitem

\bibitem[\protect\citeauthoryear{Han et~al.}{2017}]{Han2017}
\begin{barticle}
\bauthor{\bsnm{Han}, \binits{Q.}},
\bauthor{\bsnm{Chen}, \binits{N.}},
\bauthor{\bsnm{Zhang}, \binits{J.}},
\bauthor{\bsnm{Qu}, \binits{L.}}:
\batitle{Graphene/graphitic carbon nitride hybrids for catalysis}.
\bjtitle{Mater. Horiz.}
\bvolume{4},
\bfpage{832}--\blpage{850}
(\byear{2017})
\doiurl{10.1039/C7MH00379J}
\end{barticle}
\endbibitem

\bibitem[\protect\citeauthoryear{Zhang et~al.}{2018}]{Zhang2018}
\begin{barticle}
\bauthor{\bsnm{Zhang}, \binits{N.}},
\bauthor{\bsnm{Hong}, \binits{Y.}},
\bauthor{\bsnm{Yazdanparast}, \binits{S.}},
\bauthor{\bsnm{Zaeem}, \binits{M.A.}}:
\batitle{Superior structural, elastic and electronic properties of 2d titanium
  nitride mxenes over carbide mxenes: a comprehensive first principles study}.
\bjtitle{2D Materials}
\bvolume{5},
\bfpage{045004}
(\byear{2018})
\doiurl{10.1088/2053-1583/aacfb3}
\end{barticle}
\endbibitem

\bibitem[\protect\citeauthoryear{Shein and Ivanovskii}{2012}]{Shein2012}
\begin{barticle}
\bauthor{\bsnm{Shein}, \binits{I.R.}},
\bauthor{\bsnm{Ivanovskii}, \binits{A.L.}}:
\batitle{Graphene-like titanium carbides and nitrides tin+1cn, tin+1nn (n=1, 2,
  and 3) from de-intercalated max phases: First-principles probing of their
  structural, electronic properties and relative stability}.
\bjtitle{Computational Materials Science}
\bvolume{65},
\bfpage{104}--\blpage{114}
(\byear{2012})
\doiurl{10.1016/j.commatsci.2012.07.011}
\end{barticle}
\endbibitem

\bibitem[\protect\citeauthoryear{Xu et~al.}{2019}]{Xu_2019}
\begin{barticle}
\bauthor{\bsnm{Xu}, \binits{L.}},
\bauthor{\bsnm{Li}, \binits{Q.}},
\bauthor{\bsnm{Li}, \binits{X.-F.}},
\bauthor{\bsnm{Long}, \binits{M.-Q.}},
\bauthor{\bsnm{Chen}, \binits{T.}},
\bauthor{\bsnm{Peng}, \binits{B.}},
\bauthor{\bsnm{Wang}, \binits{L.-L.}},
\bauthor{\bsnm{Yang}, \binits{Y.}},
\bauthor{\bsnm{Shuai}, \binits{C.}}:
\batitle{Rationally designed 2d/2d sic/g-c3n4 photocatalysts for hydrogen
  production}.
\bjtitle{Catal. Sci. Technol.}
\bvolume{9},
\bfpage{3896}--\blpage{3906}
(\byear{2019})
\doiurl{10.1039/C9CY00329K}
\end{barticle}
\endbibitem

\bibitem[\protect\citeauthoryear{Bafekry et~al.}{2020}]{Bafekry_2020}
\begin{barticle}
\bauthor{\bsnm{Bafekry}, \binits{A.}},
\bauthor{\bsnm{Stampfl}, \binits{C.}},
\bauthor{\bsnm{Akgenc}, \binits{B.}},
\bauthor{\bsnm{Ghergherehchi}, \binits{M.}}:
\batitle{Control of c3n4 and c4n3 carbon nitride nanosheets’ electronic and
  magnetic properties through embedded atoms}.
\bjtitle{Phys. Chem. Chem. Phys.}
\bvolume{22},
\bfpage{2249}--\blpage{2261}
(\byear{2020})
\doiurl{10.1039/C9CP06031F}
\end{barticle}
\endbibitem

\bibitem[\protect\citeauthoryear{Rangraz et~al.}{2021}]{Rangraz_2021}
\begin{barticle}
\bauthor{\bsnm{Rangraz}, \binits{Y.}},
\bauthor{\bsnm{Heravi}, \binits{M.M.}},
\bauthor{\bsnm{Elhampour}, \binits{A.}}:
\batitle{Recent advances on heteroatom-doped porous carbon/metal materials:
  Fascinating heterogeneous catalysts for organic transformations}.
\bjtitle{The Chemical Record}
\bvolume{21},
\bfpage{1985}--\blpage{2073}
(\byear{2021})
\doiurl{10.1002/tcr.202100124}
\end{barticle}
\endbibitem

\bibitem[\protect\citeauthoryear{Praus et~al.}{2022}]{Praus2022}
\begin{barticle}
\bauthor{\bsnm{Praus}, \binits{P.}},
\bauthor{\bsnm{Řeháčková}, \binits{L.}},
\bauthor{\bsnm{Čížek}, \binits{J.}},
\bauthor{\bsnm{Smýkalová}, \binits{A.}},
\bauthor{\bsnm{Koštejn}, \binits{M.}},
\bauthor{\bsnm{Pavlovský}, \binits{J.}},
\bauthor{\bsnm{Filip~Edelmannová}, \binits{M.}},
\bauthor{\bsnm{Kočí}, \binits{K.}}:
\batitle{Synthesis of vacant graphitic carbon nitride in argon atmosphere and
  its utilization for photocatalytic hydrogen generation}.
\bjtitle{Scientific Reports}
\bvolume{12},
\bfpage{13622}
(\byear{2022})
\doiurl{10.1038/s41598-022-17940-3}
\end{barticle}
\endbibitem

\bibitem[\protect\citeauthoryear{Lee et~al.}{2016}]{Lee2016}
\begin{barticle}
\bauthor{\bsnm{Lee}, \binits{J.-U.}},
\bauthor{\bsnm{Lee}, \binits{S.}},
\bauthor{\bsnm{Ryoo}, \binits{J.H.}},
\bauthor{\bsnm{Kang}, \binits{S.}},
\bauthor{\bsnm{Kim}, \binits{T.Y.}},
\bauthor{\bsnm{Kim}, \binits{P.}},
\bauthor{\bsnm{Park}, \binits{C.-H.}},
\bauthor{\bsnm{Park}, \binits{J.-G.}},
\bauthor{\bsnm{Cheong}, \binits{H.}}:
\batitle{Ising-type magnetic ordering in atomically thin feps3}.
\bjtitle{Nano Letters}
\bvolume{16},
\bfpage{7433}--\blpage{7438}
(\byear{2016})
\doiurl{10.1021/acs.nanolett.6b03052}
\end{barticle}
\endbibitem

\bibitem[\protect\citeauthoryear{Huang et~al.}{2017}]{Huang2017}
\begin{barticle}
\bauthor{\bsnm{Huang}, \binits{B.}},
\bauthor{\bsnm{Clark}, \binits{G.}},
\bauthor{\bsnm{Navarro-Moratalla}, \binits{E.}},
\bauthor{\bsnm{Klein}, \binits{D.R.}},
\bauthor{\bsnm{Cheng}, \binits{R.}},
\bauthor{\bsnm{Seyler}, \binits{K.L.}},
\bauthor{\bsnm{Zhong}, \binits{D.}},
\bauthor{\bsnm{Schmidgall}, \binits{E.}},
\bauthor{\bsnm{McGuire}, \binits{M.A.}},
\bauthor{\bsnm{Cobden}, \binits{D.H.}},
\bauthor{\bsnm{Yao}, \binits{W.}},
\bauthor{\bsnm{Xiao}, \binits{D.}},
\bauthor{\bsnm{Jarillo-Herrero}, \binits{P.}},
\bauthor{\bsnm{Xu}, \binits{X.}}:
\batitle{Layer-dependent ferromagnetism in a van der waals crystal down to the
  monolayer limit}.
\bjtitle{Nature}
\bvolume{546},
\bfpage{270}--\blpage{273}
(\byear{2017})
\doiurl{10.1038/nature22391}
\end{barticle}
\endbibitem

\bibitem[\protect\citeauthoryear{Liu et~al.}{2018}]{Zhong-Liu2018}
\begin{barticle}
\bauthor{\bsnm{Liu}, \binits{Z.-L.}},
\bauthor{\bsnm{Wu}, \binits{X.}},
\bauthor{\bsnm{Shao}, \binits{Y.}},
\bauthor{\bsnm{Qi}, \binits{J.}},
\bauthor{\bsnm{Cao}, \binits{Y.}},
\bauthor{\bsnm{Huang}, \binits{L.}},
\bauthor{\bsnm{Liu}, \binits{C.}},
\bauthor{\bsnm{Wang}, \binits{J.-O.}},
\bauthor{\bsnm{Zheng}, \binits{Q.}},
\bauthor{\bsnm{Zhu}, \binits{Z.-L.}},
\bauthor{\bsnm{Ibrahim}, \binits{K.}},
\bauthor{\bsnm{Wang}, \binits{Y.-L.}},
\bauthor{\bsnm{Gao}, \binits{H.-J.}}:
\batitle{Epitaxially grown monolayer vse2: an air-stable magnetic
  two-dimensional material with low work function at edges}.
\bjtitle{Science Bulletin}
\bvolume{63},
\bfpage{419}--\blpage{425}
(\byear{2018})
\doiurl{10.1016/j.scib.2018.03.008}
\end{barticle}
\endbibitem

\bibitem[\protect\citeauthoryear{Gong et~al.}{2017}]{Gong2017}
\begin{barticle}
\bauthor{\bsnm{Gong}, \binits{C.}},
\bauthor{\bsnm{Li}, \binits{L.}},
\bauthor{\bsnm{Li}, \binits{Z.}},
\bauthor{\bsnm{Ji}, \binits{H.}},
\bauthor{\bsnm{Stern}, \binits{A.}},
\bauthor{\bsnm{Xia}, \binits{Y.}},
\bauthor{\bsnm{Cao}, \binits{T.}},
\bauthor{\bsnm{Bao}, \binits{W.}},
\bauthor{\bsnm{Wang}, \binits{C.}},
\bauthor{\bsnm{Wang}, \binits{Y.}},
\bauthor{\bsnm{Qiu}, \binits{Z.Q.}},
\bauthor{\bsnm{Cava}, \binits{R.J.}},
\bauthor{\bsnm{Louie}, \binits{S.G.}},
\bauthor{\bsnm{Xia}, \binits{J.}},
\bauthor{\bsnm{Zhang}, \binits{X.}}:
\batitle{Discovery of intrinsic ferromagnetism in two-dimensional van der waals
  crystals}.
\bjtitle{Nature}
\bvolume{546},
\bfpage{265}--\blpage{269}
(\byear{2017})
\doiurl{10.1038/nature22060}
\end{barticle}
\endbibitem

\bibitem[\protect\citeauthoryear{Fei et~al.}{2018}]{Fei2018}
\begin{barticle}
\bauthor{\bsnm{Fei}, \binits{Z.}},
\bauthor{\bsnm{Huang}, \binits{B.}},
\bauthor{\bsnm{Malinowski}, \binits{P.}},
\bauthor{\bsnm{Wang}, \binits{W.}},
\bauthor{\bsnm{Song}, \binits{T.}},
\bauthor{\bsnm{Sanchez}, \binits{J.}},
\bauthor{\bsnm{Yao}, \binits{W.}},
\bauthor{\bsnm{Xiao}, \binits{D.}},
\bauthor{\bsnm{Zhu}, \binits{X.}},
\bauthor{\bsnm{May}, \binits{A.F.}},
\bauthor{\bsnm{Wu}, \binits{W.}},
\bauthor{\bsnm{Cobden}, \binits{D.H.}},
\bauthor{\bsnm{Chu}, \binits{J.-H.}},
\bauthor{\bsnm{Xu}, \binits{X.}}:
\batitle{Two-dimensional itinerant ferromagnetism in atomically thin fe3gete2}.
\bjtitle{Nature Materials}
\bvolume{17},
\bfpage{778}--\blpage{782}
(\byear{2018})
\doiurl{10.1038/s41563-018-0149-7}
\end{barticle}
\endbibitem

\bibitem[\protect\citeauthoryear{Tomar et~al.}{2019}]{Shalini2019}
\begin{barticle}
\bauthor{\bsnm{Tomar}, \binits{S.}},
\bauthor{\bsnm{Ghosh}, \binits{B.}},
\bauthor{\bsnm{Mardanya}, \binits{S.}},
\bauthor{\bsnm{Rastogi}, \binits{P.}},
\bauthor{\bsnm{Bhadoria}, \binits{B.S.}},
\bauthor{\bsnm{Chauhan}, \binits{Y.S.}},
\bauthor{\bsnm{Agarwal}, \binits{A.}},
\bauthor{\bsnm{Bhowmick}, \binits{S.}}:
\batitle{Intrinsic magnetism in monolayer transition metal trihalides: A
  comparative study}.
\bjtitle{Journal of Magnetism and Magnetic Materials}
\bvolume{489},
\bfpage{165384}
(\byear{2019})
\doiurl{10.1016/j.jmmm.2019.165384}
\end{barticle}
\endbibitem

\bibitem[\protect\citeauthoryear{Gao et~al.}{2016}]{Gao2016}
\begin{barticle}
\bauthor{\bsnm{Gao}, \binits{G.}},
\bauthor{\bsnm{Ding}, \binits{G.}},
\bauthor{\bsnm{Li}, \binits{J.}},
\bauthor{\bsnm{Yao}, \binits{K.}},
\bauthor{\bsnm{Wu}, \binits{M.}},
\bauthor{\bsnm{Qian}, \binits{M.}}:
\batitle{Monolayer mxenes: promising half-metals and spin gapless
  semiconductors}.
\bjtitle{Nanoscale}
\bvolume{8},
\bfpage{8986}--\blpage{8994}
(\byear{2016})
\doiurl{10.1039/C6NR01333C}
\end{barticle}
\endbibitem

\bibitem[\protect\citeauthoryear{Jiang et~al.}{2021}]{Jiang2021}
\begin{barticle}
\bauthor{\bsnm{Jiang}, \binits{X.}},
\bauthor{\bsnm{Liu}, \binits{Q.}},
\bauthor{\bsnm{Xing}, \binits{J.}},
\bauthor{\bsnm{Liu}, \binits{N.}},
\bauthor{\bsnm{Guo}, \binits{Y.}},
\bauthor{\bsnm{Liu}, \binits{Z.}},
\bauthor{\bsnm{Zhao}, \binits{J.}}:
\batitle{Recent progress on 2d magnets: Fundamental mechanism, structural
  design and modification}.
\bjtitle{Applied Physics Reviews}
\bvolume{8},
\bfpage{031305}
(\byear{2021})
\doiurl{10.1063/5.0039979}
\end{barticle}
\endbibitem

\bibitem[\protect\citeauthoryear{Wang et~al.}{2009}]{XWang_2009}
\begin{barticle}
\bauthor{\bsnm{Wang}, \binits{X.}},
\bauthor{\bsnm{Chen}, \binits{X.}},
\bauthor{\bsnm{Thomas}, \binits{A.}},
\bauthor{\bsnm{Fu}, \binits{X.}},
\bauthor{\bsnm{Antonietti}, \binits{M.}}:
\batitle{Metal-containing carbon nitride compounds: A new functional
  organic–metal hybrid material}.
\bjtitle{Advanced Materials}
\bvolume{21},
\bfpage{1609}--\blpage{1612}
(\byear{2009})
\doiurl{10.1002/adma.200802627}
\end{barticle}
\endbibitem

\bibitem[\protect\citeauthoryear{Zhou and Lin}{2018}]{Yungang_2018}
\begin{barticle}
\bauthor{\bsnm{Zhou}, \binits{Y.}},
\bauthor{\bsnm{Lin}, \binits{X.}}:
\batitle{Effects of interstitial dopings of 3d transition metal atoms on
  antimonene: A first-principles study}.
\bjtitle{Applied Surface Science}
\bvolume{458},
\bfpage{572}--\blpage{579}
(\byear{2018})
\doiurl{10.1016/j.apsusc.2018.07.126}
\end{barticle}
\endbibitem

\bibitem[\protect\citeauthoryear{Liu et~al.}{2018}]{Liu_2018}
\begin{barticle}
\bauthor{\bsnm{Liu}, \binits{M.-Y.}},
\bauthor{\bsnm{Chen}, \binits{Q.-Y.}},
\bauthor{\bsnm{Huang}, \binits{Y.}},
\bauthor{\bsnm{Li}, \binits{Z.-Y.}},
\bauthor{\bsnm{Cao}, \binits{C.}},
\bauthor{\bsnm{He}, \binits{Y.}}:
\batitle{Electronic and magnetic properties of 3d transition-metal atom
  adsorbed arsenene}.
\bjtitle{Nanotechnology}
\bvolume{29},
\bfpage{095203}
(\byear{2018})
\doiurl{10.1088/1361-6528/aaa684}
\end{barticle}
\endbibitem

\bibitem[\protect\citeauthoryear{Kaloni}{2014}]{Kaloni_2014}
\begin{barticle}
\bauthor{\bsnm{Kaloni}, \binits{T.P.}}:
\batitle{Tuning the structural, electronic, and magnetic properties of
  germanene by the adsorption of 3d transition metal atoms}.
\bjtitle{The Journal of Physical Chemistry C}
\bvolume{118},
\bfpage{25200}--\blpage{25208}
(\byear{2014})
\doiurl{10.1021/jp5058644}
\end{barticle}
\endbibitem

\bibitem[\protect\citeauthoryear{Sun et~al.}{2017}]{Minglei_2017}
\begin{barticle}
\bauthor{\bsnm{Sun}, \binits{M.}},
\bauthor{\bsnm{Ren}, \binits{Q.}},
\bauthor{\bsnm{Zhao}, \binits{Y.}},
\bauthor{\bsnm{Chou}, \binits{J.-P.}},
\bauthor{\bsnm{Yu}, \binits{J.}},
\bauthor{\bsnm{Tang}, \binits{W.}}:
\batitle{Electronic and magnetic properties of 4d series transition metal
  substituted graphene: A first-principles study}.
\bjtitle{Carbon}
\bvolume{120},
\bfpage{265}--\blpage{273}
(\byear{2017})
\doiurl{10.1016/j.carbon.2017.04.060}
\end{barticle}
\endbibitem

\bibitem[\protect\citeauthoryear{Sakhraoui and Bouzid}{2022}]{SAKHRAOUI2022}
\begin{barticle}
\bauthor{\bsnm{Sakhraoui}, \binits{T.}},
\bauthor{\bsnm{Bouzid}, \binits{A.}}:
\batitle{Induced ferromagnetism on late transition metals adsorbed on antimony
  arsenide monolayer from first-principles}.
\bjtitle{Journal of Magnetism and Magnetic Materials}
\bvolume{545},
\bfpage{168658}
(\byear{2022})
\doiurl{10.1016/j.jmmm.2021.168658}
\end{barticle}
\endbibitem

\bibitem[\protect\citeauthoryear{{Isa khan} et~al.}{2024}]{Muhammad2024}
\begin{barticle}
\bauthor{\bsnm{{Isa khan}}, \binits{M.}},
\bauthor{\bsnm{khalid}, \binits{S.}},
\bauthor{\bsnm{Majid}, \binits{A.}},
\bauthor{\bsnm{Alarfaji}, \binits{S.S.}}:
\batitle{Empowering spintronics performance of 3d transition metal adsorbed
  b4c3 monolayer: A dft outlook}.
\bjtitle{Journal of Physics and Chemistry of Solids}
\bvolume{187},
\bfpage{111599}
(\byear{2024})
\doiurl{10.1016/j.jpcs.2023.111599}
\end{barticle}
\endbibitem

\bibitem[\protect\citeauthoryear{Zhao et~al.}{2024}]{Wenyu2024}
\begin{barticle}
\bauthor{\bsnm{Zhao}, \binits{W.}},
\bauthor{\bsnm{Huang}, \binits{H.}},
\bauthor{\bsnm{Yang}, \binits{M.}},
\bauthor{\bsnm{Hu}, \binits{Y.}}:
\batitle{Adsorption behaviour of transition metal atoms on pristine and
  defective two-dimensional mgal2s4 monolayer}.
\bjtitle{Materials Science in Semiconductor Processing}
\bvolume{184},
\bfpage{108816}
(\byear{2024})
\doiurl{10.1016/j.mssp.2024.108816}
\end{barticle}
\endbibitem

\bibitem[\protect\citeauthoryear{Gebredingle et~al.}{2024}]{Gebredingle2024}
\begin{barticle}
\bauthor{\bsnm{Gebredingle}, \binits{Y.}},
\bauthor{\bsnm{Kim}, \binits{H.}},
\bauthor{\bsnm{Kim}, \binits{N.}}:
\batitle{Defect states in magnetically “seasoned” wsse - adsorption and
  doping effects of magnetic transition metals x = v, cr, mn, fe, co – a
  comprehensive first-principles study}.
\bjtitle{Scientific Reports}
\bvolume{14},
\bfpage{29271}
(\byear{2024})
\doiurl{10.1038/s41598-024-77938-x}
\end{barticle}
\endbibitem

\bibitem[\protect\citeauthoryear{Majumder et~al.}{2019}]{Majumder2019}
\begin{barticle}
\bauthor{\bsnm{Majumder}, \binits{C.}},
\bauthor{\bsnm{Bhattacharya}, \binits{S.}},
\bauthor{\bsnm{Saha}, \binits{S.K.}}:
\batitle{Anomalous large negative magnetoresistance in transition-metal
  decorated graphene: Evidence for electron-hole puddles}.
\bjtitle{Phys. Rev. B}
\bvolume{99},
\bfpage{045408}
(\byear{2019})
\doiurl{10.1103/PhysRevB.99.045408}
\end{barticle}
\endbibitem

\bibitem[\protect\citeauthoryear{Liu et~al.}{2011}]{Liu2011}
\begin{barticle}
\bauthor{\bsnm{Liu}, \binits{X.}},
\bauthor{\bsnm{Wang}, \binits{C.Z.}},
\bauthor{\bsnm{Yao}, \binits{Y.X.}},
\bauthor{\bsnm{Lu}, \binits{W.C.}},
\bauthor{\bsnm{Hupalo}, \binits{M.}},
\bauthor{\bsnm{Tringides}, \binits{M.C.}},
\bauthor{\bsnm{Ho}, \binits{K.M.}}:
\batitle{Bonding and charge transfer by metal adatom adsorption on graphene}.
\bjtitle{Phys. Rev. B}
\bvolume{83},
\bfpage{235411}
(\byear{2011})
\doiurl{10.1103/PhysRevB.83.235411}
\end{barticle}
\endbibitem

\end{thebibliography}

\end{document}